\documentclass[useAMS,referee]{biom}
\def\bSig\mathbf{\Sigma}

\usepackage{natbib}
\usepackage{amsmath}
\usepackage{amssymb}
\usepackage{graphicx}
\usepackage{tikz}
\usepackage{subfigure}
\usepackage{comment}

\title[Mediation with External Summary Statistic Information]{Mediation with External Summary Statistic Information (MESSI)}

\author{Jonathan Boss$^{1,*}$\email{bossjona@umich.edu}, Wei Hao$^{1}$,
Amber Cathey$^{2}$, Barrett M. Welch$^{3,4}$, Kelly K. Ferguson$^{4}$, \\ \textbf{John D. Meeker$^{2}$, Jian Kang$^{1,**}$\email{jiankang@umich.edu}, and Bhramar Mukherjee$^{1,5,***}$\email{bhramar@umich.edu}}\\
$^{1}$Department of Biostatistics, University of Michigan, Ann Arbor, Michigan, U.S.A. \\
$^{2}$Department of Environmental Health Sciences, University of Michigan, Ann Arbor, Michigan, U.S.A. \\
$^{3}$School of Public Health, University of Nevada, Reno, Nevada, U.S.A. \\
$^{4}$Epidemiology Branch, National Institute of Environmental Health Sciences, Durham, North Carolina, U.S.A. \\
$^{5}$Department of Epidemiology, University of Michigan, Ann Arbor, Michigan, U.S.A.}

\begin{document}




\pagerange{\pageref{firstpage}--\pageref{lastpage}} 
\volume{XXX}
\pubyear{XXX}
\artmonth{XXX}


\doi{XXX}


\label{firstpage}


\begin{abstract}
Environmental health studies are increasingly measuring endogenous omics data ($\boldsymbol{M}$) to study intermediary biological pathways by which an exogenous exposure ($\boldsymbol{A}$) affects a health outcome ($\boldsymbol{Y}$), given confounders ($\boldsymbol{C}$). Mediation analysis is frequently carried out to understand such mechanisms. If intermediary pathways are of interest, then there is likely literature establishing statistical and biological significance of the total effect, defined as the effect of $\boldsymbol{A}$ on $\boldsymbol{Y}$ given $\boldsymbol{C}$. For mediation models with continuous outcomes and mediators, we show that leveraging external summary-level information on the total effect improves estimation efficiency of the natural direct and indirect effects. Moreover, the efficiency gain depends on the asymptotic partial $R^2$ between the outcome ($\boldsymbol{Y}\mid\boldsymbol{M},\boldsymbol{A},\boldsymbol{C}$) and total effect ($\boldsymbol{Y}\mid\boldsymbol{A},\boldsymbol{C}$) models, with smaller (larger) values benefiting direct (indirect) effect estimation. We robustify our estimation procedure to incongenial external information by assuming the total effect follows a random distribution. This framework allows shrinkage towards the external information if the total effects in the internal and external populations agree. We illustrate our methodology using data from the Puerto Rico Testsite for Exploring Contamination Threats, where Cytochrome p450 metabolites are hypothesized to mediate the effect of phthalate exposure on gestational age at delivery. External information on the total effect comes from a recently published pooled analysis of 16 studies. The proposed framework blends mediation analysis with emerging data integration techniques.
\end{abstract}

%



\maketitle


\section{Introduction}
\label{s:intro}


Mediation analysis is an important tool in epidemiology to elucidate the intermediary pathways by which an exposure affects an outcome \citep{baron1986, robins1992, pearl2001, vanderWeele2015, song2020}. In mediation analysis, the total effect (TE) characterizes the effect of the exposure on the outcome and is additively decomposed into the natural direct effect (NDE) and the natural indirect effect (NIE). The NDE and NIE quantify how well measures of the intermediary pathways, called mediators, explain the TE. The logical progression of mediation analysis is generally sequential, where researchers first establish that the exposure is causally related to the outcome, and then hypothesize mechanisms that may explain the causal relationship. Consequently, researchers frequently consider mediation hypotheses only if there is a well-established literature showing statistical and biological significance of the TE. The objective of this paper is to integrate available external summary-level information on the TE into mediation models, thereby improving NDE and NIE estimation for mediation analyses with individual level omics data on a limited number of participants.

The motivating example comes from the Puerto Rico Testsite for Exploring Contamination Threats (PROTECT), a prospective birth cohort study in Puerto Rico. Preterm births, defined as gestational age at delivery of less than 37 weeks, coupled with their downstream health complications, are a large concern for the Puerto Rican health care system. One widely studied risk factor for preterm deliveries is elevated exposure to a class of endocrine disrupting chemicals called phthalates \citep{ferguson2014, welch2022}. The goal of the present study is to test whether metabolites corresponding to the inflammatory pathway Cytochrome p450 ($\boldsymbol{M}$) mediate the relationship between phthalate exposure ($\boldsymbol{A}$) and gestational age at delivery ($\boldsymbol{Y}$) adjusted for confounders $(\boldsymbol{C})$. The sample size in PROTECT with exposure and mediator data is approximately 450. However, a study by \cite{welch2022}, which pools data corresponding to the TE of phthalates on birth outcomes across 16 studies, has an approximate sample size of 5000 (after omitting PROTECT). The goal of this paper is to utilize the external summary-level information on the TE from the \cite{welch2022} pooled study to improve estimation efficiency of the NDE and NIE in PROTECT.

There is no existing work explicitly incorporating external summary-level information on the TE into an internal mediation model, however there is related work on nested internal and external models in the data integration literature \citep{chatterjee2016, cheng2018, estes2018, cheng2019, gu2019, han2019, gu2021, zhai2022}. Specifically, these papers consider the situation where an external model or prediction algorithm is fit on a set of predictors and the resulting summary-level statistics or predictions are then used to inform an internal model that contains a proper superset of those predictors. External information on the TE can partially be framed in a similar manner where the external information comes from the TE model, $\boldsymbol{Y} \mid \boldsymbol{A}, \boldsymbol{C}$, and the model of interest is $\boldsymbol{Y} \mid \boldsymbol{M}, \boldsymbol{A}, \boldsymbol{C}$. However, the key difference for mediation models is that mediation models are specified from $\boldsymbol{M} \mid \boldsymbol{A}, \boldsymbol{C}$ and $\boldsymbol{Y} \mid \boldsymbol{M}, \boldsymbol{A}, \boldsymbol{C}$ models. Therefore, it is important to understand how information on $\boldsymbol{Y} \mid \boldsymbol{A}, \boldsymbol{C}$ informs parameter estimation corresponding to $\boldsymbol{Y} \mid \boldsymbol{M}, \boldsymbol{A}, \boldsymbol{C}$ and $\boldsymbol{M} \mid \boldsymbol{A}, \boldsymbol{C}$ simultaneously.

Our work has several new aspects. First, we develop a method to integrate external summary-level information on the TE into an internal mediation model through constrained maximum likelihood estimation. Second, we show that, for a continuous outcome and continuous candidate mediators, the constrained estimator is asymptotically more efficient than the unconstrained estimator for estimating the NDE and, provided that the outcome-mediator association conditional on exposure is non-zero, the NIE. More specifically, the magnitude of the asymptotic relative efficiency gains for estimating the NDE and NIE are both functions of the asymptotic partial $R^2$ between the $\boldsymbol{Y} \mid \boldsymbol{A}, \boldsymbol{C}$ and $\boldsymbol{Y} \mid \boldsymbol{M}, \boldsymbol{A}, \boldsymbol{C}$ models. Third, we robustify this mediation framework to violations of transportability assumptions by introducing a mediation model where the internal TE parameter is modeled as a random effect to deal with potential incongeniality of the external and internal TE estimates. The random effect treatment of the internal TE parameter facilitates Empirical-Bayes style shrinkage which data-adaptively shrinks more strongly towards the external TE estimate if the internal and external populations appear to have {\it similar} TEs \citep{morris1983, mukherjee2008}. Lastly, we provide corroborative evidence in PROTECT to the conclusions of \cite{aung2020}, which found a significant indirect effect of phthalate exposure on gestational age at delivery through the Cytochrome p450 pathway in the LIFECODES prospective birth cohort with participants from greater Boston area. The two cohorts are very different in demographics, socioeconomic profile, behavior and lifestyle factors, thus this replicated finding may offer a genuine biological insight. To our knowledge, this is the first paper that combines ideas from data integration and mediation analysis.

The structure of the paper is as follows. In Section \ref{s:methods} we explicitly define the problem, discuss methods for estimating model parameters in a linear mediation model with and without external information. We derive estimators of the NDE and NIE corresponding to each method. In Section \ref{s:asym_eff_res}, we compare asymptotic results corresponding to the NDE and NIE estimators defined in Section \ref{s:methods} and discuss robustness to violations of transportability. In Section \ref{s:simulations} we empirically substantiate the findings from Section \ref{s:asym_eff_res} with a comprehensive simulation study. In Section \ref{s:data_example} we apply this methodology to the PROTECT mediation analysis. Section \ref{s:discussion} offers a brief concluding discussion.

\section{Methods}
\label{s:methods}

\subsection{Notation and Model Specifications}
\label{ss:notation}

\begin{figure}[h!]
\centering
\subfigure[External Total Effect Model]  
{
\begin{tikzpicture}
 	\node (p1) at (3, 0) {$\boldsymbol{Y}$};
    \node (p3) at (-1, 0) {$\boldsymbol{A}$};
    \node (p4) at (1, -1) {$\boldsymbol{C}_E$};
    \path[->, draw, thick] (p3) -- (p1);
    \path[->, draw, thick] (p4) -- (p3);
    \path[->, draw, thick] (p4) -- (p1);
\end{tikzpicture}
}
\subfigure[Internal Mediation Model]  
{
\begin{tikzpicture}
 	\node (p1) at (4, 0) {$\boldsymbol{Y}$}; 
    \node (p2) at (1, 0) {$\boldsymbol{M} = \{\boldsymbol{M}_{1},\ldots,\boldsymbol{M}_{p_m}\}$};
    \node (p3) at (-2, 0) {$\boldsymbol{A}$};
    \node (p4) at (-4, 0) {$\boldsymbol{C}_1$};
    \node (p5) at (1, -1) {$\boldsymbol{C}_2$};
    \path[->, draw, thick] (p2) -- (p1);
    \path[->, draw, thick] (p3) -- (p2);
    \path[->, draw, thick] (p4) -- (p3);
    \path[->, draw, thick] (p5) -- (p2);
    \path[->, draw, thick] (p5) -- (p1);
    \draw[->, thick] (p3) to[bend left] (p1);
    \draw[->, thick] (p4) to[bend left] (p1);
\end{tikzpicture}
}
\caption{Directed Acyclic Graph (DAG) for internal and external mediation models where $\boldsymbol{A}$ denotes the exposure, $\boldsymbol{M}$ denotes the collection of $p_m$ candidate mediators, $\boldsymbol{Y}$ denotes the outcome, $\boldsymbol{C}_1$ denotes the outcome-exposure confounders, and $\boldsymbol{C}_2$ denotes the outcome-mediator confounders conditional on exposure. Throughout the paper, $\boldsymbol{C} \supseteq \boldsymbol{C}_1 \cup \boldsymbol{C}_2$. Here, $\boldsymbol{C}_E$ will likely have some overlap with $\boldsymbol{C}_1$, but the total effect model adjustment sets for the external and internal models are not assumed to be the same.}
\label{fig:mediation_dag}
\end{figure}
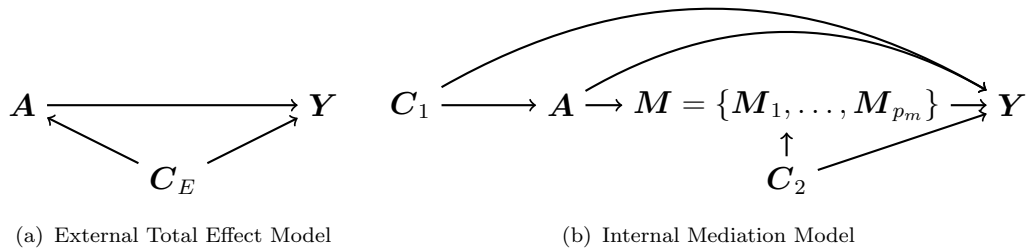

We consider a mediation analysis setting where a collection of continuous candidate mediators is hypothesized to mediate the association between a single exposure and a continuous health outcome (see Figure \ref{fig:mediation_dag}). For the internal study, we assume that we have individual-level data on $n$ observations. For observation $i$ $(i = 1,\ldots, n)$, let $Y_i$ denote the outcome,   $\boldsymbol{M}_{i\cdot}^{\top} = (M_{i1},\ldots,M_{ip_m})^{\top}$ denote a collection of $p_m$ candidate mediators, $A_i$ denote the exposure with  $E[A_i] = 0$, and  $\boldsymbol{C}_{i\cdot}^{\top}$ denote a collection of $p_c$ confounders and adjustment covariates plus the intercept term.  To be clear on the notation, $\boldsymbol{M}_{i\cdot}^{\top}$ is a $p_m \times 1$ column vector, $M_{ij}$ is the realization of the $j$-th mediator for observation $i$, and $\boldsymbol{C}_{i\cdot}^{\top}$ is a $p_c \times 1$ column vector. We do not distinguish between confounders of the outcome-exposure relationship and the outcome-mediator relationship, as in Figure \ref{fig:mediation_dag}; we assume that $\boldsymbol{C}_{i\cdot}^\top$ contains all confounders for both relationships. In our presentation, we also use matrix notation, namely $\boldsymbol{Y}=(Y_1,\ldots, Y_n)^\top$ is the $n \times 1$ column vector containing the observed outcomes, $\boldsymbol{M} = (\boldsymbol{M}^{\top}_{1\cdot},\ldots,\boldsymbol{M}^\top_{n\cdot})^{\top}$ is the $n \times p_m$ design matrix of observed mediator values, $\boldsymbol{A} = (A_1,\ldots, A_n)^\top$ is the $n \times 1$ column vector containing the observed exposures, and $\boldsymbol{C} = (\boldsymbol{C}_{1\cdot}^\top,\ldots, \boldsymbol{C}_{n\cdot}^\top)^\top$ represents the $n \times p_c$ matrix of observed confounders, plus an intercept term. The true generative model for the internal data is
\begin{align}
    [Y_i \mid \boldsymbol{M}_{i\cdot}, A_i, \boldsymbol{C}_{i\cdot}] &\sim N\Big(\boldsymbol{M}_{i\cdot}\boldsymbol{\beta}_m + A_i\beta_a + \boldsymbol{C}_{i\cdot}\boldsymbol{\beta}_c,\sigma_e^2\Big), \label{eqn:outcome_model}\\
    [\boldsymbol{M}_{i\cdot}^{\top} \mid \boldsymbol{A}_i, \boldsymbol{C}_i] &\sim N\Big(A_i\boldsymbol{\alpha}_a + \boldsymbol{\alpha}_c\boldsymbol{C}_i^{\top},\boldsymbol{\Sigma}_m\Big), \hspace{2 mm} i = 1,\ldots,n. \label{eqn:mediator_model}
\end{align}

\noindent We refer to \eqref{eqn:outcome_model} as the outcome model and \eqref{eqn:mediator_model} as the mediator model. Note that integrating out the mediators in \eqref{eqn:outcome_model} and \eqref{eqn:mediator_model} yields: 
\begin{equation}
    [Y_i \mid A_i, \boldsymbol{C}_{i\cdot}] \sim N\Big(A_i\theta_a^I + \boldsymbol{C}_{i\cdot}\boldsymbol{\theta}_c,\sigma_t^2\Big), \hspace{2 mm} i = 1,\ldots,n, \label{eqn:te_model}
\end{equation}

\noindent where $\theta_a^I = \beta_a +  \boldsymbol{\alpha}_a^{\top}\boldsymbol{\beta}_m$, $\boldsymbol{\theta}_c = \boldsymbol{\beta}_c + \boldsymbol{\alpha}_c^{\top}\boldsymbol{\beta}_m$, and $\sigma_t^2 = \sigma_e^2 + \boldsymbol{\beta}_m^{\top}\boldsymbol{\Sigma}_m\boldsymbol{\beta}_m$. We refer to \eqref{eqn:te_model} as the internal TE model. 

For the external study, we assume that we have summary-level information on the TE, $\theta_a^I$, in the form of a point estimate $\widehat{\theta}_a^{E}$ and an associated measure of uncertainty $\widehat{\text{Var}}(\widehat{\theta}_a^{E})$ based on a sample size of $n_E$ ($n \ll n_E$). Furthermore, we assume that we do not have access to the individual-level data from the external data source. The main objective of this paper is to leverage the available external summary-level information, $\widehat{\theta}_a^{E}$ and $\widehat{\text{Var}}(\widehat{\theta}_a^{E})$, to improve estimation of the NDE and the NIE in the internal study.

\subsection{Identification of Causal Effects}
\label{ss:causal_mediation}

In the potential outcomes framework, $\boldsymbol{M}_{i\cdot}^{\top}(a)$ is the counterfactual value of the mediator vector had the exposure been equal to $a$ and $Y_i(a,\boldsymbol{m})$ is the counterfactual outcome had the exposure been equal to $a$ and had the candidate mediator vector been equal to $\boldsymbol{m}$. Combining these two counterfactual quantities, $Y_i(a,\boldsymbol{M}_{i\cdot}^{\top}(a))$ is the potential outcome for exposure level $a$ and the TE, which quantifies how the exposure marginally impacts the counterfactual outcome in the internal population, is defined as $Y_i(a,\boldsymbol{M}_{i\cdot}^{\top}(a)) - Y_i(a^*,\boldsymbol{M}_{i\cdot}^{\top}(a^*))$, where the exposure changes from the reference level $a^*$ to $a$. The NDE and NIE are obtained by decomposing the TE as follows: \begin{align*}
    \text{TE}_i &= Y_i(a,\boldsymbol{M}_{i\cdot}^{\top}(a)) - Y_i(a^*,\boldsymbol{M}_{i\cdot}^{\top}(a^*))\\
    &= Y_i(a,\boldsymbol{M}_{i\cdot}^{\top}(a)) - Y_i(a,\boldsymbol{M}_{i\cdot}^{\top}(a^*)) + Y_i(a,\boldsymbol{M}_{i\cdot}^{\top}(a^*)) - Y_i(a^*,\boldsymbol{M}_{i\cdot}^{\top}(a^*))\\
    &= \text{NIE}_i + \text{NDE}_i.
\end{align*}

\noindent The NDE quantifies how the potential outcome changes as a function of the exposure level subject to identical realizations of the reference level mediator values. Conversely, the NIE quantifies how the potential outcome changes as a function of the counterfactual mediators subject to identical realizations of the comparison exposure value. The conditional independence assumptions required for identification of the average NDE and NIE from observed data are: (i) $Y_i(a,\boldsymbol{m}) \perp A_i \mid \boldsymbol{C}_i$, (ii) $Y_i(a,\boldsymbol{m}) \perp \boldsymbol{M}_{i\cdot}^{\top} \mid \{A_i, \boldsymbol{C}_i\}$, (iii) $\boldsymbol{M}_{i\cdot}^{\top}(a) \perp A_i \mid \boldsymbol{C}_i$, and (iv) $Y_i(a,\boldsymbol{m}) \perp \boldsymbol{M}_{i\cdot}^{\top}(a^*) \mid \boldsymbol{C}_i$ (for a detailed exposition see \cite{song2020}). We will assume that (i)-(iv) hold for the internal study. Under these assumptions: \begin{align*}
    \text{NDE} &= E[Y_i(a,\boldsymbol{M}_{i\cdot}^{\top}(a^*)) - Y_i(a^*,\boldsymbol{M}_{i\cdot}^{\top}(a^*)) \mid \boldsymbol{C}_i] = \beta_a(a-a^*)\\
    \text{NIE} &= E[Y_i(a,\boldsymbol{M}_{i\cdot}^{\top}(a)) - Y_i(a,\boldsymbol{M}_{i\cdot}^{\top}(a^*)) \mid \boldsymbol{C}_i] = \boldsymbol{\alpha}_a^{\top}\boldsymbol{\beta}_m(a-a^*)\\
    \text{TE} &= E[Y_i(a,\boldsymbol{M}_{i\cdot}^{\top}(a)) - Y_i(a^*,\boldsymbol{M}_{i\cdot}^{\top}(a^*)) \mid \boldsymbol{C}_i] = (\beta_a + \boldsymbol{\alpha}_a^{\top}\boldsymbol{\beta}_m)(a-a^*) = \theta_a^I(a-a^*)
\end{align*}

For observation $i'$ in the external study ($i' = 1,\ldots,n_E$), we define $Y_{E,i'}$ as the observed outcome, $A_{E,i'}$ as the observed exposure, and $\boldsymbol{C}_{E,i'\cdot}$ as the observed confounder vector, which may or may not be the same as the set of confounders in the internal study. To incorporate external summary-level information on the TE, certain methods we present in this paper require the following transportability condition:
\begin{flalign*}
    E[Y_i \mid A_i = a, \boldsymbol{C}_{i\cdot} = \boldsymbol{c}] - E[Y_i \mid A_i = a^{*}, \boldsymbol{C}_{i\cdot} = \boldsymbol{c}]&&
\end{flalign*}
\begin{flalign}
    &&= E[Y_{E,i'} \mid A_{E,i'} = a, \boldsymbol{C}_{E,i'\cdot} = \boldsymbol{c}_E] - E[Y_{E,i'} \mid A_{E,i'} = a^{*}, \boldsymbol{C}_{E,i'\cdot} = \boldsymbol{c}_E]
    \label{eqn:transport}
\end{flalign}
for all possible realizations of $a$ and $a^{*}$. Transportability condition (\ref{eqn:transport}) in our context ensures that $\theta_a^I = \theta_a^E$, where $\theta_a^E$ is the true TE in the external population.

\subsection{Maximum Likelihood Estimation without External Information}
\label{ss:unconst_est}

To establish an inferential baseline, consider a mediation analysis that ignores available external summary-level information on the TE. That is, the model specification is \eqref{eqn:outcome_model} and \eqref{eqn:mediator_model}, which we call the unconstrained model. The maximum likelihood estimator (MLE) with respect to model specification \eqref{eqn:outcome_model} and \eqref{eqn:mediator_model} is defined as \vspace{2 mm}
\begin{gather}
    \nonumber\underset{\boldsymbol{\alpha}_a,\boldsymbol{\alpha}_c,\boldsymbol{\Sigma}_m, \beta_a,\boldsymbol{\beta}_m,\boldsymbol{\beta}_c,\sigma_e^2}{\arg\min} \bigg\{\frac{1}{2}\sum_{i=1}^{n}\Big(\boldsymbol{M}_{i\cdot}^{\top} - A_i\boldsymbol{\alpha}_a - \boldsymbol{\alpha}_c\boldsymbol{C}_{i\cdot}^{\top}\Big)^{\top}\boldsymbol{\Sigma}_m^{-1}\Big(\boldsymbol{M}_{i\cdot}^{\top} - A_i\boldsymbol{\alpha}_a - \boldsymbol{\alpha}_c\boldsymbol{C}_{i\cdot}^{\top}\Big) \\+ \frac{1}{2\sigma_e^2}\Big(\boldsymbol{Y} - \boldsymbol{A}\beta_a - \boldsymbol{M}\boldsymbol{\beta}_m - \boldsymbol{C}\boldsymbol{\beta}_c\Big)^{\top}\Big(\boldsymbol{Y} - \boldsymbol{A}\beta_a - \boldsymbol{M}\boldsymbol{\beta}_m - \boldsymbol{C}\boldsymbol{\beta}_c\Big) \bigg\}.
    \label{eq:unconstrained_opt}
\end{gather}
\vspace{-4 mm}

The MLE is denoted as $(\widehat{\boldsymbol{\alpha}}_a^U, \widehat{\boldsymbol{\alpha}}_c^U, \widehat{\boldsymbol{\Sigma}}_m^U, \widehat{\beta}_a^U, \widehat{\boldsymbol{\beta}}_m^U, \widehat{\boldsymbol{\beta}}_c^U, \big\{\widehat{\sigma}_e^U\big\}^2)$, where $\widehat{\boldsymbol{\alpha}}_a^U$ is the MLE of $\boldsymbol{\alpha}_a$, $\widehat{\beta}_a^U$ is the MLE of $\beta_a$, and $\widehat{\boldsymbol{\beta}}_m^U$ is the MLE of $\boldsymbol{\beta}_m$. For the unconstrained model, the MLEs have closed-form expressions (see Web Appendix A). Going forward, $\widehat{\text{NDE}}^U = \widehat{\beta}_a^U$ and $\widehat{\text{NIE}}^U = \big\{\widehat{\boldsymbol{\alpha}}_a^U\big\}^{\top}\widehat{\boldsymbol{\beta}}_m^U$ are called the unconstrained estimators of the NDE and NIE, respectively.

\subsection{Maximum Likelihood Estimation with Congenial External Information}
\label{ss:hard_est}

Next, assume that there is an external point estimate, $\widehat{\theta}_a^E$, such that $\widehat{\theta}_a^E$ is a consistent estimate of $\theta_a^I$. This corresponds to the situation where transportability assumption (\ref{eqn:transport}) holds. When $\widehat{\theta}_a^E$ is a consistent estimate of $\theta_a^I$, we say that the external information is congenial with the internal study population. Consider the optimization problem: \vspace{2 mm}
\begin{gather}
    \nonumber\underset{\boldsymbol{\alpha}_a,\boldsymbol{\alpha}_c,\boldsymbol{\Sigma}_m,\boldsymbol{\beta}_m,\boldsymbol{\beta}_c\sigma_e^2}{\arg\min} \bigg\{\frac{1}{2}\sum_{i=1}^{n}\Big(\boldsymbol{M}_{i\cdot}^{\top} - A_i\boldsymbol{\alpha}_a - \boldsymbol{\alpha}_c\boldsymbol{C}_{i\cdot}^{\top}\Big)^{\top}\boldsymbol{\Sigma}_m^{-1}\Big(\boldsymbol{M}_{i\cdot}^{\top} - A_i\boldsymbol{\alpha}_a - \boldsymbol{\alpha}_c\boldsymbol{C}_{i\cdot}^{\top}\Big)
    \\+ \frac{1}{2\sigma_e^2}\Big(\boldsymbol{Y} - \boldsymbol{A}\beta_a^E - \boldsymbol{M}\boldsymbol{\beta}_m - \boldsymbol{C}\boldsymbol{\beta}_c\Big)^{\top}\Big(\boldsymbol{Y} - \boldsymbol{A}\beta_a^E - \boldsymbol{M}\boldsymbol{\beta}_m  - \boldsymbol{C}\boldsymbol{\beta}_c\Big) \bigg\}
    \label{eq:hard_constraint_likelihood}
\end{gather} \vspace{-6 mm}

\noindent where $\beta_a^E = \widehat{\theta}_a^E-\boldsymbol{\alpha}_a^{\top}\boldsymbol{\beta}_m$. Alternatively, \eqref{eq:hard_constraint_likelihood} can be viewed as a minimization over the negative log-likelihood corresponding to the following model specification, which we call the hard constraint model: \begin{gather*}
    \big[Y_i \mid \boldsymbol{M}_{i\cdot}, A_i, \boldsymbol{C}_{i\cdot}\big] \sim N\big(\boldsymbol{M}_{i\cdot}\boldsymbol{\beta}_m + A_i\{\widehat{\theta}_a^E - \boldsymbol{\alpha}_a^{\top}\boldsymbol{\beta}_m\} + \boldsymbol{C}_{i\cdot}\boldsymbol{\beta}_c, \sigma_e^2\big)
    \\\big[\boldsymbol{M}_{i\cdot}^{\top} \mid A_i, \boldsymbol{C}_{i\cdot}\big] \sim N(A_i\boldsymbol{\alpha}_a + \boldsymbol{\alpha}_c\boldsymbol{C}_{i\cdot}^{\top},\boldsymbol{\Sigma}_m), \hspace{2 mm} i = 1,\ldots,n.
\end{gather*} We denote the optimizer of \eqref{eq:hard_constraint_likelihood} as $(\widehat{\boldsymbol{\alpha}}_a^H, \widehat{\boldsymbol{\alpha}}_c^H, \widehat{\boldsymbol{\Sigma}}_m^H, \widehat{\boldsymbol{\beta}}_m^H, \widehat{\boldsymbol{\beta}}_c^H, \big\{\widehat{\sigma}_e^H\big\}^2)$, where $\widehat{\boldsymbol{\alpha}}_a^H$ is the estimator of $\boldsymbol{\alpha}_a$ and $\widehat{\boldsymbol{\beta}}_m^H$ is the estimator of $\boldsymbol{\beta}_m$. The purpose of (\ref{eq:hard_constraint_likelihood}) is to impose a hard constraint on TE estimation so that the estimated TE is always equal to $\big\{\widehat{\beta}_a^E\big\}^H + \big\{\widehat{\boldsymbol{\alpha}}_a^H\big\}^{\top}\widehat{\boldsymbol{\beta}}_m^H = \widehat{\theta}_a^E$. Cyclical coordinate descent is used to compute the optimizer of \eqref{eq:hard_constraint_likelihood} (see Web Appendix A). 

The approach described in this section represents the other extreme compared to the unconstrained estimator, where the estimated TE is forced to be equal to $\widehat{\theta}_a^E$, showing exact and extreme faith in the external information. Moreover, the hard constraint on the estimated TE induces information sharing between the internal mediator and outcome models through the $\boldsymbol{\alpha}_a^{\top}\boldsymbol{\beta}_m$ term in the mean function of the outcome model. Going forward, we refer to $\widehat{\text{NDE}}^H = \widehat{\theta}_a^E - \big\{\widehat{\boldsymbol{\alpha}}_a^H\big\}^{\top}\widehat{\boldsymbol{\beta}}_m^H$ and $\widehat{\text{NIE}}^H = \big\{\widehat{\boldsymbol{\alpha}}_a^H\big\}^{\top}\widehat{\boldsymbol{\beta}}_m^H$ as the hard constraint estimators of the NDE and NIE, respectively.

\subsection{Robust Soft Constraint Estimator}
\label{ss:soft_est}

The final method considers the case where $\widehat{\theta}_a^{E}$ may or may not be a consistent estimate of $\theta_a^{I}$. That is, the validity of transportability assumption (\ref{eqn:transport}) is unknown. When $\widehat{\theta}_a^{E}$ is an inconsistent estimate of $\theta_a^I$, we say that the external information is incongenial with the internal study. Transportability assumption (\ref{eqn:transport}) may be violated for a variety of reasons, including fundamentally different $\boldsymbol{Y} \mid \boldsymbol{A}, \boldsymbol{C}$ distributions in the external and internal populations, unmeasured confounding in the external TE model, and differing adjustment sets between the external and internal TE models. To address potential violations of (\ref{eqn:transport}), we treat the internal TE parameter as a random effect, $\widetilde{\theta}_a^I$, and define a random effect mediation model, which we call the soft constraint model:
\begin{align*}
    [Y_i \mid \boldsymbol{M}_{i\cdot}, A_i, \boldsymbol{C}_{i\cdot}, \widetilde{\theta}_a^I] &\sim N\Big(\boldsymbol{M}_{i\cdot}\boldsymbol{\beta}_m + A_i\{\widetilde{\theta}_a^I - \boldsymbol{\alpha}_a^{\top}\boldsymbol{\beta}_m\} + \boldsymbol{C}_{i\cdot}\boldsymbol{\beta}_c,\sigma_e^2\Big)\\
    [\boldsymbol{M}_{i\cdot}^{\top} \mid A_i, \boldsymbol{C}_i] &\sim N\Big(A_i\boldsymbol{\alpha}_a + \boldsymbol{\alpha}_c\boldsymbol{C}_i^{\top},\boldsymbol{\Sigma}_m\Big), \hspace{2 mm} i = 1,\ldots,n\\
    \widetilde{\theta}_a^I &\sim N\Big(\widehat{\theta}_a^E,s^2\widehat{\text{Var}}(\widehat{\theta}_a^{E})\Big)
\end{align*}

\noindent It is important to clarify that the soft constraint model is a working model and the true generative model of the internal data remains (\ref{eqn:outcome_model}) and (\ref{eqn:mediator_model}). The advantage of a random effects formulation is that it allows for shrinkage towards the external information without imposing inflexible hard constraints on the estimated TE. After integrating out $\widetilde{\theta}_a^I$ the soft constraint likelihood function becomes $$L(\boldsymbol{\alpha}_a,\boldsymbol{\alpha}_c,\boldsymbol{\Sigma}_M,\boldsymbol{\beta}_m,\boldsymbol{\beta}_c,\sigma_e^2 \mid \boldsymbol{Y},\boldsymbol{M},\boldsymbol{A},\boldsymbol{C}) = \int_{-\infty}^{\infty}\pi(\boldsymbol{Y}\mid\boldsymbol{M},\boldsymbol{A},\boldsymbol{C},\widetilde{\theta}_a^I)\pi(\boldsymbol{M}\mid\boldsymbol{A}, \boldsymbol{C})\pi(\widetilde{\theta}_a^I)d\widetilde{\theta}_a^I,$$ where $\pi$ is general notation for a probability density function. The maximum likelihood estimators, defined by
\begin{equation}
    \underset{\boldsymbol{\alpha}_a,\boldsymbol{\alpha}_c,\boldsymbol{\Sigma}_m,\boldsymbol{\beta}_m,\boldsymbol{\beta}_c,\sigma_e^2}{\arg\max} L(\boldsymbol{\alpha}_a,\boldsymbol{\alpha}_c,\boldsymbol{\Sigma}_M,\boldsymbol{\beta}_m,\boldsymbol{\beta}_c,\sigma_e^2 \mid \boldsymbol{Y},\boldsymbol{M},\boldsymbol{A},\boldsymbol{C}),
    \label{eq:soft_constraint_likelihood_opt}
\end{equation}
\vspace{-8 mm} 

\noindent are denoted as $(\widehat{\boldsymbol{\alpha}}_a^S,\widehat{\boldsymbol{\alpha}}_c^S,\widehat{\boldsymbol{\Sigma}}_M^S,\widehat{\boldsymbol{\beta}}_m^S,
\widehat{\boldsymbol{\beta}}_c^S,\big\{\widehat{\sigma}_e^S\big\}^2)$ and the soft constraint estimator of the NIE is $\widehat{\text{NIE}}^S = \{\widehat{\boldsymbol{\alpha}}_a^S\}^{\top}\widehat{\boldsymbol{\beta}}_m^S$. The soft constraint estimator of the TE is the posterior mean estimator corresponding to the posterior distribution $\pi(\widetilde{\theta}_a^I \mid \boldsymbol{Y}, \boldsymbol{M}, \boldsymbol{A}, \boldsymbol{C})$, with the maximum likelihood estimators, $\widehat{\boldsymbol{\alpha}}_a^S$, $\widehat{\boldsymbol{\alpha}}_c^S$, $\widehat{\boldsymbol{\Sigma}}_M^S$, $\widehat{\boldsymbol{\beta}}_m^S$,
$\widehat{\boldsymbol{\beta}}_c^S$, and $\big\{\widehat{\sigma}_e^S\big\}^2$, substituted in for their corresponding true parameter values \citep{verbeke2000}. The resultant soft constraint estimator for the NDE is the difference between the soft constraint TE and NIE estimators: 

\vspace{-6 mm}$$\widehat{\text{NDE}}^{S} = \bigg[\frac{\boldsymbol{A}^{\top}\boldsymbol{A}}{\big\{\widehat{\sigma}_e^S\big\}^2} + \frac{1}{s^2\widehat{Var}(\widehat{\theta}_a^{E})}\bigg]^{-1}\bigg[\frac{\boldsymbol{A}^{\top}(\boldsymbol{Y} - \boldsymbol{C}\widehat{\boldsymbol{\beta}}_c^S - \boldsymbol{M}\widehat{\boldsymbol{\beta}}_m^S)}{\big\{\widehat{\sigma}_e^S\big\}^2} + \frac{(\widehat{\theta}_a^E - \{\widehat{\boldsymbol{\alpha}}_a^S\}^{\top}\widehat{\boldsymbol{\beta}}_m^S)}{s^2\widehat{Var}(\widehat{\theta}_a^{E})}\bigg]$$

\vspace{2 mm}

For the soft constraint model, the Expectation-Maximization (EM) algorithm is used to solve \eqref{eq:soft_constraint_likelihood_opt}, where $\boldsymbol{Y}$, $\boldsymbol{M}$, $\boldsymbol{A}$, and $\boldsymbol{C}$ are treated as the observed data and $\widetilde{\theta}_a^I$ is treated as the unobserved latent data \citep{dempster1977}. See Web Appendix A for details on the EM algorithm implementation.

\section{Asymptotic Efficiency Results}
\label{s:asym_eff_res}
The goals of this section are to understand the efficiency gain attributable to incorporating congenial external information on the TE and to provide commentary on dealing with potentially incongenial external information. Here, $\sigma_a^2 = \text{Var}(A_i \mid \boldsymbol{C}_{i\cdot})$, which is obtained by regressing out the confounders from the exposure using a linear regression model.

\subsection{Asymptotic Distributions of the Unconstrained and Hard Constraint Estimators}
\label{ss:asym_eff_unconst_hard_comp}

\noindent \textbf{Theorem 1.} The joint asymptotic distribution of $\widehat{\boldsymbol{\alpha}}_a^{U}$, $\widehat{\beta}_a^{U}$, and $\widehat{\boldsymbol{\beta}}_m^{U}$ is, \vspace{3 mm}
$$\sqrt{n}\begin{pmatrix} \widehat{\boldsymbol{\alpha}}_a^{U} - \boldsymbol{\alpha}_a \\ \widehat{\beta}_a^{U} - \beta_a \\ \widehat{\boldsymbol{\beta}}_m^{U} - \boldsymbol{\beta}_m \end{pmatrix} \to_d N\Bigg(\boldsymbol{0},\Big\{\mathcal{I}^{U}(\boldsymbol{\alpha}_a,\beta_a,\boldsymbol{\beta}_m)\Big\}^{-1}\Bigg)$$

$$\Big\{\mathcal{I}^{U}(\boldsymbol{\alpha}_a,\beta_a,\boldsymbol{\beta}_m)\Big\}^{-1} = \begin{pmatrix} \frac{1}{\sigma_a^2}\boldsymbol{\Sigma}_m & \boldsymbol{0} & \boldsymbol{0} \\ \boldsymbol{0} & \frac{\sigma_e^2}{\sigma_a^2}\big(1 + \sigma_a^2\boldsymbol{\alpha}_a^{\top}\boldsymbol{\Sigma}_m^{-1}\boldsymbol{\alpha}_a\big) & -\sigma_e^2\boldsymbol{\alpha}_a^{\top}\boldsymbol{\Sigma}_m^{-1} \\ \boldsymbol{0} & -\sigma_e^2\boldsymbol{\Sigma}_m^{-1}\boldsymbol{\alpha}_a & \sigma_e^2\boldsymbol{\Sigma}_m^{-1} \end{pmatrix}$$ \vspace{2 mm}

\noindent Let $\text{NDE} = \beta_a$, $\text{NIE} = \boldsymbol{\alpha}_a^{\top}\boldsymbol{\beta}_m$, $\text{TE} = \text{NDE} + \text{NIE}$, and $\widehat{\text{TE}}^{U} = \widehat{\text{NDE}}^{U} + \widehat{\text{NIE}}^{U}$. Then, $$\sqrt{n}(\widehat{\text{NDE}}^{U} - \text{NDE}) \to_d N\bigg(0,\frac{\sigma_e^2}{\sigma_a^2} + \sigma_e^2\boldsymbol{\alpha}_a^{\top}\boldsymbol{\Sigma}_m^{-1}\boldsymbol{\alpha}_a\bigg),$$ $$\sqrt{n}(\widehat{\text{TE}}^{U} - \text{TE}) \to_d N\bigg(0,\frac{1}{\sigma_a^2}\Big\{\sigma_e^2 + \boldsymbol{\beta}_m^{\top}\boldsymbol{\Sigma}_m\boldsymbol{\beta}_m\Big\}\bigg),$$ and, provided that $\boldsymbol{\alpha}_a \neq \boldsymbol{0}$ or $\boldsymbol{\beta}_m \neq \boldsymbol{0}$, $$\sqrt{n}(\widehat{\text{NIE}}^{U} - \text{NIE}) \to_d N\bigg(0,\frac{1}{\sigma_a^2}\boldsymbol{\beta}_m^{\top}\boldsymbol{\Sigma}_m\boldsymbol{\beta}_m + \sigma_e^2\boldsymbol{\alpha}_a^{\top}\boldsymbol{\Sigma}_m^{-1}\boldsymbol{\alpha}_a\bigg).$$

\noindent \textit{Proof.} See Web Appendix B for details.

\noindent \textbf{Theorem 2.} Suppose that $\sqrt{n}(\widehat{\theta}_a^E-\theta_a^I) \to_p 0$ as $n \to \infty$ and $n_{E} \to \infty$. Then, \vspace{3 mm}
$$\sqrt{n}\begin{pmatrix} \widehat{\boldsymbol{\alpha}}_a^{H} - \boldsymbol{\alpha}_a \\ \widehat{\boldsymbol{\beta}}_m^{H} - \boldsymbol{\beta}_m \end{pmatrix} \to_d N\Bigg(\boldsymbol{0},\Big\{\mathcal{I}^{H}(\boldsymbol{\alpha}_a,\boldsymbol{\beta}_m)\Big\}^{-1}\Bigg)$$

$$\Big\{\mathcal{I}^{H}(\boldsymbol{\alpha}_a,\boldsymbol{\beta}_m)\Big\}^{-1} = \begin{pmatrix} \frac{1}{\sigma_a^2}\Big(\boldsymbol{\Sigma}_m^{-1} + \frac{1}{\sigma_e^2}\boldsymbol{\beta}_m\boldsymbol{\beta}_m^{\top}\Big)^{-1} & \boldsymbol{0} \\ \boldsymbol{0} & \sigma_e^2\boldsymbol{\Sigma}_m^{-1} \end{pmatrix}$$

\noindent Let $\text{NDE} = \theta_a^I - \boldsymbol{\alpha}_a^{\top}\boldsymbol{\beta}_m$, $\text{NIE} = \boldsymbol{\alpha}_a^{\top}\boldsymbol{\beta}_m$, $\text{TE} = \theta_a^I$, and $\widehat{\text{TE}}^{H} = \widehat{\theta}_a^E$. Then, provided that $\boldsymbol{\alpha}_a \neq \boldsymbol{0}$ or $\boldsymbol{\beta}_m \neq \boldsymbol{0}$, $$\sqrt{n}(\widehat{\text{NDE}}^{H} - \text{NDE}) \to_d N\bigg(0,\frac{\sigma_e^2}{\sigma_a^2}R_{\boldsymbol{M} \mid \boldsymbol{A}, \boldsymbol{C}}^2 + \sigma_e^2\boldsymbol{\alpha}_a^{\top}\boldsymbol{\Sigma}_m^{-1}\boldsymbol{\alpha}_a\bigg)$$ $$\sqrt{n}(\widehat{\text{NIE}}^{H} - \text{NIE}) \to_d N\bigg(0,\frac{1}{\sigma_a^2}\boldsymbol{\beta}_m^{\top}\boldsymbol{\Sigma}_m\boldsymbol{\beta}_m\big\{1-R_{\boldsymbol{M} \mid \boldsymbol{A}, \boldsymbol{C}}^2\big\} + \sigma_e^2\boldsymbol{\alpha}_a^{\top}\boldsymbol{\Sigma}_m^{-1}\boldsymbol{\alpha}_a\bigg),$$ where $$R_{\boldsymbol{M} \mid \boldsymbol{A}, \boldsymbol{C}}^2 = \frac{\boldsymbol{\beta}_m^{\top}\boldsymbol{\Sigma}_m\boldsymbol{\beta}_m}{\sigma_e^2+\boldsymbol{\beta}_m^{\top}\boldsymbol{\Sigma}_m\boldsymbol{\beta}_m}$$

\noindent \textit{Proof.} See Web Appendix B for details.

Theorems 1 and 2 show the asymptotic distributions of the unconstrained and hard constraint estimators, respectively. The inverted information matrices clarify that efficiency gains for NIE estimation exclusively come from improved estimation of $\boldsymbol{\alpha}_a$. When the TE and outcome models are framed as nested regression models, this result is consistent with \cite{gu2019}, which showed that leveraging external summary-level information from a reduced model only results in efficiency gains for the regression coefficients corresponding to regressors in common between the two models. Theorems 1 and 2 also show that $\boldsymbol{\beta}_m = \boldsymbol{0}$ implies no efficiency gain for NIE estimation. When $\boldsymbol{\beta}_m \neq 0$, the absolute efficiency gain corresponding to the NDE and NIE is completely dependent on the quantity, $R_{\boldsymbol{M} \mid \boldsymbol{A}, \boldsymbol{C}}^2$, the asymptotic partial $R^2$ between the TE and outcome models. If $R_{\boldsymbol{M} \mid \boldsymbol{A}, \boldsymbol{C}}^2 \approx 1$, then the inclusion of candidate mediators substantially improves model fit and consequently there are large gains for NIE estimation. Conversely, if $R_{\boldsymbol{M} \mid \boldsymbol{A}, \boldsymbol{C}}^2 \approx 0$, then inclusion of candidate mediators do not improve model fit compared to the TE model and consequently there are large gains for NDE estimation. An intuitive understanding of the latter point comes when $R_{\boldsymbol{M} \mid \boldsymbol{A}, \boldsymbol{C}}^2 = 0$, which implies that $\boldsymbol{\beta}_m = \boldsymbol{0}$ and $\text{TE} = \text{NDE}$. That is, the smaller $||\boldsymbol{\beta}_m||_2$ is, the closer the TE is to the NDE, and the external information on the TE becomes increasingly more relevant for NDE estimation. With respect to NDE and NIE estimation, a small value of $\sigma_e^2\boldsymbol{\alpha}_a^{\top}\boldsymbol{\Sigma}_m^{-1}\boldsymbol{\alpha}_a$, a scaling of the signal-to-noise ratio in the mediator model, leads to larger relative efficiency gains.

\subsection{Asymptotic Distribution when $\boldsymbol{\alpha}_a = \boldsymbol{\beta}_m = \boldsymbol{0}$}
\label{ss:asym_eff_special_case}

Theorem 3 clarifies the asymptotic distribution of the unconstrained and hard constraint estimators when $\boldsymbol{\alpha}_a = \boldsymbol{\beta}_m = \boldsymbol{0}$. There are no efficiency gains for NIE estimation in this setting because $\boldsymbol{\beta}_m = 
\boldsymbol{0}$.

\noindent \textbf{Theorem 3.} Suppose that $\boldsymbol{\alpha}_a = \boldsymbol{\beta}_m = \boldsymbol{0}$ and that there are estimators $\widehat{\boldsymbol{\alpha}}_a$ and $\widehat{\boldsymbol{\beta}}_m$ of $\boldsymbol{\alpha}_a$ and $\boldsymbol{\beta}_m$, respectively, which satisfy: $$\sqrt{n}\begin{pmatrix} \widehat{\boldsymbol{\alpha}}_a - \boldsymbol{\alpha}_a \\ \widehat{\boldsymbol{\beta}}_m - \boldsymbol{\beta}_m \end{pmatrix} \to_d N\Bigg(\begin{pmatrix} \boldsymbol{0} \\ \boldsymbol{0} \end{pmatrix}, \begin{pmatrix} \sigma_a^{-2}\boldsymbol{\Sigma}_m & \boldsymbol{0} \\ \boldsymbol{0} & \sigma_e^2\boldsymbol{\Sigma}_m^{-1} \end{pmatrix}\Bigg)$$ Then, $$n\Big(\widehat{\boldsymbol{\alpha}}_a^{\top}\widehat{\boldsymbol{\beta}}_m - \boldsymbol{\alpha}_a^{\top}\boldsymbol{\beta}_m\Big) \to_d \frac{1}{2}\sqrt{\frac{\sigma_e^2}{\sigma_a^2}}\big(\xi_1-\xi_2\big),$$ where $\xi_1$ and $\xi_2$ are independent $\chi_{p_m}^2$ random variables.

\noindent \textit{Proof.} See Web Appendix B for details.

Theorems 1-3 can be used to inform the construction of hypothesis tests and confidence intervals for the NIE and NDE. However, when determining the reference distribution, how much to weight the $\boldsymbol{\alpha}_a = \boldsymbol{\beta}_m = \boldsymbol{0}$ case relative to the $\boldsymbol{\alpha}_a \neq \boldsymbol{0}$ or $\boldsymbol{\beta}_m \neq \boldsymbol{0}$ case is unknown. One straightforward workaround is to check whether or not $\boldsymbol{\alpha}_a = \boldsymbol{\beta}_m = \boldsymbol{0}$ by using a Wald test with the reference distribution given by
$$n\begin{pmatrix} \widehat{\boldsymbol{\alpha}}_a^{\top} & \widehat{\boldsymbol{\beta}}_m^{\top} \end{pmatrix} \begin{pmatrix} \sigma_a^{-2}\boldsymbol{\Sigma}_{m} & \boldsymbol{0} \\ \boldsymbol{0} & \sigma_e^2\boldsymbol{\Sigma}_{m}^{-1} \end{pmatrix}^{-1} \begin{pmatrix} \widehat{\boldsymbol{\alpha}}_a \\ \widehat{\boldsymbol{\beta}}_m \end{pmatrix} \to_d \chi_{2p_m}^2$$

\noindent and then use the appropriate asymptotic result to construct confidence intervals.

\subsection{Robustness to Incongenial External Information}
\label{ss:asym_eff_soft}

Theorem 2 focuses on settings where transportability condition (\ref{eqn:transport}) holds, however there are likely many instances where fundamental differences across internal and external study populations lead to violations of (\ref{eqn:transport}). The desire to account for such cases motivates an estimator that is robust to departures from (\ref{eqn:transport}), but is still more efficient than the unconstrained estimator when (\ref{eqn:transport}) is satisfied. In this context, robustness refers to the fact that the estimator is as asymptotically efficient as the unconstrained estimator when (\ref{eqn:transport}) does not hold. Theorem 4 establishes that the soft constraint estimator is as efficient or more efficient than the unconstrained estimator with respect to NIE estimation when (\ref{eqn:transport}) does not hold.

\noindent \textbf{Theorem 4.} Suppose that $ns^2\widehat{\text{Var}}(\widehat{\theta}_a^{E}) \to_p \tau_a^2$, where $\tau_a^2 \in (0,\infty)$, and $\widehat{\theta}_a^E \to_p \theta_a^I$ as $n \to \infty$ and $n_{E} \to \infty$. Then, \vspace{3 mm}
$$\sqrt{n}\begin{pmatrix} \widehat{\boldsymbol{\alpha}}_a^{S} - \boldsymbol{\alpha}_a \\ \widehat{\boldsymbol{\beta}}_m^{S} - \boldsymbol{\beta}_m \end{pmatrix} \to_d N\Bigg(\boldsymbol{0},\Big\{\mathcal{I}^{S}(\boldsymbol{\alpha}_a,\boldsymbol{\beta}_m)\Big\}^{-1}\Bigg)$$

$$\Big\{\mathcal{I}^{S}(\boldsymbol{\alpha}_a,\boldsymbol{\beta}_m)\Big\}^{-1} = \begin{pmatrix} \frac{1}{\sigma_a^2}\bigg(\boldsymbol{\Sigma}_m^{-1} + \frac{1}{\tau_a^2}\Big[\frac{\sigma_a^2}{\sigma_e^2}+\frac{1}{\tau_a^2}\Big]^{-1}\frac{1}{\sigma_e^2}\boldsymbol{\beta}_m\boldsymbol{\beta}_m^{\top}\bigg)^{-1} & \boldsymbol{0} \\ \boldsymbol{0} & \sigma_e^2\boldsymbol{\Sigma}_m^{-1} \end{pmatrix}$$

\noindent Let $\text{NIE} = \boldsymbol{\alpha}_a^{\top}\boldsymbol{\beta}_m$. Then, provided that $\boldsymbol{\alpha}_a \neq \boldsymbol{0}$ or $\boldsymbol{\beta}_m \neq \boldsymbol{0}$, $$\sqrt{n}(\widehat{\text{NIE}}^{S} - \text{NIE}) \to_d N\Bigg(0,\frac{1}{\sigma_a^2}\boldsymbol{\beta}_m^{\top}\boldsymbol{\Sigma}_m\boldsymbol{\beta}_m\bigg[1+\frac{1}{\tau_a^2}\bigg(\frac{\sigma_a^2}{\sigma_e^2}+\frac{1}{\tau_a^2}\bigg)^{-1}\frac{1}{\sigma_e^2}\boldsymbol{\beta}_m^{\top}\boldsymbol{\Sigma}_m\boldsymbol{\beta}_m\bigg]^{-1}+\sigma_e^2\boldsymbol{\alpha}_a^{\top}\boldsymbol{\Sigma}_m^{-1}\boldsymbol{\alpha}_a\Bigg)$$

\noindent \textit{Proof.} See Web Appendix B for details.

Theorem 4 provides the asymptotic distribution of the soft constraint estimator of the NIE. The asymptotic variance-covariance matrix converges to the asymptotic variance-covariance matrix of the unconstrained estimator if $s^2 \to \infty$ and the asymptotic variance-covariance matrix of the hard constraint estimator if $s^2 \to 0$. Additionally, the conclusions of Theorem 3 hold for the soft constraint estimator when $\boldsymbol{\alpha}_a = \boldsymbol{\beta}_m = \boldsymbol{0}$. Inference corresponding to the soft constraint NDE estimator is more challenging because there is not an easily derivable asymptotic distribution. In this paper, for interval estimation, we use quantile-based confidence intervals via the parametric bootstrap \citep{efron1982}.

Although Theorem 4 is derived for a fixed value of $s^2$, $s^2$ can be data-adaptively estimated to robustify model parameter estimation from incongenial external information. We obtain a data adaptive estimator for $s^2$ following an empirical-Bayes argument \citep{morris1983, mukherjee2008}, where the MLE of the TE, denoted by $\widehat{\theta}_a^I$, has the conditional distribution $[\widehat{\theta}_a^I \mid \widetilde{\theta}_a^I] \sim N\big(\widetilde{\theta}_a^I,\text{Var}(\widehat{\theta}_a^{I})\big)$, coupled with the random effect distribution $\widetilde{\theta}_a^I \sim N\Big(\widehat{\theta}_a^E,s^2\widehat{\text{Var}}(\widehat{\theta}_a^{E})\Big)$. Maximizing the marginal likelihood after integrating out $\widetilde{\theta}_a^I$ yields $\widehat{s}^2 = \big\{\widehat{\text{Var}}(\widehat{\theta}_a^{E})\big\}^{-1}\max\{0,(\widehat{\theta}_a^I-\widehat{\theta}_a^E)^2 - \widehat{\text{Var}}(\widehat{\theta}_a^{I})\}$. That is, $(\widehat{\theta}_a^I-\widehat{\theta}_a^E)^2 \leq \widehat{\text{Var}}(\widehat{\theta}_a^{I})$ corresponds to the hard constraint model and $(\widehat{\theta}_a^I-\widehat{\theta}_a^E)^2 > \widehat{\text{Var}}(\widehat{\theta}_a^{I})$ corresponds to the soft constraint model. However, from a practical perspective, we recommend that if $\widehat{s}^2 = 0$, then set $\widehat{s}^2$ equal to a small value close to zero and use the soft constraint algorithm; this helps to avoid coverage issues for the parametric bootstrap confidence intervals of the NDE.

\section{Simulations}
\label{s:simulations}

\subsection{Generative Model}
\label{ss:sim_generative_model}

The purpose of the simulation section is to empirically show estimation properties of the unconstrained, hard constraint, and soft constraint estimators. We consider one set of simulation scenarios where $\theta_a^I = \theta_a^E$ and two sets of simulation scenarios where $\theta_a^I \neq \theta_a^E$. The generative model for the internal data in all simulation settings is (\ref{eqn:outcome_model}) and (\ref{eqn:mediator_model}).

For the $\theta_a^I = \theta_a^E$ simulation scenarios, which we refer to as the congenial simulation scenarios, the parameters are set as follows: $p_m = 50$, $p_c = 5$, and $(A_i, \boldsymbol{C}_i^{\top})^{\top} \sim MVN(\boldsymbol{0},\boldsymbol{\Omega})$. Here, $\boldsymbol{\Omega}$ has an exchangeable correlation structure with the correlation parameter $\rho = 0.2$ and variance parameters equal to one. We consider two values of the internal sample size, $n = 200$ and $n = 2000$, with external sample sizes 10, 100, and 1000 times greater than the internal sample sizes. For the regression coefficient parameters in the mediator model, we fix $\boldsymbol{\alpha}_c$ so that it is a matrix of $0.1$'s and set $\boldsymbol{\alpha}_a = (0.6,\ldots,0.6,0,\ldots,0)^{\top}$ where $0.6$ and $0$ are repeated $10$ and $40$ times, respectively. The error variance-covariance matrix in the mediator model, $\boldsymbol{\Sigma}_m$, has a block exchangeable correlation structure with correlation parameters within blocks set to $0.3$ and the correlation parameters across blocks set to $0.2$. The error variance parameters in $\boldsymbol{\Sigma}_m$ are determined by pre-specified values of $R_{\boldsymbol{A} \mid \boldsymbol{C}}^2$, where $R_{\boldsymbol{A} \mid \boldsymbol{C}}^2 = (\boldsymbol{\alpha}_a)_1^2/\{\boldsymbol{\Sigma}_{m,jj}+(\boldsymbol{\alpha}_a)_1^2\}$ and $\boldsymbol{\Sigma}_{m,jj}$ is the $j$-th entry along the diagonal of $\boldsymbol{\Sigma}_m$. We consider two options: $R_{\boldsymbol{A} \mid \boldsymbol{C}}^2 = 0.05$ and $R_{\boldsymbol{A} \mid \boldsymbol{C}}^2 = 0.2$. For the regression coefficient parameters in the outcome model, $\boldsymbol{\beta}_c = (0.1,\ldots,0.1)^{\top}$, $\boldsymbol{\beta}_m = (0.1,\ldots,0.1,0,\ldots,0,0.1,\ldots,0.1,0,\ldots,0)^{\top}$, where the $0.1$'s are repeated $5$ times and the $0$'s are repeated $5$ and $35$ times, respectively, and $\beta_a = \theta_a^I - \boldsymbol{\alpha}_a^{\top}\boldsymbol{\beta}_m$. Here, $\theta_a^I = 1$. The error variance in the outcome model, $\sigma_e^2$, is determined based on $R_{\boldsymbol{M} \mid \boldsymbol{A},\boldsymbol{C}}^2$. In this case, we consider three options: $R_{\boldsymbol{M} \mid \boldsymbol{A},\boldsymbol{C}}^2 = 0.2$, $R_{\boldsymbol{M} \mid \boldsymbol{A},\boldsymbol{C}}^2 = 0.5$, and $R_{\boldsymbol{M} \mid \boldsymbol{A},\boldsymbol{C}}^2 = 0.8$. The purpose of varying $R_{\boldsymbol{A} \mid \boldsymbol{C}}^2$ and $R_{\boldsymbol{M} \mid \boldsymbol{A},\boldsymbol{C}}^2$ is because our asymptotic variance results from Section \ref{s:asym_eff_res} suggest that these quantities govern the asymptotic relative efficiency gains for NDE and NIE estimation. To generate external datasets from the external TE model, we use the same parameter values as the internal TE model but with different sample sizes of $n_E = 10n$, $n_E = 100n$, and $n_E = 1000n$. The simulated external TE estimate is then obtained by calculating the MLE on the simulated external data, $\widehat{\theta}_a^E$ and $\widehat{\text{Var}}(\widehat{\theta}_a^E)$.

For the $\theta_a^I \neq \theta_a^E$ simulation scenarios, we consider the same simulation parameters as the congenial external model simulation settings, with one exception; we either set $\theta_a^E = 2$ or randomly generate an external TE parameter such that $\theta_a^E \sim N(1,0.1)$. The former scenario considers a case where the external population is notably different from the internal population, potentially due to transportability violations. The latter scenario considers the average performance across a distribution of possible population-level external TE realizations, including both congenial and incongenial settings with the internal TE. The goal of these simulations is to determine which methods are robust to discordant external and internal targets for the TE. We refer to the scenarios with $\theta_a^E = 2$ as incongenial settings and  those with $\theta_a^E \sim N(1,0.1)$ as random settings.

\subsection{Comparison Methods and Evaluation Metrics}
\label{ss:sim_comp_methods_eval_metrics}

We consider three estimators in all simulation scenarios: the unconstrained estimator, the soft constraint estimator with $\widehat{s}^2 = \big\{\widehat{\text{Var}}(\widehat{\theta}_a^{E})\big\}^{-1}\max\{0,(\widehat{\theta}_a^I-\widehat{\theta}_a^E)^2 - \widehat{\text{Var}}(\widehat{\theta}_a^{I})\}$, and the hard constraint estimator. For brevity, we refer to the soft constraint estimator as the soft constraint empirical-Bayes (EB) estimator. As a benchmark for the maximal possible efficiency gain attainable from leveraging external information on the TE, we also consider the hard constraint estimator with the true $\theta_a^I$ enforced as the hard constraint on the TE, although this is not implementable in practice. We evaluate these estimators based on their relative root mean-squared error (RMSE) for NDE and NIE estimation compared with their unconstrained equivalents. Note that in the random simulation settings this is a root integrated mean-squared error (RIMSE) metric, where the mean-squared error is integrated over the generative distribution of $\theta_a^E$. Moreover, we evaluate the empirical coverage probabilities corresponding to 95\% asymptotic confidence intervals. Since $\boldsymbol{\alpha}_a \neq 0$ and $\boldsymbol{\beta}_m \neq 0$, we do not use the asymptotic results from Section \ref{ss:asym_eff_special_case} to construct confidence intervals. All RMSE and RIMSE estimates are based on 2000 simulation replicates.

\subsection{Results}
\label{ss:sim_results}

\begin{figure}[!ht]
    \centering
    \includegraphics[scale=1.0, height = 0.9\textheight, width = 1.0\linewidth]{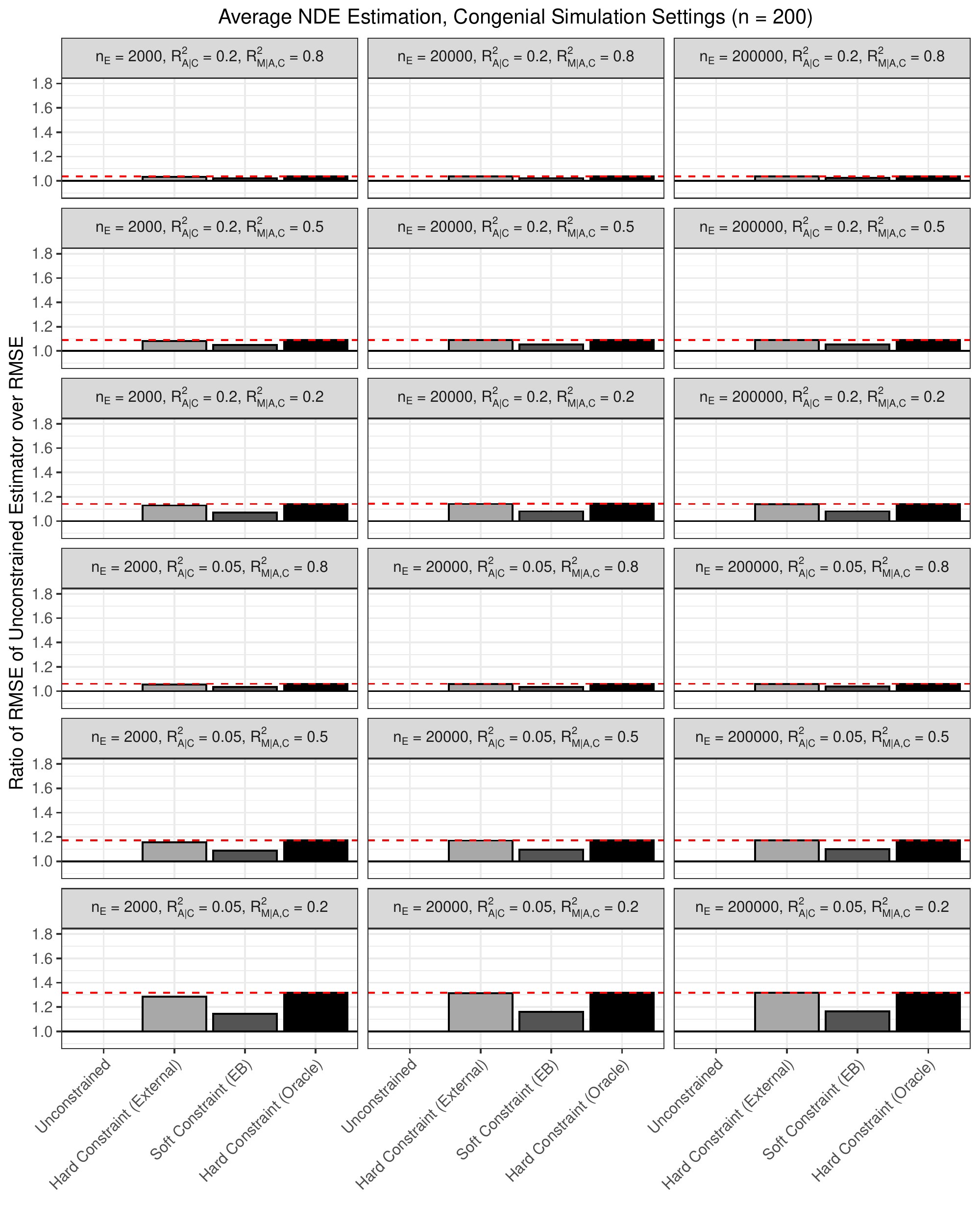}
    \caption{Relative root mean-squared error (RMSE) corresponding to Natural Direct Effect (NDE) estimation for the congenial simulation scenarios ($n = 200$). The red, horizontal dashed line indicates the upper bound on the possible gain in estimation efficiency, as determined by the hard constraint estimator with the oracle constraint.}
    \label{fig:rmse_n200_nde_unpenalized_correct}
\end{figure}

\begin{figure}[!ht]
    \centering
    \includegraphics[scale=1.0, height = 0.9\textheight, width = 1.0\linewidth]{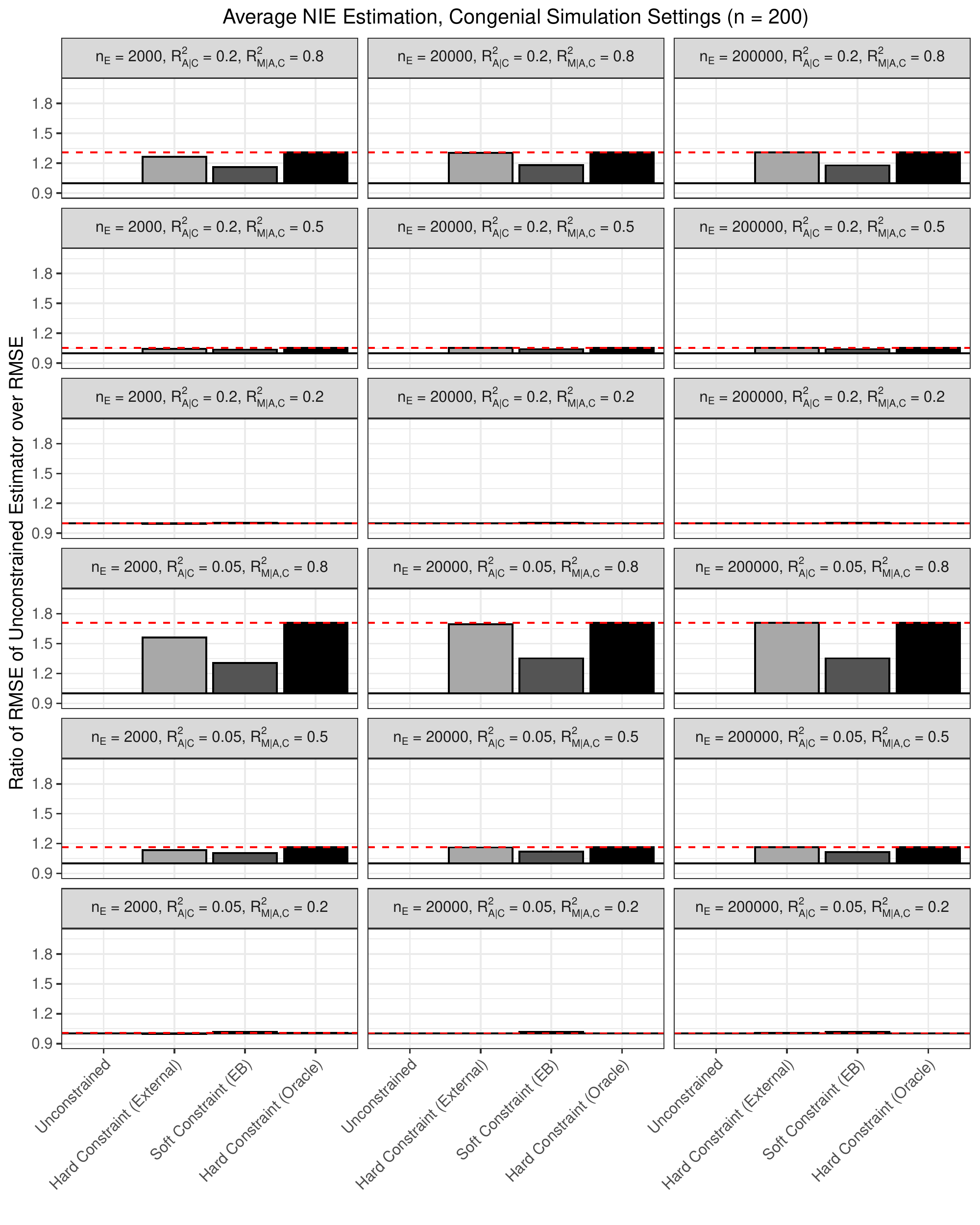}
    \caption{Relative root mean-squared error (RMSE) corresponding to Natural Indirect Effect (NIE) estimation for the congenial simulation scenarios ($n = 200$). The red, horizontal dashed line indicates the upper bound on the possible gain in estimation efficiency, as determined by the hard constraint estimator with the oracle constraint.}
    \label{fig:rmse_n200_nie_unpenalized_correct}
\end{figure}

Figures \ref{fig:rmse_n200_nde_unpenalized_correct} and \ref{fig:rmse_n200_nie_unpenalized_correct} show the relative RMSE for NDE and NIE estimation in the congenial simulation settings where $n = 200$. In general, the hard and soft constraint estimators demonstrate smaller RMSE than the unconstrained estimator, with the hard constraint estimator mostly having the best performance. More specifically, larger relative efficiency gains for NDE estimation occur when $R_{\boldsymbol{A}\mid\boldsymbol{C}}^2$ and $R_{\boldsymbol{M}\mid\boldsymbol{A},\boldsymbol{C}}^2$ are small. For example, when $n_E = 20000$, $R_{\boldsymbol{A}\mid\boldsymbol{C}}^2 = 0.05$, and $R_{\boldsymbol{M}\mid\boldsymbol{A},\boldsymbol{C}}^2 = 0.2$, the RMSE of the unconstrained estimator is 31.4\% higher than that of the hard constraint estimator and 16.1\% higher than that of the soft constraint estimator. However, when $n_E = 20000$, $R_{\boldsymbol{A}\mid\boldsymbol{C}}^2 = 0.2$, and $R_{\boldsymbol{M}\mid\boldsymbol{A},\boldsymbol{C}}^2 = 0.8$, the RMSE of the unconstrained estimator is only 3.7\% higher than that of the hard constraint estimator and 2.3\% higher than that of the soft constraint estimator. Conversely, larger relative efficiency gains for NIE estimation occur when $R_{\boldsymbol{A}\mid\boldsymbol{C}}^2$ is small and $R_{\boldsymbol{M}\mid\boldsymbol{A},\boldsymbol{C}}^2$ is large. As an example, when $n_E = 20000$, $R_{\boldsymbol{A}\mid\boldsymbol{C}}^2 = 0.05$, and $R_{\boldsymbol{M}\mid\boldsymbol{A},\boldsymbol{C}}^2 = 0.8$, the RMSE of the unconstrained estimator is 69.5\% higher than that of the hard constraint estimator and 35.2\% higher than that of the soft constraint estimator. However, when $n_E = 20000$, $R_{\boldsymbol{A}\mid\boldsymbol{C}}^2 = 0.2$, and $R_{\boldsymbol{M}\mid\boldsymbol{A},\boldsymbol{C}}^2 = 0.2$, the RMSE of the unconstrained estimator is only 0.3\% lower than that of the hard constraint estimator and 0.4\% higher than that of the soft constraint estimator. Therefore, these findings empirically corroborate the conclusions of Theorem 1, Theorem 2, and Theorem 4. Additionally, Figures \ref{fig:rmse_n200_nde_unpenalized_correct} and \ref{fig:rmse_n200_nie_unpenalized_correct} show that as $n_E$ increases for a fixed value of $n$, the hard constraint estimator approaches the upper bound of the achievable relative RMSE as measured by the dashed horizontal line at hard constraint (oracle). For the $n = 2000$ congenial simulations, trends for the relative RMSE are the same as the trends for the $n = 200$ congenial simulation scenarios (see Web Figures 1 and 2).

Web Figures 3 and 4 show the empirical coverage probabilities of the 95\% asymptotic confidence intervals in the $n = 2000$ congenial simulation settings. For the NDE, all methods have approximately 95\% empirical coverage probabilities across all simulation settings. For the NIE, all methods achieve nominal coverage rates when $R_{\boldsymbol{M}\mid\boldsymbol{A},\boldsymbol{C}}^2$ is small, although the hard and soft constraint methods tend to be slightly anti-conservative when $R_{\boldsymbol{M}\mid\boldsymbol{A},\boldsymbol{C}}^2$ is large and $R_{\boldsymbol{A}\mid\boldsymbol{C}}^2$ is small. Web Figures 5 and 6 show the empirical coverage probabilities of the 95\% asymptotic confidence intervals in the $n = 200$ congenial simulation settings. For the NDE, the asymptotic normality based confidence intervals for the unconstrained and hard constraint estimators exhibit some degree of undercoverage, suggesting that a larger internal sample size is needed for the asymptotic confidence intervals derived from Theorems 1 and 2 to achieve nominal coverage rates. Conversely, the parametric bootstrap-based confidence intervals for NDE inference in the soft constraint method result in nominal coverage rates. For the NIE, all methods generally exhibit slight undercoverage in all settings, except for the unconstrained method when $R_{\boldsymbol{A}\mid\boldsymbol{C}}^2 = 0.05$.

\begin{figure}[!ht]
    \centering
    \includegraphics[scale=1.0, height = 0.9\textheight, width = 1.0\linewidth]{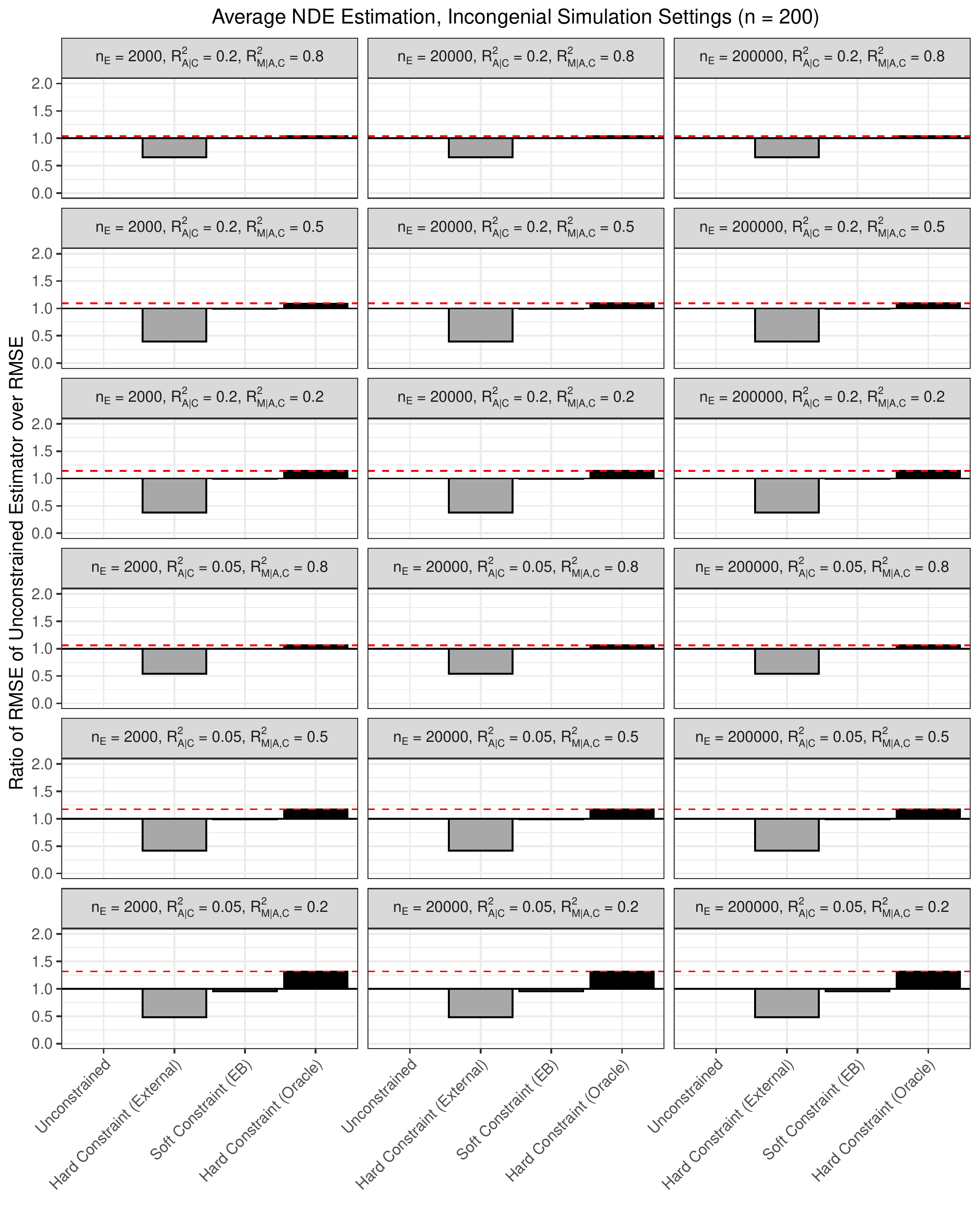}
    \caption{Relative root mean-squared error (RMSE) corresponding to Natural Direct Effect (NDE) estimation for the incongenial simulation scenarios ($n = 200$). The red, horizontal dashed line indicates the upper bound on the possible gain in estimation efficiency, as determined by the hard constraint estimator with the oracle constraint.}
    \label{fig:rmse_n200_nde_unpenalized_incorrect_fixed}
\end{figure}

\begin{figure}[!ht]
    \centering
    \includegraphics[scale=1.0, height = 0.9\textheight, width = 1.0\linewidth]{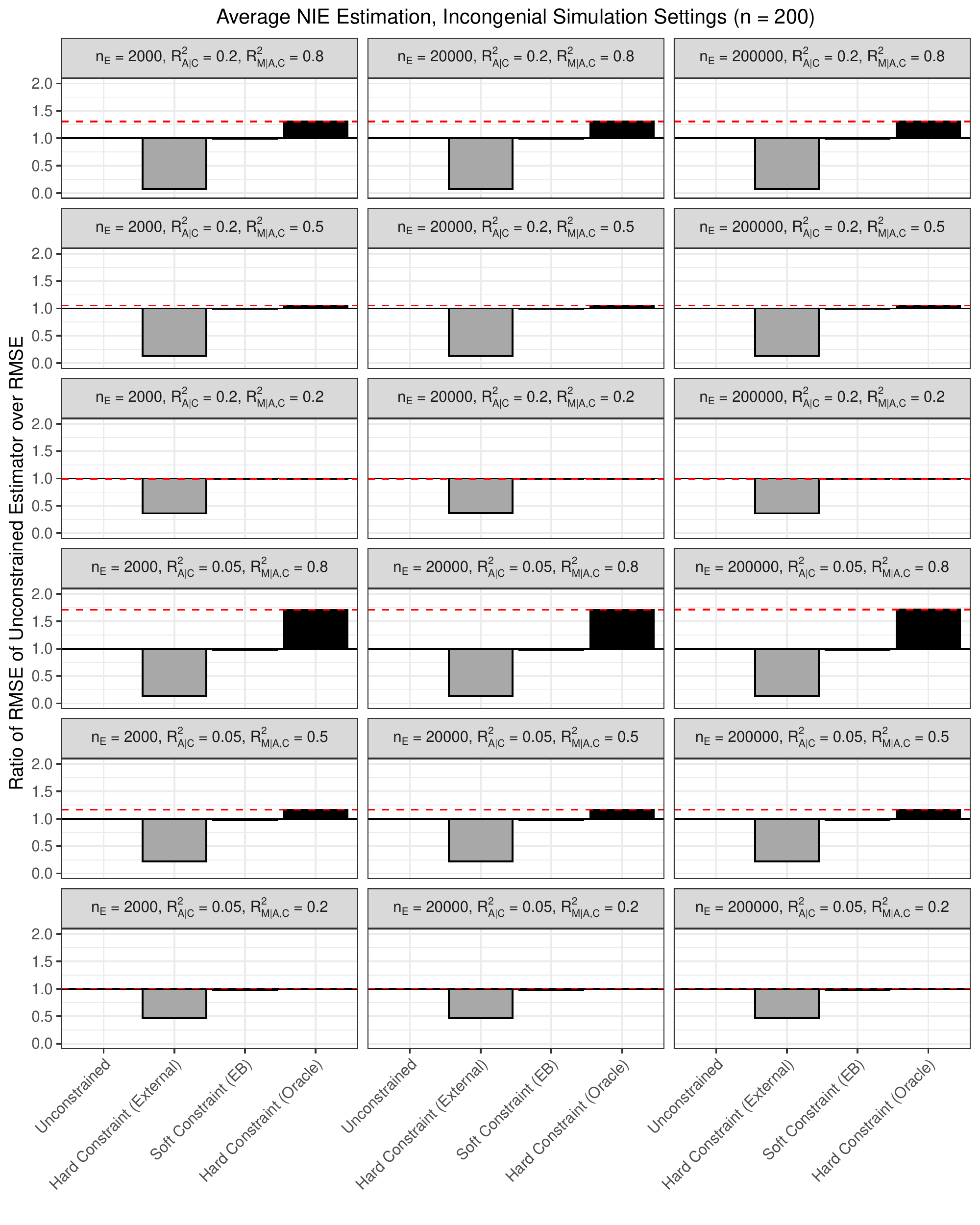}
    \caption{Relative root mean-squared error (RMSE) corresponding to Natural Indirect Effect (NIE) estimation for the incongenial simulation scenarios ($n = 200$). The red, horizontal dashed line indicates the upper bound on the possible gain in estimation efficiency, as determined by the hard constraint estimator with the oracle constraint.}
    \label{fig:rmse_n200_nie_unpenalized_incorrect_fixed}
\end{figure}

Figures \ref{fig:rmse_n200_nde_unpenalized_incorrect_fixed} and \ref{fig:rmse_n200_nie_unpenalized_incorrect_fixed} show the relative RMSE for NDE and NIE estimation in the $n = 200$ incongenial simulation settings. The unconstrained estimator has a 34.8\% - 62.4\% lower RMSE for NDE estimation and a 53.6\% - 92.7\% lower RMSE for NIE estimation compared to the hard constraint estimator. For example, when $n_E = 20000$, $R_{\boldsymbol{A}\mid\boldsymbol{C}}^2 = 0.2$, and $R_{\boldsymbol{M}\mid\boldsymbol{A},\boldsymbol{C}}^2 = 0.5$, the RMSE of the unconstrained estimator is 60.5\% lower and 86.2\% lower than that of the hard constraint estimator for NDE and NIE estimation, respectively. However, the soft constraint estimator has nearly identical RMSE to the unconstrained estimator, indicating no loss in estimation efficiency; when $n_E = 20000$, $R_{\boldsymbol{A}\mid\boldsymbol{C}}^2 = 0.2$, and $R_{\boldsymbol{M}\mid\boldsymbol{A},\boldsymbol{C}}^2 = 0.5$, the RMSE of the unconstrained estimator is 0.2\% lower and 0.3\% lower than that of the soft constraint for NDE and NIE estimation, respectively. Hence, the soft constraint (EB) estimator recovers the performance of the unconstrained estimator when the external information is incongenial. Moreover, for the random simulation settings, the conclusions are similar to the incongenial simulations settings, although the trends are less extreme because $\theta_a^E \sim N(1, 0.1)$ is almost always closer to $\theta_a^I = 1$ than $\theta_a^E = 2$ (see Web Figures 7-8). With respect to coverage, both the soft constraint (EB) and unconstrained methods achieve the nominal coverage rate when $n = 2000$ (see Web Figures 9-12). This suggests that coverage for the soft constraint (EB) confidence intervals is also robust to incongenial external information.

\section{Data Example}
\label{s:data_example}

PROTECT is a prospective birth cohort study in Puerto Rico that aims to better understand how environmental chemical exposures adversely impact birth outcomes. Women are followed-up at three visits throughout pregnancy, with visit 1 taking place at a median of 18 weeks, visit 2 at a median of 22 weeks, and visit 3 at a median of 26 weeks. In the proposed analysis, gestational age at delivery is the outcome of interest and urinary phthalate metabolites at visits 1 and 2 are the exposure of interest. Phthalates are used to make plastics more durable and flexible and exposure in humans usually occurs through ingestion of contaminated food and water, the use of personal care products, and physical contact with household items such as polyvinyl flooring and shower curtains \citep{ferguson2014, boss2018}. Numerous studies in the United States have shown that higher exposure to phthalates is significantly associated with shorter gestational age at delivery \citep{welch2022}.

Though current research is sparse, there is evidence of altered Cytochrome p450 metabolites among women that had spontaneous preterm deliveries compared to those who had full-term deliveries \citep{aung2019, borkowski2020}.  There is also evidence that cytochrome p450 partially mediates the effect of a phthalate risk score on gestational age at delivery \citep{aung2020}. In PROTECT, 18 Cytochrome p450 metabolites are measured at the third visit. Therefore, the proposed analysis investigates a mediation hypothesis, where the effect of log-transformed, specific-gravity adjusted phthalate metabolites at the first and second visits on gestational age at delivery is mediated by log-transformed Cytochrome p450 metabolites at the third visit, adjusted for maternal age, education, and pre-pregnancy body mass index. The phthalate metabolites of interest in this analysis are Monobutyl phthalate (MBP), Monobenzyl phthalate (MBzP), and Monoisobutyl phthalate (MiBP), which are selected based on their significant TEs as reported in eTable 13 in \cite{welch2022}. The external summary-level information on the TE is obtained by re-generating the eTable 13 models from \cite{welch2022} excluding the PROTECT study. Depending on the visit number and phthalate metabolite, the external sample size ranges between 4890 and 4944 and the internal sample size ranges between 445 and 456 (see Web Table 1 for descriptive statistics).

\begin{figure}[!ht]
    \centering
    \includegraphics[scale=1.0, height = 0.55\textheight, width = 1.0\linewidth]{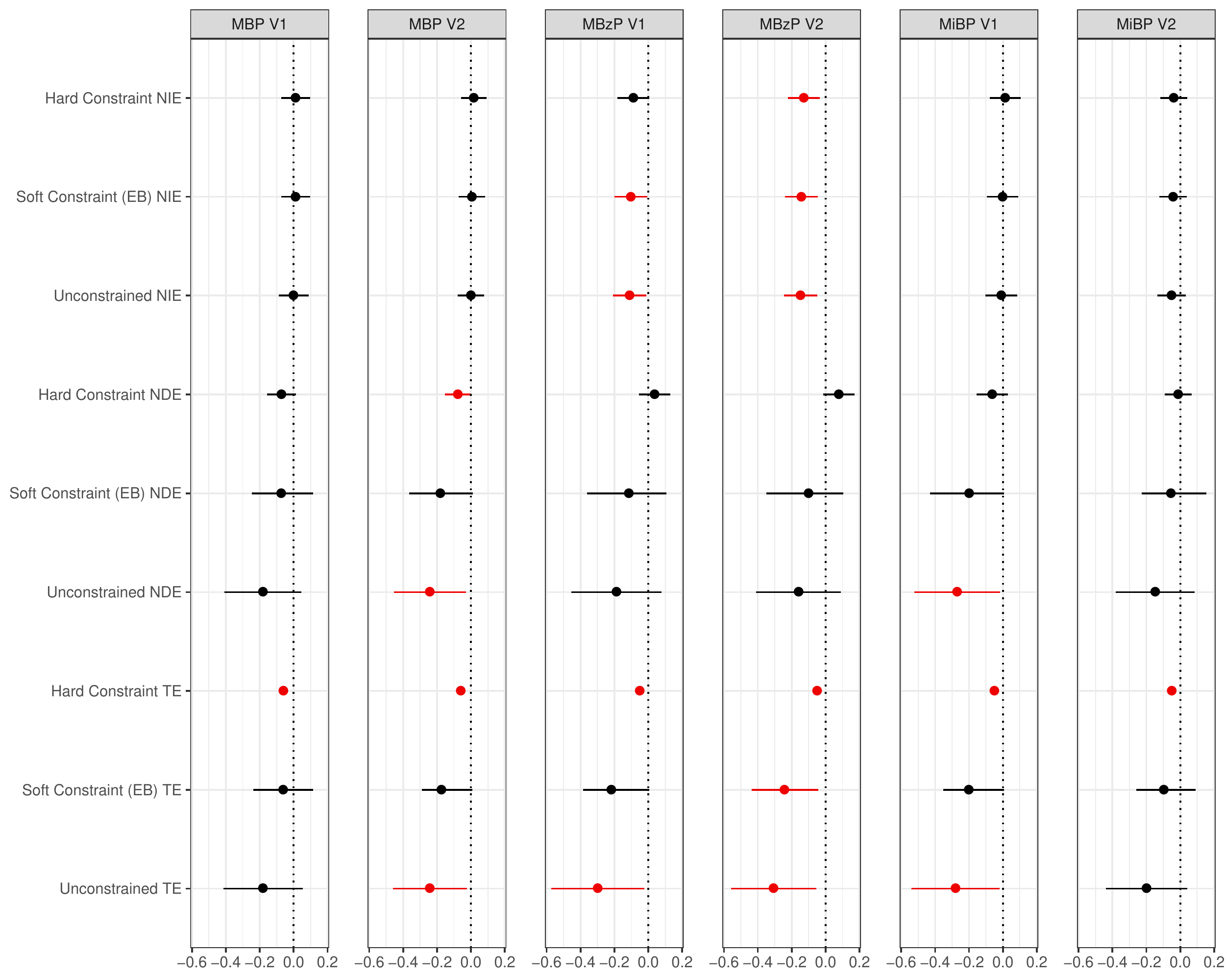}
    \caption{Results for the PROTECT mediation analysis, with the unconstrained, soft constraint, and hard constraint methods. Recall that MBP, MBzP, and MiBP are individual phthalate metbolites and V1 and V2 indicate that the results correspond to visit one and visit two, respectively. External sample size for MBP is 4944 and the external sample sizes for MBzP and MiBP are 4890. The internal sample sizes slightly differ for each phthalate metabolite and visit pair, however they all range between 445 and 456. All models are adjusted for maternal age, education, and maternal pre-pregnancy body mass index.}
    \label{fig:protect_results}
\end{figure}

Figure \ref{fig:protect_results} presents the results of the mediation analyses using the unconstrained, soft constraint (EB), and hard constraint methods. MBzP is the only phthalate metabolite for which at least one method identifies a significant NIE. Interestingly, both the unconstrained and soft constraint methods decompose the MBzP TE in such a way that the estimated TE, NDE, and NIE are all negative, implying that Cytochrome p450 metabolites may partially explain the mechanism by which MBzP exposure shortens gestational age at delivery. For example, according to the unconstrained and soft constraint (EB) methods, the Cytochrome p450 metabolites are estimated to mediate 48.2\% and 58.8\% of the relationship between MBzP and gestational age at delivery at visit 2, respectively. The hard constraint interval lengths for the TE, NDE, and NIE are uniformly narrower than the soft constraint (EB) interval lengths, which in turn are uniformly narrower than the unconstrained interval lengths. For example, the unconstrained method for MiBP at visit 2 has interval lengths of 0.479 for the TE, 0.465 for the NDE, and 0.166 for the NIE, the soft constraint (EB) method yields interval lengths of 0.352 for the TE, 0.382 for the NDE, and 0.161 for the NIE, and the hard constraint method yields interval lengths of 0 for the TE, 0.159 for the NDE, and 0.159 for NIE. Larger reductions in the interval lengths are observed for the NDE compared to the NIE because $\widehat{R}_{\boldsymbol{M}\mid\boldsymbol{A},\boldsymbol{C}}^2$ ranges between $0.08$ and $0.10$, as estimated by the unconstrained method. Also note that the hard constraint method has a TE interval length of 0 because the hard constraint model guarantees $\widehat{\text{TE}}^H = \widehat{\theta}_a^E$. Our results provide corroborating evidence to the findings of the LIFECODES study, which also identified a significant NIE associated with Cytochrome p450 metabolites \citep{aung2020}.

\section{Discussion}
\label{s:discussion}

In this paper, we show that external summary-level information on the TE can be used to improve NDE and NIE estimation in an internal mediation analysis (see Web Table 2 for a summary of the different methods presented in this paper). When the signal-to-noise ratio in the mediator model is low, large $R_{\boldsymbol{M}\mid\boldsymbol{A},\boldsymbol{C}}^2$ results in substantially more efficient NIE estimation and small $R_{\boldsymbol{M}\mid\boldsymbol{A},\boldsymbol{C}}^2$ results in substantially more efficient NDE estimation. Smaller values of $R_{\boldsymbol{M}\mid\boldsymbol{A},\boldsymbol{C}}^2$ and the signal-to-noise ratio are more common in practice, so we generally expect see large improvements for NDE estimation. Furthermore, when the TEs in the internal and external populations differ, we can then employ EB estimation strategies to robustify shrinkage towards the external TE estimate. One major limitation is the generalizability of our results when mediators and outcomes are non-continuous or when the internal mediation model is missspecified. When the outcome or mediators are non-continuous or there is misspecification of the mean structure of the mediator and outcome models, such as when exposure-mediator interaction is present, then the expression for the TE becomes a function of the confounders and the exposure level, making leveraging external estimates less straightforward. More work is needed to build a general framework for incorporating external information on the TE into a broader class of mediation models.

One additional technical challenge that we did not fully address in the paper is how to handle the $\boldsymbol{\alpha}_a = \boldsymbol{\beta}_m = \boldsymbol{0}$ case when constructing asymptotic confidence intervals for the NDE and NIE. While we recommended using a Wald test to test whether the null hypothesis $\boldsymbol{\alpha}_a = \boldsymbol{\beta}_m = \boldsymbol{0}$ holds as a workaround, there are likely more principled ways to construct the appropriate asymptotic reference distribution for confidence interval construction based on mixture distributions. The major challenge is that the relative weight to assign the asymptotic reference distributions from Theorems 1, 2, and 4 compared to the reference distribution from Theorem 3 is unknown, and therefore needs to be estimated \citep{liu2022}. There is existing work in the mediation analysis literature which discusses this issue in the context of large-scale causal mediation effect identification, namely through the construction of the Divide-Aggregate Composite-Null (DACT) test, however this solves the problem by running many single-mediator tests to estimate the probability of each of the three cases in the composite null rather than trying to estimate the probability that $\boldsymbol{\alpha}_a = \boldsymbol{\beta}_m = \boldsymbol{0}$ \citep{liu2022}. In our simulations we assumed that $\boldsymbol{\alpha}_a \neq \boldsymbol{0}$ or $\boldsymbol{\beta}_m \neq \boldsymbol{0}$ and directly used the asymptotic normality results from Theorems 1, 2, and 4 as a way to check our theoretical results. In the data example, we used the Wald test to test $\boldsymbol{\alpha}_a = \boldsymbol{\beta}_m = \boldsymbol{0}$ versus $\boldsymbol{\alpha}_a \neq \boldsymbol{0}$ or $ \boldsymbol{\beta}_m \neq \boldsymbol{0}$ for all methods, all of which rejected the null hypothesis at the $0.05$ level, and therefore directly used Theorem 1, Theorem 2, and Theorem 4 to construct confidence intervals. Since the main aim of the paper is to comment on the relative efficiency gains attributable to leveraging the external summary-level information on the TE, we leave this topic as future work.


\backmatter


\section*{Acknowledgements}

Funding for JDM and AC from NIH R01ES031591, R01ES032203, P42ES017198, P30ES017885, UH3OD023251. KKF was supported by the Intramural Research Program of the National Institute of Environmental Health Sciences, National Institutes of Health (ZIA ES103321). The research of BM was supported by NSF DMS 1712933, NIH NCI UG3 CA267907. The research of JK was partially supported by NIH R01DA048993, R01MH105561, R01GM124061 and NSF IIS2123777. The authors would additionally like to thank all participants of PROTECT and the Pooled Phthalate Exposure and Preterm Birth Study Group for supporting supplemental analyses necessary for the data example. \vspace*{-8pt}


%

\newcommand{\newblock}{}
\bibliographystyle{biom} \bibliography{main_bib.bib}


\appendix


%

\label{lastpage}

\end{document}


\maketitle

\section*{Web Appendix A}
\label{app:algorithms}

\subsection*{Unconstrained (Closed-Form Optimization)}

Suppose that $\boldsymbol{M}_{i\cdot}^{\top} \sim N(A_i\boldsymbol{\alpha}_a + \boldsymbol{\alpha}_c\boldsymbol{C}_{i\cdot}^{\top},\boldsymbol{\Sigma}_m)$ and $Y_i \sim N(\boldsymbol{M}_{i\cdot}\boldsymbol{\beta}_m + A_i\beta_a + \boldsymbol{C}_{i\cdot}\boldsymbol{\beta}_c, \sigma_e^2)$ for $i = 1,\ldots,n$. Let $\boldsymbol{Z} = (\boldsymbol{A},\boldsymbol{C},\boldsymbol{M})$. Then, $$\widehat{\boldsymbol{\beta}} = \big(\boldsymbol{Z}^{\top}\boldsymbol{Z}\big)^{-1}\boldsymbol{Z}^{\top}\boldsymbol{Y}$$

$$\widehat{\sigma}_e^2 = \frac{1}{n}(\boldsymbol{Y} - \boldsymbol{Z}\widehat{\boldsymbol{\beta}})^{\top}(\boldsymbol{Y} - \boldsymbol{Z}\widehat{\boldsymbol{\beta}})$$

\noindent Let $$\boldsymbol{X} = (\boldsymbol{A},\boldsymbol{C}), \hspace{2 mm} \boldsymbol{\alpha} = \begin{pmatrix} \boldsymbol{\alpha}_a^{\top} \\ \boldsymbol{\alpha}_c^{\top} \end{pmatrix}.$$ Then, $$\widehat{\boldsymbol{\alpha}} = \big(\boldsymbol{X}^{\top}\boldsymbol{X}\big)^{-1}\boldsymbol{X}^{\top}\boldsymbol{M}$$
$$\widehat{\boldsymbol{\Sigma}}_m = \frac{1}{n}\sum_{i=1}^{n}\big(\boldsymbol{M}_{i\cdot}^{\top} - \widehat{\boldsymbol{\alpha}}^{\top}\boldsymbol{X}_{i\cdot}^{\top}\big)\big(\boldsymbol{M}_{i\cdot}^{\top} - \widehat{\boldsymbol{\alpha}}^{\top}\boldsymbol{X}_{i\cdot}^{\top}\big)^{\top}$$

\subsection*{Hard Constraint (Cyclical Coordinate Descent)}

Suppose that $\boldsymbol{M}_{i\cdot}^{\top} \sim N(A_i\boldsymbol{\alpha}_a + \boldsymbol{\alpha}_c\boldsymbol{C}_{i\cdot}^{\top},\boldsymbol{\Sigma}_m)$ and $Y_i \sim N(\boldsymbol{M}_{i\cdot}\boldsymbol{\beta}_m + A_i(\widehat{\theta}_a^E - \boldsymbol{\alpha}_a^{\top}\boldsymbol{\beta}_m) + \boldsymbol{C}_{i\cdot}\boldsymbol{\beta}_c, \sigma_e^2)$ for $i = 1,\ldots,n$. Then the cyclical coordinate descent updates are as follows:

$$\widetilde{\boldsymbol{\alpha}}_c \leftarrow \big(\boldsymbol{C}^{\top}\boldsymbol{C}\big)^{-1}\boldsymbol{C}^{\top}\big(\boldsymbol{M} - \boldsymbol{A}\widetilde{\boldsymbol{\alpha}}_a^{\top}\big)$$

$$\widetilde{\boldsymbol{\alpha}}_a \leftarrow \frac{1}{\boldsymbol{A}^{\top}\boldsymbol{A}}\bigg(\widetilde{\boldsymbol{\Sigma}}_m^{-1} + \frac{1}{\widetilde{\sigma}_e^2}\widetilde{\boldsymbol{\beta}}_m\widetilde{\boldsymbol{\beta}}_m^{\top}\bigg)^{-1}\sum_{i=1}^{n}\Bigg[A_i\widetilde{\boldsymbol{\Sigma}}_m^{-1}\big(\boldsymbol{M}_{i\cdot}^{\top} - \widetilde{\boldsymbol{\alpha}}_c^{\top}\boldsymbol{C}_{i\cdot}^{\top}\big) - \frac{1}{\widetilde{\sigma}_e^2}A_i\big(Y_i - \widehat{\theta}_a^EA_i - \boldsymbol{M}_{i\cdot}\widetilde{\boldsymbol{\beta}}_m - \boldsymbol{C}_{i\cdot}\widetilde{\boldsymbol{\beta}}_c\big)\widetilde{\boldsymbol{\beta}}_m\Bigg]$$

$$\widetilde{\boldsymbol{\Sigma}}_m \leftarrow \frac{1}{n}\sum_{i=1}^{n}\big(\boldsymbol{M}_{i\cdot}^{\top} - \widetilde{\boldsymbol{\alpha}}^{\top}\boldsymbol{X}_{i\cdot}^{\top}\big)\big(\boldsymbol{M}_{i\cdot}^{\top} - \widetilde{\boldsymbol{\alpha}}^{\top}\boldsymbol{X}_{i\cdot}^{\top}\big)^{\top}, \hspace{2 mm} \boldsymbol{X} = (\boldsymbol{A},\boldsymbol{C}), \hspace{2 mm} \boldsymbol{\alpha} = \begin{pmatrix} \boldsymbol{\alpha}_a^{\top} \\ \boldsymbol{\alpha}_c^{\top} \end{pmatrix}$$

$$\widetilde{\boldsymbol{\beta}}_c \leftarrow \big(\boldsymbol{C}^{\top}\boldsymbol{C}\big)^{-1}\boldsymbol{C}^{\top}\big(\boldsymbol{Y} - \{\widehat{\theta}_a^E - \widetilde{\boldsymbol{\alpha}}_a^{\top}\widetilde{\boldsymbol{\beta}}_m\}\boldsymbol{A} - \boldsymbol{M}\widetilde{\boldsymbol{\beta}}_m\big)$$

$$\widetilde{\boldsymbol{\beta}}_m \leftarrow \bigg(\sum_{i=1}^{n}\big(\boldsymbol{M}_{i\cdot}^{\top} - A_i\widetilde{\boldsymbol{\alpha}}_a\big)\big(\boldsymbol{M}_{i\cdot}^{\top} - A_i\widetilde{\boldsymbol{\alpha}}_a\big)^{\top}\bigg)^{-1}\sum_{i=1}^{n}\big(Y_i - A_i\widehat{\theta}_a^E - \boldsymbol{C}_{i\cdot}\widetilde{\boldsymbol{\beta}}_c\big)\big(\boldsymbol{M}_{i\cdot}^{\top} - A_i\widetilde{\boldsymbol{\alpha}}_a\big)$$

$$\widetilde{\sigma}_e^2 \leftarrow \frac{1}{n}\sum_{i=1}^{n}\big(Y_i - \{\widehat{\theta}_a^E - \widetilde{\boldsymbol{\alpha}}_a^{\top}\widetilde{\boldsymbol{\beta}}_m\}A_i - \boldsymbol{M}_{i\cdot}\widetilde{\boldsymbol{\beta}}_m - \boldsymbol{C}_{i\cdot}\widetilde{\boldsymbol{\beta}}_c\big)^2$$

\noindent Note that the cyclic coordinate descent algorithm is similar to Gibbs sampling, in that you use the always use the most recently updated values for all as you update each parameter. The initial values for the algorithm come from the unconstrained optimization algorithm.

\subsection*{Expectation-Maximization (EM) Algorithm (Soft Constraint)}

Within the EM algorithm framework we are treating $\boldsymbol{Y}$, $\boldsymbol{M}$, $\boldsymbol{A}$, and $\boldsymbol{C}$ as our observed data and $\widetilde{\theta}_a^I$ as the unobserved latent data. Let $\mathcal{P} = \{\boldsymbol{\alpha}_a,\boldsymbol{\alpha}_c,\boldsymbol{\Sigma}_M,\boldsymbol{\beta}_m,\boldsymbol{\beta}_c,\sigma_e^2\}$. That is, we wish to maximize the marginal likelihood: $$L(\mathcal{P} \mid \boldsymbol{Y},\boldsymbol{M},\boldsymbol{A},\boldsymbol{C}) = \int_{-\infty}^{\infty}\pi(\boldsymbol{Y}\mid\boldsymbol{M},\boldsymbol{A},\boldsymbol{C},\widetilde{\theta}_a^I)\pi(\boldsymbol{M}\mid\boldsymbol{A}, \boldsymbol{C})\pi(\widetilde{\theta}_a^I)d\widetilde{\theta}_a^I$$

\noindent Moreover, we have that $$\pi(\widetilde{\theta}_a^I \mid \boldsymbol{Y},\boldsymbol{M},\boldsymbol{A},\boldsymbol{C}) \propto \pi(\boldsymbol{Y} \mid \boldsymbol{M},\boldsymbol{A},\boldsymbol{C}, \widetilde{\theta}_a^I)\pi(\widetilde{\theta}_a^I)$$ which implies that $$[\widetilde{\theta}_a^I \mid \boldsymbol{Y},\boldsymbol{M},\boldsymbol{A},\boldsymbol{C}] \sim N\Bigg(\bigg[\frac{\boldsymbol{A}^{\top}\boldsymbol{A}}{\sigma_e^2}+\frac{1}{s^2\widehat{\text{Var}}(\widehat{\theta}_a^{E})}\bigg]^{-1}\bigg[\frac{\boldsymbol{A}^{\top}\boldsymbol{Y}^*}{\sigma_e^2}+\frac{\widehat{\theta}_a^E}{s^2\widehat{\text{Var}}(\widehat{\theta}_a^{E})}\bigg],\bigg[\frac{\boldsymbol{A}^{\top}\boldsymbol{A}}{\sigma_e^2}+\frac{1}{s^2\widehat{\text{Var}}(\widehat{\theta}_a^{E})}\bigg]^{-1}\Bigg)$$

\noindent where $\boldsymbol{Y}^* = \boldsymbol{Y}-\boldsymbol{M}\boldsymbol{\beta}_m-\boldsymbol{C}\boldsymbol{\beta}_c+\boldsymbol{A}\boldsymbol{\alpha}_a^{\top}\boldsymbol{\beta}_m$. The complete data log-likelihood is:
$$l(\mathcal{P} \mid \boldsymbol{Y},\boldsymbol{M},\boldsymbol{A},\boldsymbol{C},\widetilde{\theta}_a^I) = -\frac{np_m}{2}\log(2\pi) - \frac{n}{2}\log(|\boldsymbol{\Sigma}_m|)$$ $$- \frac{1}{2}\sum_{i=1}^{n}(\boldsymbol{M}_{i\cdot}^{\top} - \boldsymbol{\alpha}_c\boldsymbol{C}_{i\cdot}^{\top} - A_i\boldsymbol{\alpha}_a)^{\top}\boldsymbol{\Sigma}_m^{-1}(\boldsymbol{M}_{i\cdot}^{\top} - \boldsymbol{\alpha}_c\boldsymbol{C}_{i\cdot}^{\top} - A_i\boldsymbol{\alpha}_a)$$ $$- \frac{n}{2}\log(2\pi) - \frac{n}{2}\log(\sigma_e^2) - \frac{1}{2\sigma_e^2}\sum_{i=1}^{n}(Y_i - \boldsymbol{C}_{i\cdot}\boldsymbol{\beta}_c - \boldsymbol{M}_{i\cdot}\boldsymbol{\beta}_m - A_i\{\widetilde{\theta}_a^I - \boldsymbol{\alpha}_a^{\top}\boldsymbol{\beta}_m\})^2$$ $$- \frac{1}{2}\log\big(2{\pi}s^2\widehat{\text{Var}}(\widehat{\theta}_a^{E})\big) - \frac{1}{2s^2\widehat{\text{Var}}(\widehat{\theta}_a^{E})}(\widetilde{\theta}_a^I - \widehat{\theta}_a^E)^2$$

\noindent The E-Step is therefore: $$E_{\widetilde{\theta}_a^I \mid \boldsymbol{Y}, \boldsymbol{M}, \boldsymbol{A}, \boldsymbol{C}, \mathcal{P}^{(t)}}\big[l(\mathcal{P} \mid \boldsymbol{Y},\boldsymbol{M},\boldsymbol{A},\boldsymbol{C},\widetilde{\theta}_a^I)\big] = -\frac{np_m}{2}\log(2\pi) - \frac{n}{2}\log(|\boldsymbol{\Sigma}_m|)$$ $$- \frac{1}{2}\sum_{i=1}^{n}(\boldsymbol{M}_{i\cdot}^{\top} - \boldsymbol{\alpha}_c\boldsymbol{C}_{i\cdot}^{\top} - A_i\boldsymbol{\alpha}_a)^{\top}\boldsymbol{\Sigma}_m^{-1}(\boldsymbol{M}_{i\cdot}^{\top} - \boldsymbol{\alpha}_c\boldsymbol{C}_{i\cdot}^{\top} - A_i\boldsymbol{\alpha}_a)$$ $$- \frac{n}{2}\log(2\pi) - \frac{n}{2}\log(\sigma_e^2) - \frac{1}{2\sigma_e^2}\sum_{i=1}^{n}E_{\widetilde{\theta}_a^I \mid \boldsymbol{Y}, \boldsymbol{M}, \boldsymbol{A}, \boldsymbol{C}, \mathcal{P}^{(t)}}\big[(Y_i - \boldsymbol{C}_{i\cdot}\boldsymbol{\beta}_c - \boldsymbol{M}_{i\cdot}\boldsymbol{\beta}_m - A_i\{\widetilde{\theta}_a^I - \boldsymbol{\alpha}_a^{\top}\boldsymbol{\beta}_m\})^2\big]$$ $$- \frac{1}{2}\log\big(2{\pi}s^2\widehat{\text{Var}}(\widehat{\theta}_a^{E})\big) - \frac{1}{2s^2\widehat{\text{Var}}(\widehat{\theta}_a^{E})}E_{\widetilde{\theta}_a^I \mid \boldsymbol{Y}, \boldsymbol{M}, \boldsymbol{A}, \boldsymbol{C}, \mathcal{P}^{(t)}}\big[(\widetilde{\theta}_a^I - \widehat{\theta}_a^E)^2\big]$$

$$E_{\widetilde{\theta}_a^I \mid \boldsymbol{Y}, \boldsymbol{M}, \boldsymbol{A}, \boldsymbol{C}, \mathcal{P}^{(t)}}\big[l(\mathcal{P} \mid \boldsymbol{Y},\boldsymbol{M},\boldsymbol{A},\boldsymbol{C},\widetilde{\theta}_a^I)\big] = -\frac{np_m}{2}\log(2\pi) - \frac{n}{2}\log(|\boldsymbol{\Sigma}_m|)$$ $$- \frac{1}{2}\sum_{i=1}^{n}(\boldsymbol{M}_{i\cdot}^{\top} - \boldsymbol{\alpha}_c\boldsymbol{C}_{i\cdot}^{\top} - A_i\boldsymbol{\alpha}_a)^{\top}\boldsymbol{\Sigma}_m^{-1}(\boldsymbol{M}_{i\cdot}^{\top} - \boldsymbol{\alpha}_c\boldsymbol{C}_{i\cdot}^{\top} - A_i\boldsymbol{\alpha}_a)$$ $$- \frac{n}{2}\log(2\pi\sigma_e^2) - \frac{1}{2}\log\big(2{\pi}s^2\widehat{\text{Var}}(\widehat{\theta}_a^{E})\big)$$ $$-\frac{1}{2s^2\widehat{\text{Var}}(\widehat{\theta}_a^{E})}E_{\widetilde{\theta}_a^I \mid \boldsymbol{Y}, \boldsymbol{M}, \boldsymbol{A}, \boldsymbol{C}, \mathcal{P}^{(t)}}\big[\{\widetilde{\theta}_a^I\}^2\big] + \frac{\widehat{\theta}_a^E}{s^2\widehat{\text{Var}}(\widehat{\theta}_a^{E})}E_{\widetilde{\theta}_a^I \mid \boldsymbol{Y}, \boldsymbol{M}, \boldsymbol{A}, \boldsymbol{C}, \mathcal{P}^{(t)}}\big[\widetilde{\theta}_a^I\big] - \frac{(\widehat{\theta}_a^E)^2}{2s^2\widehat{\text{Var}}(\widehat{\theta}_a^{E})}$$ $$-\frac{1}{2\sigma_e^2}\big(\boldsymbol{Y}-\boldsymbol{M}\boldsymbol{\beta}_m-\boldsymbol{C}\boldsymbol{\beta}_c+\boldsymbol{A}\boldsymbol{\alpha}_a^{\top}\boldsymbol{\beta}_m\big)^{\top}\big(\boldsymbol{Y}-\boldsymbol{M}\boldsymbol{\beta}_m-\boldsymbol{C}\boldsymbol{\beta}_c+\boldsymbol{A}\boldsymbol{\alpha}_a^{\top}\boldsymbol{\beta}_m\big)$$ $$+\frac{1}{\sigma_e^2}E_{\widetilde{\theta}_a^I \mid \boldsymbol{Y}, \boldsymbol{M}, \boldsymbol{A}, \boldsymbol{C}, \mathcal{P}^{(t)}}\big[\widetilde{\theta}_a^I\big]\boldsymbol{A}^{\top}\big(\boldsymbol{Y}-\boldsymbol{M}\boldsymbol{\beta}_m-\boldsymbol{C}\boldsymbol{\beta}_c+\boldsymbol{A}\boldsymbol{\alpha}_a^{\top}\boldsymbol{\beta}_m\big) - \frac{1}{2\sigma_e^2}E_{\widetilde{\theta}_a^I \mid \boldsymbol{Y}, \boldsymbol{M}, \boldsymbol{A}, \boldsymbol{C}, \mathcal{P}^{(t)}}\big[\{\widetilde{\theta}_a^I\}^2\big]\boldsymbol{A}^{\top}\boldsymbol{A}$$

\noindent Note that we can calculate $$E_{\widetilde{\theta}_a^I \mid \boldsymbol{Y}, \boldsymbol{M}, \boldsymbol{A}, \boldsymbol{C}, \mathcal{P}^{(t)}}\big[\widetilde{\theta}_a^I\big] \text{ and } E_{\widetilde{\theta}_a^I \mid \boldsymbol{Y}, \boldsymbol{M}, \boldsymbol{A}, \boldsymbol{C}, \mathcal{P}^{(t)}}\big[\{\widetilde{\theta}_a^I\}^2\big]$$ from $$[\widetilde{\theta}_a^I \mid \boldsymbol{Y},\boldsymbol{M},\boldsymbol{A},\boldsymbol{C}] \sim N\Bigg(\bigg[\frac{\boldsymbol{A}^{\top}\boldsymbol{A}}{\sigma_e^2}+\frac{1}{s^2\widehat{\text{Var}}(\widehat{\theta}_a^{E})}\bigg]^{-1}\bigg[\frac{\boldsymbol{A}^{\top}\boldsymbol{Y}^*}{\sigma_e^2}+\frac{\widehat{\theta}_a^E}{s^2\widehat{\text{Var}}(\widehat{\theta}_a^{E})}\bigg],\bigg[\frac{\boldsymbol{A}^{\top}\boldsymbol{A}}{\sigma_e^2}+\frac{1}{s^2\widehat{\text{Var}}(\widehat{\theta}_a^{E})}\bigg]^{-1}\Bigg)$$

\noindent For the M-Step we will use a cyclical coordinate descent algorithm with the following updates:

$$\widetilde{\boldsymbol{\alpha}}_c \leftarrow \big(\boldsymbol{C}^{\top}\boldsymbol{C}\big)^{-1}\boldsymbol{C}^{\top}\big(\boldsymbol{M} - \boldsymbol{A}\widetilde{\boldsymbol{\alpha}}_a^{\top}\big)$$

$$\widetilde{\boldsymbol{\alpha}}_a \leftarrow \frac{1}{\boldsymbol{A}^{\top}\boldsymbol{A}}\bigg(\widetilde{\boldsymbol{\Sigma}}_m^{-1} + \frac{1}{\widetilde{\sigma}_e^2}\widetilde{\boldsymbol{\beta}}_m\widetilde{\boldsymbol{\beta}}_m^{\top}\bigg)^{-1}\Bigg[- \frac{1}{\widetilde{\sigma}_e^2}\boldsymbol{A}^{\top}\big(\boldsymbol{Y} - \boldsymbol{M}\widetilde{\boldsymbol{\beta}}_m - \boldsymbol{C}\widetilde{\boldsymbol{\beta}}_c - E_{\widetilde{\theta}_a^I \mid \boldsymbol{Y}, \boldsymbol{M}, \boldsymbol{A}, \boldsymbol{C}, \mathcal{P}^{(t)}}\big[\widetilde{\theta}_a^I\big]\boldsymbol{A}\big)\widetilde{\boldsymbol{\beta}}_m +\sum_{i=1}^{n}A_i\widetilde{\boldsymbol{\Sigma}}_m^{-1}\big(\boldsymbol{M}_{i\cdot}^{\top} - \widetilde{\boldsymbol{\alpha}}_c^{\top}\boldsymbol{C}_{i\cdot}^{\top}\big)\Bigg]$$

$$\widetilde{\boldsymbol{\Sigma}}_m \leftarrow \frac{1}{n}\sum_{i=1}^{n}\big(\boldsymbol{M}_{i\cdot}^{\top} - \widetilde{\boldsymbol{\alpha}}^{\top}\boldsymbol{X}_{i\cdot}^{\top}\big)\big(\boldsymbol{M}_{i\cdot}^{\top} - \widetilde{\boldsymbol{\alpha}}^{\top}\boldsymbol{X}_{i\cdot}^{\top}\big)^{\top}, \hspace{2 mm} \boldsymbol{X} = (\boldsymbol{A},\boldsymbol{C}), \hspace{2 mm} \boldsymbol{\alpha} = \begin{pmatrix} \boldsymbol{\alpha}_a^{\top} \\ \boldsymbol{\alpha}_c^{\top} \end{pmatrix}$$

$$\widetilde{\boldsymbol{\beta}}_c \leftarrow \big(\boldsymbol{C}^{\top}\boldsymbol{C}\big)^{-1}\boldsymbol{C}^{\top}\big(\boldsymbol{Y} - \{E_{\widetilde{\theta}_a^I \mid \boldsymbol{Y}, \boldsymbol{M}, \boldsymbol{A}, \boldsymbol{C}, \mathcal{P}^{(t)}}\big[\widetilde{\theta}_a^I\big] - \widetilde{\boldsymbol{\alpha}}_a^{\top}\widetilde{\boldsymbol{\beta}}_m\}\boldsymbol{A} - \boldsymbol{M}\widetilde{\boldsymbol{\beta}}_m\big)$$

$$\widetilde{\boldsymbol{\beta}}_m \leftarrow \bigg[\Big(\boldsymbol{M} - \boldsymbol{A}\widetilde{\boldsymbol{\alpha}}_a^{\top}\Big)^{\top}\Big(\boldsymbol{M} - \boldsymbol{A}\widetilde{\boldsymbol{\alpha}}_a^{\top}\Big)\bigg]^{-1}\Big(\boldsymbol{M} - \boldsymbol{A}\widetilde{\boldsymbol{\alpha}}_a^{\top}\Big)^{\top}\Big(\boldsymbol{Y} - \boldsymbol{C}\widetilde{\boldsymbol{\beta}}_c - \boldsymbol{A}E_{\widetilde{\theta}_a^I \mid \boldsymbol{Y}, \boldsymbol{M}, \boldsymbol{A}, \boldsymbol{C}, \mathcal{P}^{(t)}}\big[\widetilde{\theta}_a^I\big]\Big)$$

$$\widetilde{\sigma}_e^2 \leftarrow \frac{1}{n}\big(\boldsymbol{Y} - \boldsymbol{M}\widetilde{\boldsymbol{\beta}}_m - \boldsymbol{C}\widetilde{\boldsymbol{\beta}}_c + \boldsymbol{A}\widetilde{\boldsymbol{\alpha}}_a^{\top}\widetilde{\boldsymbol{\beta}}_m\big)^{\top}\big(\boldsymbol{Y} - \boldsymbol{M}\widetilde{\boldsymbol{\beta}}_m - \boldsymbol{C}\widetilde{\boldsymbol{\beta}}_c + \boldsymbol{A}\widetilde{\boldsymbol{\alpha}}_a^{\top}\widetilde{\boldsymbol{\beta}}_m\big)$$ $$-\frac{2}{n}E_{\widetilde{\theta}_a^I \mid \boldsymbol{Y}, \boldsymbol{M}, \boldsymbol{A}, \boldsymbol{C}, \mathcal{P}^{(t)}}\big[\widetilde{\theta}_a^I\big]\boldsymbol{A}^{\top}\big(\boldsymbol{Y} - \boldsymbol{M}\widetilde{\boldsymbol{\beta}}_m - \boldsymbol{C}\widetilde{\boldsymbol{\beta}}_c + \boldsymbol{A}\widetilde{\boldsymbol{\alpha}}_a^{\top}\widetilde{\boldsymbol{\beta}}_m\big)$$ $$+\frac{\boldsymbol{A}^{\top}\boldsymbol{A}}{n}E_{\widetilde{\theta}_a^I \mid \boldsymbol{Y}, \boldsymbol{M}, \boldsymbol{A}, \boldsymbol{C}, \mathcal{P}^{(t)}}\big[\{\widetilde{\theta}_a^I\}^2\big]$$

\newpage

\section*{Web Appendix B}

\subsection*{Proof of Theorem 1}

Suppose that $[\boldsymbol{M}_{i\cdot}^{\top} \mid A_i] \sim N(\boldsymbol{\alpha}_c + A_i\boldsymbol{\alpha}_a,\boldsymbol{\Sigma}_m)$ and $[Y_i \mid \boldsymbol{M}_{i\cdot}, A_i] \sim N(\beta_c + \boldsymbol{M}_{i\cdot}\boldsymbol{\beta}_m + A_i\beta_a, \sigma_e^2)$ is the true generative model where $i = 1,\ldots,n$. That is, without loss of generality, we are assuming that there are no confounders, but that there are intercept terms in both the outcome and mediator models. This implies that $\sigma_a^2 = \text{Var}(A_i)$. Moreover, assume that $E[A_i] = 0$. Let $\mathcal{D} = \{\boldsymbol{Y},\boldsymbol{M},\boldsymbol{A},\boldsymbol{C}\}$ denote the data. Then the log-likelihood is: $$l(\boldsymbol{\alpha}_c,\boldsymbol{\alpha}_a,\beta_c,\boldsymbol{\beta}_m,\beta_a \mid \mathcal{D}) = -\frac{np_m}{2}\log(2\pi) - \frac{n}{2}\log(|\boldsymbol{\Sigma}_m|) - \frac{1}{2}\sum_{i=1}^{n}(\boldsymbol{M}_{i\cdot}^{\top} - \boldsymbol{\alpha}_c - A_i\boldsymbol{\alpha}_a)^{\top}\boldsymbol{\Sigma}_m^{-1}(\boldsymbol{M}_{i\cdot}^{\top} - \boldsymbol{\alpha}_c - A_i\boldsymbol{\alpha}_a)$$ $$- \frac{n}{2}\log(2\pi) - \frac{n}{2}\log(\sigma_e^2) - \frac{1}{2\sigma_e^2}\sum_{i=1}^{n}(Y_i - \beta_c - \boldsymbol{M}_{i\cdot}\boldsymbol{\beta}_m - A_i\beta_a)^2$$

\noindent The first order derivatives are:

$$\frac{{\partial}l}{\partial\boldsymbol{\alpha}_a} = \sum_{i=1}^{n}A_i\boldsymbol{\Sigma}_m^{-1}\big[\boldsymbol{M}_{i\cdot}^{\top} - \boldsymbol{\alpha}_c - A_i\boldsymbol{\alpha}_a\big]$$

$$\frac{{\partial}l}{\partial\boldsymbol{\alpha}_c} = \sum_{i=1}^{n}\boldsymbol{\Sigma}_m^{-1}\big[\boldsymbol{M}_{i\cdot}^{\top} - \boldsymbol{\alpha}_c - A_i\boldsymbol{\alpha}_a\big]$$

$$\frac{{\partial}l}{\partial\boldsymbol{\Sigma}_m^{-1}} = \frac{n}{2}\boldsymbol{\Sigma}_m - \frac{1}{2}\sum_{i=1}^{n}\big[\boldsymbol{M}_{i\cdot}^{\top} - \boldsymbol{\alpha}_c - A_i\boldsymbol{\alpha}_a\big]\big[\boldsymbol{M}_{i\cdot}^{\top} - \boldsymbol{\alpha}_c - A_i\boldsymbol{\alpha}_a\big]^{\top}$$

$$\frac{{\partial}l}{\partial\boldsymbol{\beta}_m} = \frac{1}{\sigma_e^2}\sum_{i=1}^{n}\boldsymbol{M}_{i\cdot}^{\top}\big[Y_i - \beta_c - \boldsymbol{M}_{i\cdot}\boldsymbol{\beta}_m - A_i\beta_a\big]$$

$$\frac{{\partial}l}{\partial\beta_a} = \frac{1}{\sigma_e^2}\sum_{i=1}^{n}A_i\big[Y_i - \beta_c - \boldsymbol{M}_{i\cdot}\boldsymbol{\beta}_m - A_i\beta_a\big]$$

$$\frac{{\partial}l}{\partial\beta_c} = \frac{1}{\sigma_e^2}\sum_{i=1}^{n}\big[Y_i - \beta_c - \boldsymbol{M}_{i\cdot}\boldsymbol{\beta}_m - A_i\beta_a\big]$$

$$\frac{{\partial}l}{\partial\sigma_e^2} = -\frac{n}{2\sigma_e^2} + \frac{1}{2\sigma_e^4}\sum_{i=1}^{n}\big[Y_i - \beta_c - \boldsymbol{M}_{i\cdot}\boldsymbol{\beta}_m - A_i\beta_a\big]^2$$

\noindent The second order derivatives are:

$$\frac{{\partial}^2l}{\partial\boldsymbol{\alpha}_a\partial\boldsymbol{\alpha}_a^{\top}} = -\boldsymbol{\Sigma}_m^{-1}\sum_{i=1}^{n}A_i^2 \implies -\frac{1}{n}E\bigg[\frac{{\partial}^2l}{\partial\boldsymbol{\alpha}_a\partial\boldsymbol{\alpha}_a^{\top}}\bigg] \to \sigma_a^2\boldsymbol{\Sigma}_m^{-1}$$

$$\frac{{\partial}^2l}{\partial\boldsymbol{\alpha}_c\partial\boldsymbol{\alpha}_c^{\top}} = -n\boldsymbol{\Sigma}_m^{-1} \implies -\frac{1}{n}E\bigg[\frac{{\partial}^2l}{\partial\boldsymbol{\alpha}_c\partial\boldsymbol{\alpha}_c^{\top}}\bigg] = \boldsymbol{\Sigma}_m^{-1}$$

$$\frac{{\partial}^2l}{\partial\boldsymbol{\alpha}_a\partial\boldsymbol{\alpha}_c^{\top}} = -\boldsymbol{\Sigma}_m^{-1}\sum_{i=1}^{n}A_i \implies -\frac{1}{n}E\bigg[\frac{{\partial}^2l}{\partial\boldsymbol{\alpha}_a\partial\boldsymbol{\alpha}_c^{\top}}\bigg] \to \boldsymbol{0}$$

$$\frac{{\partial}^2l}{\partial\boldsymbol{\alpha}_c\partial\boldsymbol{\alpha}_a^{\top}} = -\boldsymbol{\Sigma}_m^{-1}\sum_{i=1}^{n}A_i \implies -\frac{1}{n}E\bigg[\frac{{\partial}^2l}{\partial\boldsymbol{\alpha}_c\partial\boldsymbol{\alpha}_a^{\top}}\bigg] \to \boldsymbol{0}$$

$$\frac{{\partial}^2l}{\partial\boldsymbol{\beta}_m\partial\boldsymbol{\beta}_m^{\top}} = -\frac{1}{\sigma_e^2}\sum_{i=1}^{n}\boldsymbol{M}_{i\cdot}^{\top}\boldsymbol{M}_{i\cdot} \implies -\frac{1}{n}E\bigg[\frac{{\partial}^2l}{\partial\boldsymbol{\beta}_m\partial\boldsymbol{\beta}_m^{\top}}\bigg] \to \frac{1}{\sigma_e^2}(\boldsymbol{\Sigma}_m + \boldsymbol{\alpha}_c\boldsymbol{\alpha}_c^{\top}) + \frac{\sigma_a^2}{\sigma_e^2}\boldsymbol{\alpha}_a\boldsymbol{\alpha}_a^{\top}$$

$$\frac{{\partial}^2l}{\partial\beta_a^2} = -\frac{1}{\sigma_e^2}\sum_{i=1}^{n}A_i^2 \implies -\frac{1}{n}E\bigg[\frac{{\partial}^2l}{\partial\beta_a^2}\bigg] \to \frac{\sigma_a^2}{\sigma_e^2}$$

$$\frac{{\partial}^2l}{\partial\beta_c^2} = -\frac{n}{\sigma_e^2} \implies -\frac{1}{n}E\bigg[\frac{{\partial}^2l}{\partial\beta_c^2}\bigg] = \frac{1}{\sigma_e^2}$$

$$\frac{{\partial}^2l}{\partial\boldsymbol{\beta}_m\partial\beta_a} = -\frac{1}{\sigma_e^2}\sum_{i=1}^{n}A_i\boldsymbol{M}_{i\cdot}^{\top} \implies -\frac{1}{n}E\bigg[\frac{{\partial}^2l}{\partial\boldsymbol{\beta}_m\partial\beta_a}\bigg] \to \frac{\sigma_a^2}{\sigma_e^2}\boldsymbol{\alpha}_a$$

$$\frac{{\partial}^2l}{\partial\beta_a\partial\boldsymbol{\beta}_m^{\top}} = -\frac{1}{\sigma_e^2}\sum_{i=1}^{n}A_i\boldsymbol{M}_{i\cdot} \implies -\frac{1}{n}E\bigg[\frac{{\partial}^2l}{\partial\beta_a\partial\boldsymbol{\beta}_m^{\top}}\bigg] \to \frac{\sigma_a^2}{\sigma_e^2}\boldsymbol{\alpha}_a^{\top}$$

$$\frac{{\partial}^2l}{\partial\boldsymbol{\beta}_m\partial\beta_c} = -\frac{1}{\sigma_e^2}\sum_{i=1}^{n}\boldsymbol{M}_{i\cdot}^{\top} \implies -\frac{1}{n}E\bigg[\frac{{\partial}^2l}{\partial\boldsymbol{\beta}_m\partial\beta_c}\bigg] \to \frac{1}{\sigma_e^2}\boldsymbol{\alpha}_c$$

$$\frac{{\partial}^2l}{\partial\beta_c\partial\boldsymbol{\beta}_m^{\top}} = -\frac{1}{\sigma_e^2}\sum_{i=1}^{n}\boldsymbol{M}_{i\cdot} \implies -\frac{1}{n}E\bigg[\frac{{\partial}^2l}{\partial\beta_c\partial\boldsymbol{\beta}_m^{\top}}\bigg] \to \frac{1}{\sigma_e^2}\boldsymbol{\alpha}_c^{\top}$$

$$\frac{{\partial}^2l}{\partial\beta_a\partial\beta_c} = -\frac{1}{\sigma_e^2}\sum_{i=1}^{n}A_i \implies -\frac{1}{n}E\bigg[\frac{{\partial}^2l}{\partial\beta_a\partial\beta_c}\bigg] \to 0$$

\noindent Note that:

$$-\frac{1}{n}E\bigg[\frac{{\partial}^2l}{\partial\boldsymbol{\alpha}_a\partial(\boldsymbol{\Sigma}_m^{-1})_{kl}}\bigg] \to \boldsymbol{0}, \hspace{2 mm} 1 \leq k,l \leq p_m$$

$$-\frac{1}{n}E\bigg[\frac{{\partial}^2l}{\partial\boldsymbol{\alpha}_c\partial(\boldsymbol{\Sigma}_m^{-1})_{kl}}\bigg] \to \boldsymbol{0}, \hspace{2 mm} 1 \leq k,l \leq p_m$$

$$-\frac{1}{n}E\bigg[\frac{{\partial}^2l}{\partial\beta_a\partial\sigma_e^2}\bigg] = 0$$

$$-\frac{1}{n}E\bigg[\frac{{\partial}^2l}{\partial\beta_c\partial\sigma_e^2}\bigg] = 0$$

$$-\frac{1}{n}E\bigg[\frac{{\partial}^2l}{\partial\boldsymbol{\beta}_m\partial\sigma_e^2}\bigg] = \boldsymbol{0}$$

\noindent Because the outcome and mediator models are independent of one another, then we can consider the Fisher information matrix for the outcome model:

$$\mathcal{I}^{U}(\beta_c,\beta_a,\boldsymbol{\beta}_m) = \begin{pmatrix} \frac{1}{\sigma_e^2} & 0 & \frac{1}{\sigma_e^2}\boldsymbol{\alpha}_c^{\top} \\ 0 & \frac{\sigma_a^2}{\sigma_e^2} & \frac{\sigma_a^2}{\sigma_e^2}\boldsymbol{\alpha}_a^{\top} \\ \frac{1}{\sigma_e^2}\boldsymbol{\alpha}_c & \frac{\sigma_a^2}{\sigma_e^2}\boldsymbol{\alpha}_a & \frac{1}{\sigma_e^2}(\boldsymbol{\Sigma}_m + \boldsymbol{\alpha}_c\boldsymbol{\alpha}_c^{\top}) + \frac{\sigma_a^2}{\sigma_e^2}\boldsymbol{\alpha}_a\boldsymbol{\alpha}_a^{\top} \end{pmatrix}$$

\noindent Using the block inversion formula for $2 \times 2$ block matricies, we can see that:

$$\Big(\mathcal{I}^{U}(\beta_a,\boldsymbol{\beta}_m)\Big)^{-1} = \begin{pmatrix} \frac{\sigma_a^2}{\sigma_e^2} & \frac{\sigma_a^2}{\sigma_e^2}\boldsymbol{\alpha}_a^{\top} \\ \frac{\sigma_a^2}{\sigma_e^2}\boldsymbol{\alpha}_a & \frac{1}{\sigma_e^2}\boldsymbol{\Sigma}_m + \frac{\sigma_a^2}{\sigma_e^2}\boldsymbol{\alpha}_a\boldsymbol{\alpha}_a^{\top} \end{pmatrix}^{-1}$$ 

\noindent Again, using the block inversion formula for $2 \times 2$ block matricies, we can see that:

$$\Big(\mathcal{I}^{U}(\beta_a,\boldsymbol{\beta}_m)\Big)^{-1} = \begin{pmatrix} \frac{\sigma_e^2}{\sigma_a^2}\big(1 + \sigma_a^2\boldsymbol{\alpha}_a^{\top}\boldsymbol{\Sigma}_m^{-1}\boldsymbol{\alpha}_a\big) & -\sigma_e^2\boldsymbol{\alpha}_a^{\top}\boldsymbol{\Sigma}_m^{-1} \\ -\sigma_e^2\boldsymbol{\Sigma}_m^{-1}\boldsymbol{\alpha}_a & \sigma_e^2\boldsymbol{\Sigma}_m^{-1} \end{pmatrix}$$

\noindent Therefore, conclude that:

$$\Big(\mathcal{I}^{U}(\boldsymbol{\alpha}_a,\beta_a,\boldsymbol{\beta}_m)\Big)^{-1} = \begin{pmatrix} \frac{1}{\sigma_a^2}\boldsymbol{\Sigma}_m & \boldsymbol{0} & \boldsymbol{0} \\ \boldsymbol{0} & \frac{\sigma_e^2}{\sigma_a^2}\big(1 + \sigma_a^2\boldsymbol{\alpha}_a^{\top}\boldsymbol{\Sigma}_m^{-1}\boldsymbol{\alpha}_a\big) & -\sigma_e^2\boldsymbol{\alpha}_a^{\top}\boldsymbol{\Sigma}_m^{-1} \\ \boldsymbol{0} & -\sigma_e^2\boldsymbol{\Sigma}_m^{-1}\boldsymbol{\alpha}_a & \sigma_e^2\boldsymbol{\Sigma}_m^{-1} \end{pmatrix}$$

\noindent The asymptotic distribution for the unconstrained estimator of the NIE can then be computed using the multivariate delta method provided that $\boldsymbol{\alpha}_a \neq \boldsymbol{0}$ or $\boldsymbol{\beta}_m \neq \boldsymbol{0}$.

\vspace{2 mm}

\noindent To extend this result to the model with confounders \begin{align*}
    [\boldsymbol{Y} \mid \boldsymbol{M}, \boldsymbol{A}, \boldsymbol{C}] &\sim N\Big(\boldsymbol{M}\boldsymbol{\beta}_m + \boldsymbol{A}\beta_a + \boldsymbol{C}\boldsymbol{\beta}_c,\sigma_e^2\boldsymbol{I}\Big), \\
    [\boldsymbol{M}_{i\cdot}^{\top} \mid \boldsymbol{A}_i, \boldsymbol{C}_i] &\sim N\Big(A_i\boldsymbol{\alpha}_a + \boldsymbol{\alpha}_c\boldsymbol{C}_i^{\top},\boldsymbol{\Sigma}_m\Big), \hspace{2 mm} i = 1,\ldots,n.
\end{align*} we note that the model with confounders is equivalent to a model with no confounders after the confounders are regressed out from the outcome, mediators, and exposure. That is, we can use the residuals from linear regression models $[\boldsymbol{Y} \mid \boldsymbol{C}]$, $[\boldsymbol{M} \mid \boldsymbol{C}]$, and $[\boldsymbol{A} \mid \boldsymbol{C}]$ as the new outcome, mediators, and exposure and run the unconstrained model. Therefore, the only difference in the asymptotic normality expression is that the exposure are now the residuals from the $[\boldsymbol{A} \mid \boldsymbol{C}]$ linear regression model, implying that $\sigma_a^2 = \text{Var}(A_i \mid \boldsymbol{C}_{i\cdot})$.

\newpage

\subsection*{Proof of Theorem 2}

Suppose that $[\boldsymbol{M}_{i\cdot}^{\top} \mid A_i] \sim N(\boldsymbol{\alpha}_c + A_i\boldsymbol{\alpha}_a,\boldsymbol{\Sigma}_m)$ and $[Y_i \mid \boldsymbol{M}_{i\cdot}, A_i] \sim N(\beta_c + \boldsymbol{M}_{i\cdot}\boldsymbol{\beta}_m + A_i\beta_a, \sigma_e^2)$ is the true generative model where $i = 1,\ldots,n$. That is, without loss of generality, we are assuming that there are no confounders, but that there are intercept terms in both the outcome and mediator models. This implies that $\sigma_a^2 = \text{Var}(A_i)$. Moreover, assume that $E[A_i] = 0$. Let $\mathcal{D} = \{\boldsymbol{Y},\boldsymbol{M},\boldsymbol{A},\boldsymbol{C}\}$ denote the data. Suppose that we fit the hard constraint model to our data. Then the log-likelihood is:
 $$l(\boldsymbol{\alpha}_c,\boldsymbol{\alpha}_a,\boldsymbol{\Sigma}_m,\boldsymbol{\beta}_m,\beta_c,\sigma_e^2 \mid \mathcal{D}) = -\frac{np_m}{2}\log(2\pi) - \frac{n}{2}\log(|\boldsymbol{\Sigma}_m|) - \frac{1}{2}\sum_{i=1}^{n}(\boldsymbol{M}_{i\cdot}^{\top} - \boldsymbol{\alpha}_c - A_i\boldsymbol{\alpha}_a)^{\top}\boldsymbol{\Sigma}_m^{-1}(\boldsymbol{M}_{i\cdot}^{\top} - \boldsymbol{\alpha}_c - A_i\boldsymbol{\alpha}_a)$$ $$- \frac{n}{2}\log(2\pi) - \frac{n}{2}\log(\sigma_e^2) - \frac{1}{2\sigma_e^2}\sum_{i=1}^{n}(Y_i - \beta_c - \boldsymbol{M}_{i\cdot}\boldsymbol{\beta}_m - A_i\{\widehat{\theta}_a^E - \boldsymbol{\alpha}_a^{\top}\boldsymbol{\beta}_m\})^2$$

\noindent The first order derivatives are:
$$\frac{{\partial}l}{\partial\boldsymbol{\alpha}_a} = \sum_{i=1}^{n}A_i\boldsymbol{\Sigma}_m^{-1}\big[\boldsymbol{M}_{i\cdot}^{\top} - \boldsymbol{\alpha}_c - A_i\boldsymbol{\alpha}_a\big] - \frac{1}{\sigma_e^2}\sum_{i=1}^{n}A_i\boldsymbol{\beta}_m\big[Y_i - \beta_c - A_i(\widehat{\theta}_a^E - \boldsymbol{\alpha}_a^{\top}\boldsymbol{\beta}_m) - \boldsymbol{M}_{i\cdot}\boldsymbol{\beta}_m\big]$$

$$\frac{{\partial}l}{\partial\boldsymbol{\alpha}_c} = \sum_{i=1}^{n}\boldsymbol{\Sigma}_m^{-1}\big[\boldsymbol{M}_{i\cdot}^{\top} - \boldsymbol{\alpha}_c - A_i\boldsymbol{\alpha}_a\big]$$

$$\frac{{\partial}l}{\partial\boldsymbol{\Sigma}_m^{-1}} = \frac{n}{2}\boldsymbol{\Sigma}_m - \frac{1}{2}\sum_{i=1}^{n}\big[\boldsymbol{M}_{i\cdot}^{\top} - \boldsymbol{\alpha}_c - A_i\boldsymbol{\alpha}_a\big]\big[\boldsymbol{M}_{i\cdot}^{\top} - \boldsymbol{\alpha}_c - A_i\boldsymbol{\alpha}_a\big]^{\top}$$

$$\frac{{\partial}l}{\partial\boldsymbol{\beta}_m} = \frac{1}{\sigma_e^2}\sum_{i=1}^{n}\big[Y_i - \beta_c - A_i(\widehat{\theta}_a^E - \boldsymbol{\alpha}_a^{\top}\boldsymbol{\beta}_m) - \boldsymbol{M}_{i\cdot}\boldsymbol{\beta}_m\big]\big[\boldsymbol{M}_{i\cdot}^{\top} - A_i\boldsymbol{\alpha}_a\big]$$

$$\frac{{\partial}l}{\partial\beta_c} = \frac{1}{\sigma_e^2}\sum_{i=1}^{n}\big[Y_i - \beta_c - A_i(\widehat{\theta}_a^E - \boldsymbol{\alpha}_a^{\top}\boldsymbol{\beta}_m) - \boldsymbol{M}_{i\cdot}\boldsymbol{\beta}_m\big]$$

$$\frac{{\partial}l}{\partial\sigma_e^2} = -\frac{n}{2\sigma_e^2} + \frac{1}{2\sigma_e^4}\sum_{i=1}^{n}\big[Y_i - \beta_c - A_i(\widehat{\theta}_a^E - \boldsymbol{\alpha}_a^{\top}\boldsymbol{\beta}_m) - \boldsymbol{M}_{i\cdot}\boldsymbol{\beta}_m\big]^2$$

\noindent The second order derivatives are:
$$\frac{{\partial}^2l}{\partial\boldsymbol{\alpha}_a\partial\boldsymbol{\alpha}_a^{\top}} = -\boldsymbol{\Sigma}_m^{-1}\sum_{i=1}^{n}A_i^2 - \frac{1}{\sigma_e^2}\boldsymbol{\beta}_m\boldsymbol{\beta}_m^{\top}\sum_{i=1}^{n}A_i^2 \implies -\frac{1}{n}E\bigg[\frac{{\partial}^2l}{\partial\boldsymbol{\alpha}_a\partial\boldsymbol{\alpha}_a^{\top}}\bigg] \to \sigma_a^2\boldsymbol{\Sigma}_m^{-1} + \frac{\sigma_a^2}{\sigma_e^2}\boldsymbol{\beta}_m\boldsymbol{\beta}_m^{\top}$$

$$\frac{{\partial}^2l}{\partial\boldsymbol{\alpha}_c\partial\boldsymbol{\alpha}_c^{\top}} = -n\boldsymbol{\Sigma}_m^{-1} \implies -\frac{1}{n}E\bigg[\frac{{\partial}^2l}{\partial\boldsymbol{\alpha}_c\partial\boldsymbol{\alpha}_c^{\top}}\bigg] = \boldsymbol{\Sigma}_m^{-1}$$

$$\frac{{\partial}^2l}{\partial\boldsymbol{\alpha}_a\partial\boldsymbol{\alpha}_c^{\top}} = -\boldsymbol{\Sigma}_m^{-1}\sum_{i=1}^{n}A_i \implies -\frac{1}{n}E\bigg[\frac{{\partial}^2l}{\partial\boldsymbol{\alpha}_a\partial\boldsymbol{\alpha}_c^{\top}}\bigg] \to \boldsymbol{0}$$

$$\frac{{\partial}^2l}{\partial\boldsymbol{\alpha}_c\partial\boldsymbol{\alpha}_a^{\top}} = -\boldsymbol{\Sigma}_m^{-1}\sum_{i=1}^{n}A_i \implies -\frac{1}{n}E\bigg[\frac{{\partial}^2l}{\partial\boldsymbol{\alpha}_c\partial\boldsymbol{\alpha}_a^{\top}}\bigg] \to \boldsymbol{0}$$

$$\frac{{\partial}^2l}{\partial\boldsymbol{\beta}_m\partial\boldsymbol{\beta}_m^{\top}} = -\frac{1}{\sigma_e^2}\sum_{i=1}^{n}\big[\boldsymbol{M}_{i\cdot}^{\top} - A_i\boldsymbol{\alpha}_a\big]\big[\boldsymbol{M}_{i\cdot} - A_i\boldsymbol{\alpha}_a^{\top}\big] \implies -\frac{1}{n}E\bigg[\frac{{\partial}^2l}{\partial\boldsymbol{\beta}_m\partial\boldsymbol{\beta}_m^{\top}}\bigg] \to \frac{1}{\sigma_e^2}(\boldsymbol{\Sigma}_m + \boldsymbol{\alpha}_c\boldsymbol{\alpha}_c^{\top})$$

$$\frac{{\partial}^2l}{\partial\beta_c^2} = -\frac{n}{\sigma_e^2} \implies -\frac{1}{n}E\bigg[\frac{{\partial}^2l}{\partial\beta_c^2}\bigg] = \frac{1}{\sigma_e^2}$$

$$\frac{{\partial}^2l}{\partial\boldsymbol{\beta}_m\partial\beta_c} = -\frac{1}{\sigma_e^2}\sum_{i=1}^{n}\big[\boldsymbol{M}_{i\cdot}^{\top} - A_i\boldsymbol{\alpha}_a\big] \implies -\frac{1}{n}E\bigg[\frac{{\partial}^2l}{\partial\boldsymbol{\beta}_m\partial\beta_c}\bigg] \to \frac{1}{\sigma_e^2}\boldsymbol{\alpha}_c$$

$$\frac{{\partial}^2l}{\partial\beta_c\partial\boldsymbol{\beta}_m^{\top}} = -\frac{1}{\sigma_e^2}\sum_{i=1}^{n}\big[\boldsymbol{M}_{i\cdot} - A_i\boldsymbol{\alpha}_a^{\top}\big] \implies -\frac{1}{n}E\bigg[\frac{{\partial}^2l}{\partial\beta_c\partial\boldsymbol{\beta}_m^{\top}}\bigg] \to \frac{1}{\sigma_e^2}\boldsymbol{\alpha}_c^{\top}$$

$$\frac{{\partial}^2l}{\partial\boldsymbol{\alpha}_a\partial\boldsymbol{\beta}_m^{\top}} = -\frac{1}{\sigma_e^2}\sum_{i=1}^{n}\Big[A_i(Y_i - \beta_c - A_i(\widehat{\theta}_a^E - \boldsymbol{\alpha}_a^{\top}\boldsymbol{\beta}_m) - \boldsymbol{M}_{i\cdot}\boldsymbol{\beta}_m)\boldsymbol{I} + A_i\boldsymbol{\beta}_m(A_i\boldsymbol{\alpha}_a^{\top} - \boldsymbol{M}_{i\cdot})\Big]$$ $$\implies -\frac{1}{n}E\bigg[\frac{{\partial}^2l}{\partial\boldsymbol{\alpha}_a\partial\boldsymbol{\beta}_m^{\top}}\bigg] \to \boldsymbol{0}$$

$$\frac{{\partial}^2l}{\partial\boldsymbol{\beta}_m\partial\boldsymbol{\alpha}_a^{\top}} = -\frac{1}{\sigma_e^2}\sum_{i=1}^{n}\Big[A_i(Y_i - \beta_c - A_i(\widehat{\theta}_a^E - \boldsymbol{\alpha}_a^{\top}\boldsymbol{\beta}_m) - \boldsymbol{M}_{i\cdot}\boldsymbol{\beta}_m)\boldsymbol{I} + A_i(A_i\boldsymbol{\alpha}_a - \boldsymbol{M}_{i\cdot}^{\top})\boldsymbol{\beta}_m^{\top}\Big]$$ $$\implies -\frac{1}{n}E\bigg[\frac{{\partial}^2l}{\partial\boldsymbol{\beta}_m\partial\boldsymbol{\alpha}_a^{\top}}\bigg] \to \boldsymbol{0}$$

\noindent Note that:

$$-\frac{1}{n}E\bigg[\frac{{\partial}^2l}{\partial\boldsymbol{\alpha}_a\partial(\boldsymbol{\Sigma}_m^{-1})_{kl}}\bigg] \to \boldsymbol{0}, \hspace{2 mm} 1 \leq k,l \leq p_m$$

$$-\frac{1}{n}E\bigg[\frac{{\partial}^2l}{\partial\boldsymbol{\alpha}_c\partial(\boldsymbol{\Sigma}_m^{-1})_{kl}}\bigg] \to \boldsymbol{0}, \hspace{2 mm} 1 \leq k,l \leq p_m$$

$$-\frac{1}{n}E\bigg[\frac{{\partial}^2l}{\partial\boldsymbol{\alpha}_a\partial\sigma_e^2}\bigg] = \boldsymbol{0}$$

$$-\frac{1}{n}E\bigg[\frac{{\partial}^2l}{\partial\beta_c\partial\sigma_e^2}\bigg] = 0$$

$$-\frac{1}{n}E\bigg[\frac{{\partial}^2l}{\partial\boldsymbol{\beta}_m\partial\sigma_e^2}\bigg] = \boldsymbol{0}$$

\noindent Using the block inversion formula for $2 \times 2$ block matricies, we can see that:

$$\Big(\mathcal{I}^{H}(\boldsymbol{\alpha}_a,\boldsymbol{\beta}_m)\Big)^{-1} = \begin{pmatrix} \frac{1}{\sigma_a^2}\Big(\boldsymbol{\Sigma}_m^{-1} + \frac{1}{\sigma_e^2}\boldsymbol{\beta}_m\boldsymbol{\beta}_m^{\top}\Big)^{-1} & \boldsymbol{0} \\ \boldsymbol{0} & \sigma_e^2\boldsymbol{\Sigma}_m^{-1} \end{pmatrix}$$

The asymptotic distribution for the hard constraint estimator of the NIE can then be computed using the multivariate delta method provided that $\boldsymbol{\alpha}_a \neq \boldsymbol{0}$ or $\boldsymbol{\beta}_m \neq \boldsymbol{0}$. Because the hard constraint estimator of the NDE is just a location shift of the hard constraint estimator of the NIE, then provided that $\sqrt{n}(\widehat{\theta}_a^E - \theta_a^I) \to_p 0$, Slutsky's Theorem tells us that the asymptotic distribution for the hard constraint estimator of the NDE is the same as the asymptotic distribution for the hard constraint estimator of the NIE.

\vspace{2 mm}

To extend this result to the model with confounders we use the same trick that was used in the proof of Theorem 1, namely that the model with confounders is equivalent to a model with no confounders after the confounders are regressed out from the outcome, mediators, and exposure via linear regression models. Therefore, we again see that $\sigma_a^2 = \text{Var}(A_i \mid \boldsymbol{C}_{i\cdot})$, when confounders are present.

\newpage

\subsection*{Proof of Theorem 3}

We have that $$\sqrt{n}\begin{pmatrix} \widehat{\boldsymbol{\alpha}}_a - \boldsymbol{\alpha}_a \\ \widehat{\boldsymbol{\beta}}_m - \boldsymbol{\beta}_m \end{pmatrix} \to_d N\Bigg(\begin{pmatrix} \boldsymbol{0} \\ \boldsymbol{0} \end{pmatrix}, \begin{pmatrix} \sigma_a^{-2}\boldsymbol{\Sigma}_{m} & \boldsymbol{0} \\ \boldsymbol{0} & \sigma_e^2\boldsymbol{\Sigma}_m^{-1} \end{pmatrix}\Bigg)$$

\noindent Using the second order Taylor expansion around $\boldsymbol{\alpha}_a = \boldsymbol{0}$ and $\boldsymbol{\beta}_m = \boldsymbol{0}$, we can show that $$n\Big(\widehat{\boldsymbol{\alpha}}_a^{\top}\widehat{\boldsymbol{\beta}}_m - \boldsymbol{\alpha}_a^{\top}\boldsymbol{\beta}_m\Big) = \frac{n}{2}\begin{pmatrix} \widehat{\boldsymbol{\alpha}}_a^{\top} - \boldsymbol{\alpha}_a^{\top} & \widehat{\boldsymbol{\beta}}_m^{\top} - \boldsymbol{\beta}_m^{\top} \end{pmatrix} \begin{pmatrix} \boldsymbol{0} & \boldsymbol{I} \\ \boldsymbol{I} & \boldsymbol{0} \end{pmatrix} \begin{pmatrix} \widehat{\boldsymbol{\alpha}}_a - \boldsymbol{\alpha}_a \\ \widehat{\boldsymbol{\beta}}_m - \boldsymbol{\beta}_m \end{pmatrix}.$$ In this expression $0$ is a $p_m \times p_m$ matrix of zeros and $I$ is a $p_m \times p_m$ identity matrix. Because this is quadratic form of an asymptotically normal random vector (where the matrix in the quadratic form is a symmetric matrix), then we may apply the continuous mapping theorem for convergence in distribution. That is, we just need to work with the asymptotic distribution when determining the asymptotic distribution of the quadratic form. Define $$\boldsymbol{Z} = \sqrt{n}\begin{pmatrix} \sigma_a^{-2}\boldsymbol{\Sigma}_m & \boldsymbol{0} \\ \boldsymbol{0} & \sigma_e^2\boldsymbol{\Sigma}_m^{-1} \end{pmatrix}^{-1/2}\begin{pmatrix} \widehat{\boldsymbol{\alpha}}_a - \boldsymbol{\alpha}_a \\ \widehat{\boldsymbol{\beta}}_m - \boldsymbol{\beta}_m \end{pmatrix},$$ where $$\begin{pmatrix} \sigma_a^{-2}\boldsymbol{\Sigma}_m & \boldsymbol{0} \\ \boldsymbol{0} & \sigma_e^2\boldsymbol{\Sigma}_m^{-1} \end{pmatrix}^{-1/2},$$ is the inverse of the matrix square root of $$\begin{pmatrix} \sigma_a^{-2}\boldsymbol{\Sigma}_m & \boldsymbol{0} \\ \boldsymbol{0} & \sigma_e^2\boldsymbol{\Sigma}_m^{-1} \end{pmatrix}$$ Note that the matrix square root always exists for symmetric, positive definite matrices. Then we can write $$\frac{n}{2}\begin{pmatrix} \widehat{\boldsymbol{\alpha}}_a^{\top} - \boldsymbol{\alpha}_a^{\top} & \widehat{\boldsymbol{\beta}}_m^{\top} - \boldsymbol{\beta}_m^{\top} \end{pmatrix} \begin{pmatrix} \boldsymbol{0} & \boldsymbol{I} \\ \boldsymbol{I} & \boldsymbol{0} \end{pmatrix} \begin{pmatrix} \widehat{\boldsymbol{\alpha}}_a - \boldsymbol{\alpha}_a \\ \widehat{\boldsymbol{\beta}}_m - \boldsymbol{\beta}_m \end{pmatrix}$$ $$= \frac{1}{2}\boldsymbol{Z}^{\top} \begin{pmatrix} \sigma_a^{-2}\boldsymbol{\Sigma}_m & \boldsymbol{0} \\ \boldsymbol{0} & \sigma_e^2\boldsymbol{\Sigma}_m^{-1} \end{pmatrix}^{1/2} \begin{pmatrix} \boldsymbol{0} & \boldsymbol{I} \\ \boldsymbol{I} & \boldsymbol{0} \end{pmatrix}\begin{pmatrix} \sigma_a^{-2}\boldsymbol{\Sigma}_m & \boldsymbol{0} \\ \boldsymbol{0} & \sigma_e^2\boldsymbol{\Sigma}_m^{-1} \end{pmatrix}^{1/2}\boldsymbol{Z}$$

\noindent Because $$\begin{pmatrix} \sigma_a^{-2}\boldsymbol{\Sigma}_m & \boldsymbol{0} \\ \boldsymbol{0} & \sigma_e^2\boldsymbol{\Sigma}_m^{-1} \end{pmatrix}^{1/2} \begin{pmatrix} \boldsymbol{0} & \boldsymbol{I} \\ \boldsymbol{I} & \boldsymbol{0} \end{pmatrix}\begin{pmatrix} \sigma_a^{-2}\boldsymbol{\Sigma}_m & \boldsymbol{0} \\ \boldsymbol{0} & \sigma_e^2\boldsymbol{\Sigma}_m^{-1} \end{pmatrix}^{1/2} = \begin{pmatrix} \boldsymbol{0} & \frac{\sqrt{\sigma_e^2}}{\sqrt{\sigma_a^2}}\boldsymbol{I} \\ \frac{\sqrt{\sigma_e^2}}{\sqrt{\sigma_a^2}}\boldsymbol{I} & \boldsymbol{0} \end{pmatrix}$$ is a symmetric matrix, then we can obtain an eigendecomposition $$\begin{pmatrix} \boldsymbol{0} & \frac{\sqrt{\sigma_e^2}}{\sqrt{\sigma_a^2}}\boldsymbol{I} \\ \frac{\sqrt{\sigma_e^2}}{\sqrt{\sigma_a^2}}\boldsymbol{I} & \boldsymbol{0} \end{pmatrix} = \boldsymbol{P}^{\top}\boldsymbol{\Lambda}\boldsymbol{P},$$ where $\boldsymbol{P}$ is an orthogonal matrix and $\boldsymbol{\Lambda}$ is a diagonal matrix. Then the expression becomes $$\frac{1}{2}(\boldsymbol{P}\boldsymbol{Z})^{\top}\boldsymbol{\Lambda}(\boldsymbol{P}\boldsymbol{Z})$$ Note that here $$\boldsymbol{Z} \to_d N(\boldsymbol{0},\boldsymbol{I})$$ and because $\boldsymbol{P}$ is an orthogonal transformation of $\boldsymbol{Z}$, then $$\boldsymbol{P}\boldsymbol{Z} \to_d N(\boldsymbol{0},\boldsymbol{I})$$

\noindent Therefore, when $\boldsymbol{\alpha}_a = \boldsymbol{\beta}_m = \boldsymbol{0}$, we conclude that $$n\Big(\widehat{\boldsymbol{\alpha}}_a^{\top}\widehat{\boldsymbol{\beta}}_m - \boldsymbol{\alpha}_a^{\top}\boldsymbol{\beta}_m\Big) \to_d \frac{1}{2}\sum_{j=1}^{2p_m}\lambda_j\chi_1^2,$$ where the $\lambda_j$ are the eigenvalues of $$\sqrt{\frac{\sigma_e^2}{\sigma_a^2}} \begin{pmatrix} \boldsymbol{0} & \boldsymbol{I} \\ \boldsymbol{I} & \boldsymbol{0} \end{pmatrix},$$ which has $p_m$ eigenvalues equal to $\sqrt{\sigma_e^2/\sigma_a^2}$ and $p_m$ eigenvalues equal to $-\sqrt{\sigma_e^2/\sigma_a^2}$. Therefore, $$\frac{1}{2}\sum_{j=1}^{2p_m}\lambda_j\chi_1^2 = \frac{1}{2}\sqrt{\frac{\sigma_e^2}{\sigma_a^2}}\Big(\chi_{p_m}^2-\chi_{p_m}^2\Big),$$ since the $\chi_1^2$ random variables are independent.

\newpage

\subsection*{Proof of Theorem 4}

Suppose that $[\boldsymbol{M}_{i\cdot}^{\top} \mid A_i] \sim N(A_i\boldsymbol{\alpha}_a,\boldsymbol{\Sigma}_m)$ and $[Y_i \mid \boldsymbol{M}_{i\cdot}, A_i] \sim N(\boldsymbol{M}_{i\cdot}\boldsymbol{\beta}_m + A_i\beta_a, \sigma_e^2)$ is the true generative model where $i = 1,\ldots,n$. That is, without loss of generality, we are assuming that there are no confounders. This implies that $\sigma_a^2 = \text{Var}(A_i)$. Moreover, assume that $E[A_i] = 0$. Let $\mathcal{D} = \{\boldsymbol{Y},\boldsymbol{M},\boldsymbol{A},\boldsymbol{C}\}$ denote the data. Suppose that we fit the soft constraint model to our data with a fixed value $s^2$. Then the likelihood is:
$$L(\boldsymbol{\alpha}_a,\boldsymbol{\Sigma}_M,\boldsymbol{\beta}_m,\sigma_e^2 \mid \boldsymbol{Y},\boldsymbol{M},\boldsymbol{A}) = \int_{-\infty}^{\infty}\pi(\boldsymbol{Y}\mid\boldsymbol{M},\boldsymbol{A},\widetilde{\theta}_a^I)\pi(\boldsymbol{M}\mid\boldsymbol{A})\pi(\widetilde{\theta}_a^I)d\widetilde{\theta}_a^I$$ $$= \frac{\pi(\boldsymbol{M} \mid \boldsymbol{A})}{(2\pi\sigma_e^2)^{n/2}\sqrt{2{\pi}s^2\widehat{\text{Var}}(\widehat{\theta}_a^E)}}\exp\bigg[-\frac{1}{2\sigma_e^2}\big(\boldsymbol{Y} - \boldsymbol{M}\boldsymbol{\beta}_m + \boldsymbol{\alpha}_a^{\top}\boldsymbol{\beta}_m\boldsymbol{A}\big)^{\top}\big(\boldsymbol{Y} - \boldsymbol{M}\boldsymbol{\beta}_m + \boldsymbol{\alpha}_a^{\top}\boldsymbol{\beta}_m\boldsymbol{A}\big)\bigg]$$ $$\times \exp\bigg[-\frac{(\widehat{\theta}_a^E)^2}{2s^2\widehat{\text{Var}}(\widehat{\theta}_a^E)}\bigg]\exp\Bigg[\frac{1}{2}\bigg(\frac{\boldsymbol{A}^{\top}\boldsymbol{A}}{\sigma_e^2} + \frac{1}{s^2\widehat{\text{Var}}(\widehat{\theta}_a^E)}\bigg)^{-1}\bigg(\frac{\boldsymbol{A}^{\top}\big(\boldsymbol{Y} - \boldsymbol{M}\boldsymbol{\beta}_m + \boldsymbol{\alpha}_a^{\top}\boldsymbol{\beta}_m\boldsymbol{A}\big)}{\sigma_e^2} + \frac{\widehat{\theta}_a^E}{s^2\widehat{\text{Var}}(\widehat{\theta}_a^E)}\bigg)^2\Bigg]$$ $$\times \sqrt{2\pi\bigg(\frac{\boldsymbol{A}^{\top}\boldsymbol{A}}{\sigma_e^2} + \frac{1}{s^2\widehat{\text{Var}}(\widehat{\theta}_a^E)}\bigg)^{-1}}$$

\noindent Therefore, the log-likelihood function is: $$l(\boldsymbol{\alpha}_a,\boldsymbol{\Sigma}_M,\boldsymbol{\beta}_m,\sigma_e^2 \mid \boldsymbol{Y},\boldsymbol{M},\boldsymbol{A}) = -\frac{np_m}{2} - \frac{n}{2}\log{|\boldsymbol{\Sigma}_m|} - \frac{1}{2}\sum_{i=1}^{n}\big(\boldsymbol{M}_{i\cdot}^{\top} - A_i\boldsymbol{\alpha}_a\big)^{\top}\boldsymbol{\Sigma}_m^{-1}\big(\boldsymbol{M}_{i\cdot}^{\top} - A_i\boldsymbol{\alpha}_a\big)$$ $$-\frac{n}{2}\log\big(2\pi\sigma_e^2\big) - \frac{1}{2}\log\big(2{\pi}s^2\big) - \frac{1}{2\sigma_e^2}\big(\boldsymbol{Y} - \boldsymbol{M}\boldsymbol{\beta}_m + \boldsymbol{\alpha}_a^{\top}\boldsymbol{\beta}_m\boldsymbol{A}\big)^{\top}\big(\boldsymbol{Y} - \boldsymbol{M}\boldsymbol{\beta}_m + \boldsymbol{\alpha}_a^{\top}\boldsymbol{\beta}_m\boldsymbol{A}\big) - \frac{(\widehat{\theta}_a^E)^2}{2s^2\widehat{\text{Var}}(\widehat{\theta}_a^E)}$$ $$+ \frac{1}{2}\bigg(\frac{\boldsymbol{A}^{\top}\boldsymbol{A}}{\sigma_e^2} + \frac{1}{s^2\widehat{\text{Var}}(\widehat{\theta}_a^E)}\bigg)^{-1}\bigg(\frac{\boldsymbol{A}^{\top}\big(\boldsymbol{Y} - \boldsymbol{M}\boldsymbol{\beta}_m + \boldsymbol{\alpha}_a^{\top}\boldsymbol{\beta}_m\boldsymbol{A}\big)}{\sigma_e^2} + \frac{\widehat{\theta}_a^E}{s^2\widehat{\text{Var}}(\widehat{\theta}_a^E)}\bigg)^2$$ $$+ \frac{1}{2}\log(2\pi) - \frac{1}{2}\log\bigg(\frac{\boldsymbol{A}^{\top}\boldsymbol{A}}{\sigma_e^2} + \frac{1}{s^2\widehat{\text{Var}}(\widehat{\theta}_a^E)}\bigg)$$

\noindent The first order derivatives are:
$$\frac{{\partial}l}{{\partial}\boldsymbol{\alpha}_a} = \sum_{i=1}^{n}A_i\boldsymbol{\Sigma}_m^{-1}\big(\boldsymbol{M}_{i\cdot}^{\top} - A_i\boldsymbol{\alpha}_a\big) - \frac{1}{\sigma_e^2}\boldsymbol{A}^{\top}\big(\boldsymbol{Y} - \boldsymbol{M}\boldsymbol{\beta}_m + \boldsymbol{\alpha}_a^{\top}\boldsymbol{\beta}_m\boldsymbol{A}\big)\boldsymbol{\beta}_m$$ $$+ \bigg(\frac{\boldsymbol{A}^{\top}\boldsymbol{A}}{\sigma_e^2} + \frac{1}{s^2\widehat{\text{Var}}(\widehat{\theta}_a^E)}\bigg)^{-1}\bigg(\frac{\boldsymbol{A}^{\top}\big(\boldsymbol{Y} - \boldsymbol{M}\boldsymbol{\beta}_m + \boldsymbol{\alpha}_a^{\top}\boldsymbol{\beta}_m\boldsymbol{A}\big)}{\sigma_e^2} + \frac{\widehat{\theta}_a^E}{s^2\widehat{\text{Var}}(\widehat{\theta}_a^E)}\bigg)\frac{\boldsymbol{A}^{\top}\boldsymbol{A}}{\sigma_e^2}\boldsymbol{\beta}_m$$

$$\frac{{\partial}l}{{\partial}\boldsymbol{\beta}_m} = \frac{1}{\sigma_e^2}\big(\boldsymbol{M} - \boldsymbol{A}\boldsymbol{\alpha}_a^{\top}\big)^{\top}\big(\boldsymbol{Y} - \boldsymbol{M}\boldsymbol{\beta}_m + \boldsymbol{\alpha}_a^{\top}\boldsymbol{\beta}_m\boldsymbol{A}\big)$$ $$- \bigg(\frac{\boldsymbol{A}^{\top}\boldsymbol{A}}{\sigma_e^2} + \frac{1}{s^2\widehat{\text{Var}}(\widehat{\theta}_a^E)}\bigg)^{-1}\bigg(\frac{\boldsymbol{A}^{\top}\big(\boldsymbol{Y} - \boldsymbol{M}\boldsymbol{\beta}_m + \boldsymbol{\alpha}_a^{\top}\boldsymbol{\beta}_m\boldsymbol{A}\big)}{\sigma_e^2} + \frac{\widehat{\theta}_a^E}{s^2\widehat{\text{Var}}(\widehat{\theta}_a^E)}\bigg)\frac{\big(\boldsymbol{M} - \boldsymbol{A}\boldsymbol{\alpha}_a^{\top}\big)^{\top}\boldsymbol{A}}{\sigma_e^2}$$

\noindent The second order derivatives are:
$$\frac{{\partial}^2l}{{\partial}\boldsymbol{\alpha}_a{\partial}\boldsymbol{\alpha}_a^{\top}} = -\boldsymbol{A}^{\top}\boldsymbol{A}\Bigg[\boldsymbol{\Sigma}_m^{-1}+\frac{1}{\sigma_e^2}\frac{1}{s^2\widehat{\text{Var}}(\widehat{\theta}_a^E)}\bigg(\frac{\boldsymbol{A}^{\top}\boldsymbol{A}}{\sigma_e^2}+\frac{1}{s^2\widehat{\text{Var}}(\widehat{\theta}_a^E)}\bigg)^{-1}\boldsymbol{\beta}_m\boldsymbol{\beta}_m^{\top}\Bigg]$$ $$\implies -\frac{1}{n}E\bigg[\frac{{\partial}^2l}{{\partial}\boldsymbol{\alpha}_a{\partial}\boldsymbol{\alpha}_a^{\top}}\bigg] \to \sigma_a^2\Bigg[\boldsymbol{\Sigma}_m^{-1}+\frac{1}{\tau_a^2}\bigg(\frac{\sigma_a^2}{\sigma_e^2}+\frac{1}{\tau_a^2}\bigg)^{-1}\frac{1}{\sigma_e^2}\boldsymbol{\beta}_m\boldsymbol{\beta}_m^{\top}\Bigg]$$

$$\frac{{\partial}^2l}{{\partial}\boldsymbol{\alpha}_a{\partial}\boldsymbol{\beta}_m^{\top}} = \frac{\boldsymbol{A}^{\top}\boldsymbol{A}}{\sigma_e^2}\bigg(\frac{\boldsymbol{A}^{\top}\boldsymbol{A}}{\sigma_e^2} + \frac{1}{s^2\widehat{\text{Var}}(\widehat{\theta}_a^E)}\bigg)^{-1}\frac{\widehat{\theta}_a^E}{s^2\widehat{\text{Var}}(\widehat{\theta}_a^E)}\boldsymbol{I}$$ $$+ \frac{1}{s^2\widehat{\text{Var}}(\widehat{\theta}_a^E)}\bigg(\frac{\boldsymbol{A}^{\top}\boldsymbol{A}}{\sigma_e^2} + \frac{1}{s^2\widehat{\text{Var}}(\widehat{\theta}_a^E)}\bigg)^{-1}\Bigg[\frac{1}{\sigma_e^2}\boldsymbol{\beta}_m\boldsymbol{A}^{\top}\big(\boldsymbol{M}-\boldsymbol{A}\boldsymbol{\alpha}_a^{\top}\big) - \frac{1}{\sigma_e^2}\boldsymbol{A}^{\top}\big(\boldsymbol{Y} - \boldsymbol{M}\boldsymbol{\beta}_m + \boldsymbol{\alpha}_a^{\top}\boldsymbol{\beta}_m\boldsymbol{A}\big)\boldsymbol{I}\Bigg]$$ $$\implies -\frac{1}{n}E\bigg[\frac{{\partial}^2l}{{\partial}\boldsymbol{\alpha}_a{\partial}\boldsymbol{\beta}_m^{\top}}\bigg] = -\frac{1}{n}E\bigg[\frac{{\partial}^2l}{{\partial}\boldsymbol{\beta}_m{\partial}\boldsymbol{\alpha}_a^{\top}}\bigg] \to \boldsymbol{0}$$

$$\frac{{\partial}^2l}{{\partial}\boldsymbol{\beta}_m{\partial}\boldsymbol{\beta}_m^{\top}} = -\frac{1}{\sigma_e^2}\big(\boldsymbol{M} - \boldsymbol{A}\boldsymbol{\alpha}_a^{\top}\big)^{\top}\big(\boldsymbol{M} - \boldsymbol{A}\boldsymbol{\alpha}_a^{\top}\big) - \bigg(\frac{\boldsymbol{A}^{\top}\boldsymbol{A}}{\sigma_e^2} + \frac{1}{s^2\widehat{\text{Var}}(\widehat{\theta}_a^E)}\bigg)^{-1}\frac{1}{\sigma_e^2}\big(\boldsymbol{M} - \boldsymbol{A}\boldsymbol{\alpha}_a^{\top}\big)^{\top}\boldsymbol{A}\boldsymbol{A}^{\top}\big(\boldsymbol{M} - \boldsymbol{A}\boldsymbol{\alpha}_a^{\top}\big)^{\top}\frac{1}{\sigma_e^2}$$ $$\implies -\frac{1}{n}E\bigg[\frac{{\partial}^2l}{{\partial}\boldsymbol{\beta}_m{\partial}\boldsymbol{\beta}_m^{\top}}\bigg] \to \frac{1}{\sigma_e^2}\boldsymbol{\Sigma}_m$$

\noindent Note that:

$$-\frac{1}{n}E\bigg[\frac{{\partial}^2l}{\partial\boldsymbol{\alpha}_a\partial(\boldsymbol{\Sigma}_m^{-1})_{kl}}\bigg] \to \boldsymbol{0}, \hspace{2 mm} 1 \leq k,l \leq p_m$$

$$-\frac{1}{n}E\bigg[\frac{{\partial}^2l}{\partial\boldsymbol{\beta}_m\partial(\boldsymbol{\Sigma}_m^{-1})}\bigg] = \boldsymbol{0}, \hspace{2 mm} 1 \leq k,l \leq p_m$$

$$-\frac{1}{n}E\bigg[\frac{{\partial}^2l}{\partial\boldsymbol{\alpha}_a\partial(\sigma_e^2)}\bigg] = \boldsymbol{0}$$

$$-\frac{1}{n}E\bigg[\frac{{\partial}^2l}{\partial\boldsymbol{\beta}_m\partial(\sigma_e^2)}\bigg] = \boldsymbol{0}$$

\noindent Then we conclude that $$\Big\{\mathcal{I}^{S}(\boldsymbol{\alpha}_a,\boldsymbol{\beta}_m)\Big\}^{-1} = \begin{pmatrix} \frac{1}{\sigma_a^2}\bigg(\boldsymbol{\Sigma}_m^{-1} + \frac{1}{\tau_a^2}\Big[\frac{\sigma_a^2}{\sigma_e^2}+\frac{1}{\tau_a^2}\Big]^{-1}\frac{1}{\sigma_e^2}\boldsymbol{\beta}_m\boldsymbol{\beta}_m^{\top}\bigg)^{-1} & \boldsymbol{0} \\ \boldsymbol{0} & \sigma_e^2\boldsymbol{\Sigma}_m^{-1} \end{pmatrix}$$

\noindent The asymptotic distribution for the soft constraint estimator of the NIE can then be computed using the multivariate delta method provided that $\boldsymbol{\alpha}_a \neq \boldsymbol{0}$ or $\boldsymbol{\beta}_m \neq \boldsymbol{0}$.

\vspace{2 mm}

To extend this result to the model with confounders we use the same trick that was used in the proof of Theorem 1, namely that the model with confounders is equivalent to a model with no confounders after the confounders are regressed out from the outcome, mediators, and exposure via linear regression models. Therefore, we again see that $\sigma_a^2 = \text{Var}(A_i \mid \boldsymbol{C}_{i\cdot})$, when confounders are present.

\newpage

\section*{Supporting Figures and Tables}

\begin{table}[!htbp]
\begin{center}
{
\resizebox{\columnwidth}{!}{
\begin{tabular}{clcccc}
\hline
\textbf{Visit} & \textbf{Covariate} & \textbf{Total} & \textbf{Preterm} & \textbf{Full-term} & \textbf{P-Value} \\ \hline
    1 & Pre-Pregnancy BMI ($\text{kg}/\text{m}^2$) & 25.9 (5.9) & 26.4 (7.3) & 25.9 (5.7) & 0.607 \\
     & Maternal Age (years) & 26.9 (5.5) & 26.4 (6.0) & 27.0 (5.4) & 0.528 \\
     & Education & & & & 0.225 \\
     & \hspace{2 mm} GED/Equivalent or Less & 97 (21.6) & 16 (30.2) & 81 (20.5) & \\
     & \hspace{2 mm} Some College & 154 (34.3) & 18 (34.0) & 136 (34.3) & \\
     & \hspace{2 mm} Bachelor's Degree or Higher & 198 (44.1) & 19 (35.8) & 179 (45.2) & \\
     & & & & & \\
    2 & Pre-Pregnancy BMI ($\text{kg}/\text{m}^2$) & 26.1 (6.0) & 26.5 (7.2) & 26.0 (5.8) & 0.678 \\
     & Maternal Age (years) & 26.8 (5.6) & 26.3 (6.1) & 26.9 (5.5) & 0.525 \\
     & Education & & & & 0.634 \\
     & \hspace{2 mm} GED/Equivalent or Less & 97 (21.3) & 13 (25.0) & 84 (20.8) & \\
     & \hspace{2 mm} Some College & 157 (34.4) & 19  (36.5) & 138 (34.2) & \\
     & \hspace{2 mm} Bachelor's Degree or Higher & 202 (44.3) & 20 (38.5) & 182 (45.0) & \\
  \hline
\end{tabular}
}}
\end{center}
\caption{Descriptive Statistics for subset of the PROTECT Cohort with at least one of MBP, MiBP, and MBzP measured at visit X and eicosanoid measures at visit 3. Sample size at visit 1 is 449 total participants (396 full-term deliveries and 53 preterm deliveries). Sample size at visit 2 is 456 total participants (404 full-term deliveries and 52 preterm deliveries). P-values corresponding to differences between preterm and full-term deliveries for continuous and categorical variables come from t-tests and chi-squared tests, respectively.}
\end{table}

\newpage

\begin{table}[!htbp]
\begin{center}
{
\resizebox{\columnwidth}{!}{
\begin{tabular}{llll}
\hline
\textbf{Method} &\textbf{Internal TE Model} &  \textbf{External TE Model} & \textbf{Notes on Bias and Estimation Efficiency} \\ \hline
 Unconstrained & Correctly Specified & None & Estimators are unbiased, but they are also \\
 & & & the least efficient when $\theta_a^I$ and $\theta_a^E$ are close. \\
 & & & Use when $\theta_a^I$ and $\theta_a^E$ are known to be different. \\
 & & & \\
 Hard Constraint & Correctly Specified & Equation (4) Must Hold & Estimators may be asymptotically biased if $\theta_a^I \neq \theta_a^E$. \\
 & & & Most asymptotically efficient method when $\theta_a^I = \theta_a^E$. \\
 & & & Only consider using if $(\widehat{\theta}_a^E-\widehat{\theta}_a^I)^2 \leq \widehat{\text{Var}}(\widehat{\theta}_a^I)$. \\
 & & & \\
 Soft Constraint (EB) & Correctly Specified  & None & Estimators are minimally biased regardless of how \\
 & & & close $\theta_a^I$ and $\theta_a^E$ are. Asymptotically more efficient than \\
  & & & the unconstrained estimators when $\theta_a^I$ and $\theta_a^E$ are close. \\
   & & & As efficient as the unconstrained estimators when $\theta_a^I$ \\
   & & & and $\theta_a^E$ are substantially different. Less efficient than \\
    & & & the hard constraint estimators when $\theta_a^E$ and $\theta_a^I$ are close. \\
    & & & Use this method as the default method. \\
  \hline
\end{tabular}
}}
\end{center}
\caption{Summary of the methods presented in the paper and when to use each. TE, total effect.}
\end{table}

\newpage

\begin{figure}[!ht]
    \centering
    \includegraphics[scale=1.0, height = 0.9\textheight, width = 1.0\linewidth]{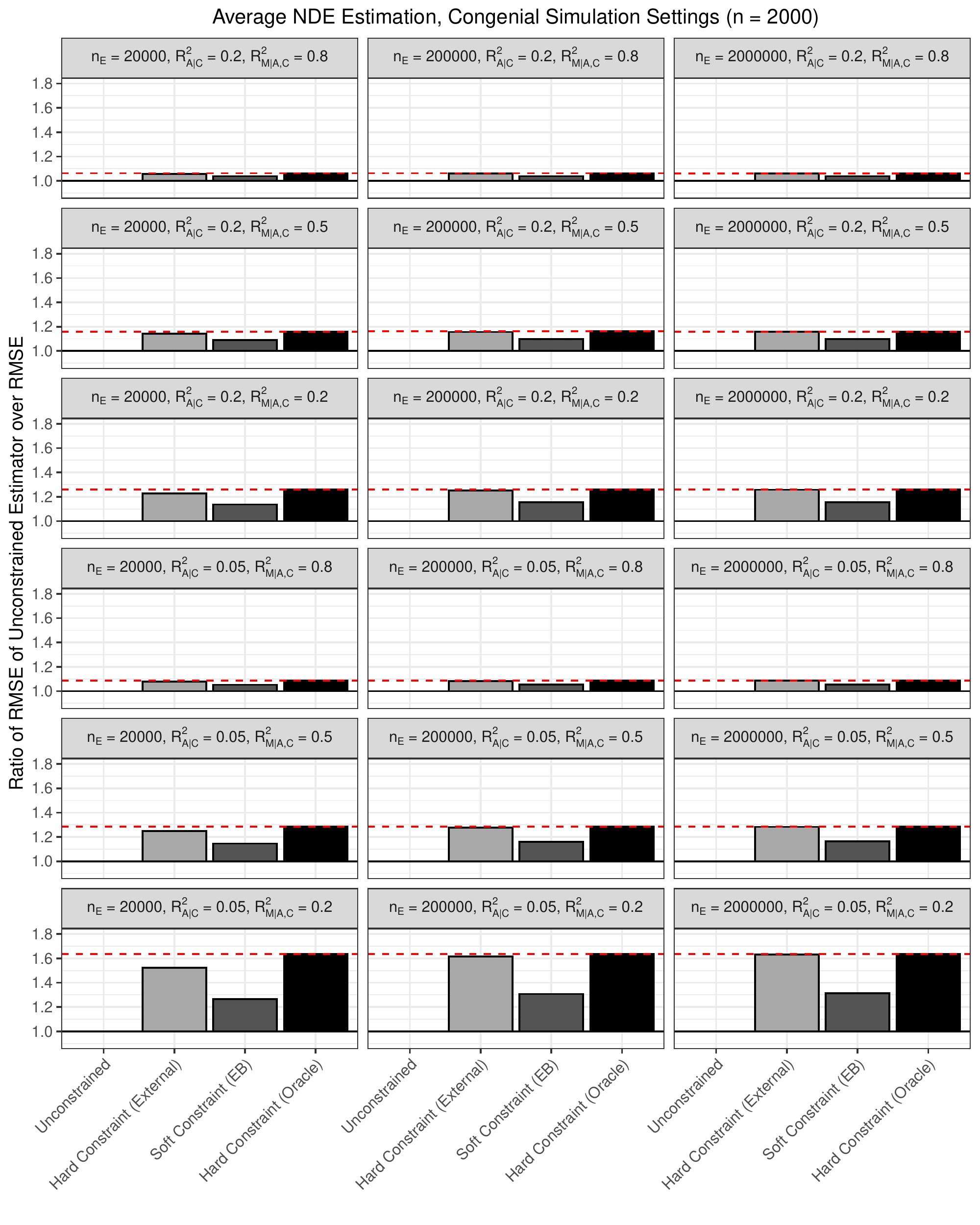}
    \caption{Relative root mean-squared error (RMSE) corresponding to Natural Direct Effect (NDE) estimation for the congenial simulation scenarios ($n = 2000$). The red, horizontal dashed line indicates the upper bound on the possible gain in estimation efficiency, as determined by the hard constraint estimator with the oracle constraint.}
    \label{fig:rmse_n2000_nde_unpenalized_correct}
\end{figure}

\newpage

\begin{figure}[!ht]
    \centering
    \includegraphics[scale=1.0, height = 0.9\textheight, width = 1.0\linewidth]{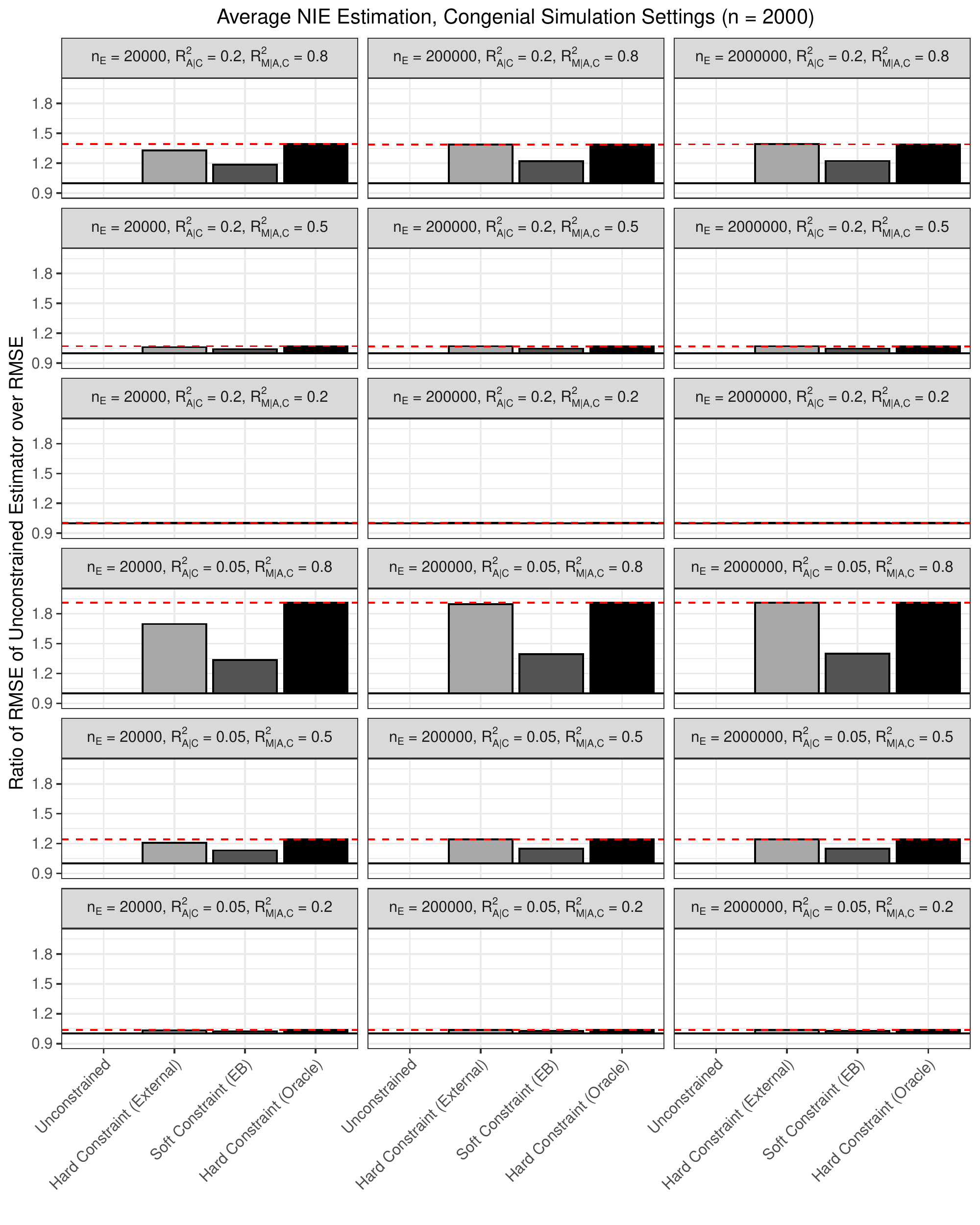}
    \caption{Relative root mean-squared error (RMSE) corresponding to Natural Indirect Effect (NIE) estimation for the congenial simulation scenarios ($n = 2000$). The red, horizontal dashed line indicates the upper bound on the possible gain in estimation efficiency, as determined by the hard constraint estimator with the oracle constraint.}
    \label{fig:rmse_n2000_nie_unpenalized_correct}
\end{figure}

\newpage

\begin{figure}[!ht]
    \centering
    \includegraphics[scale=1.0, height = 0.9\textheight, width = 1.0\linewidth]{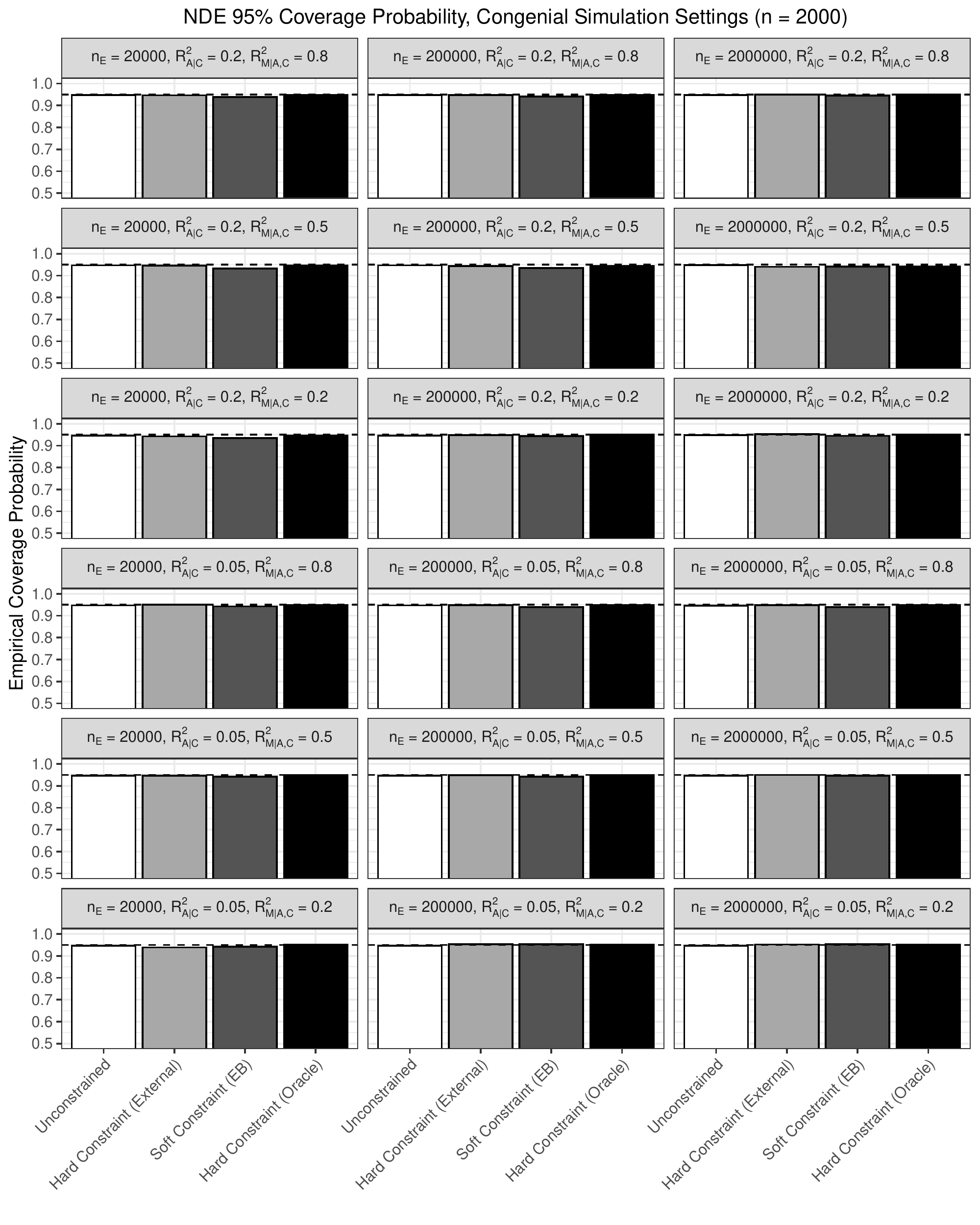}
    \caption{Empirical coverage probability corresponding to Natural Direct Effect (NDE) estimation for the congenial simulation scenarios ($n = 2000$). The horizontal dashed line indicates the nominal coverage rate of 0.95.}
    \label{fig:cp95_n2000_nde_unpenalized_correct}
\end{figure}

\newpage

\begin{figure}[!ht]
    \centering
    \includegraphics[scale=1.0, height = 0.9\textheight, width = 1.0\linewidth]{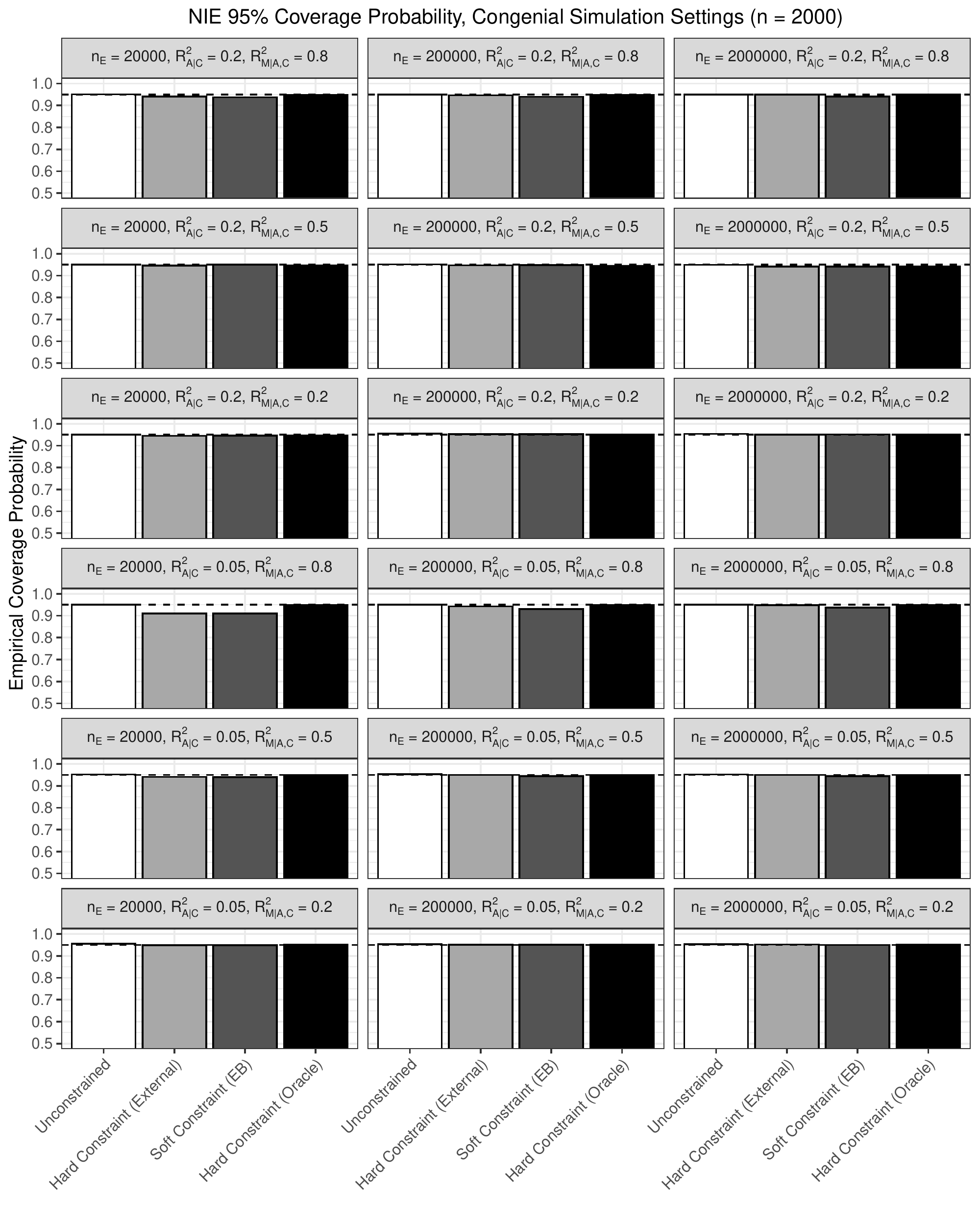}
    \caption{Empirical coverage probability corresponding to Natural Indirect Effect (NIE) estimation for the congenial simulation scenarios ($n = 2000$). The horizontal dashed line indicates the nominal coverage rate of 0.95.}
    \label{fig:cp95_n2000_nie_unpenalized_correct}
\end{figure}

\newpage

\begin{figure}[!ht]
    \centering
    \includegraphics[scale=1.0, height = 0.9\textheight, width = 1.0\linewidth]{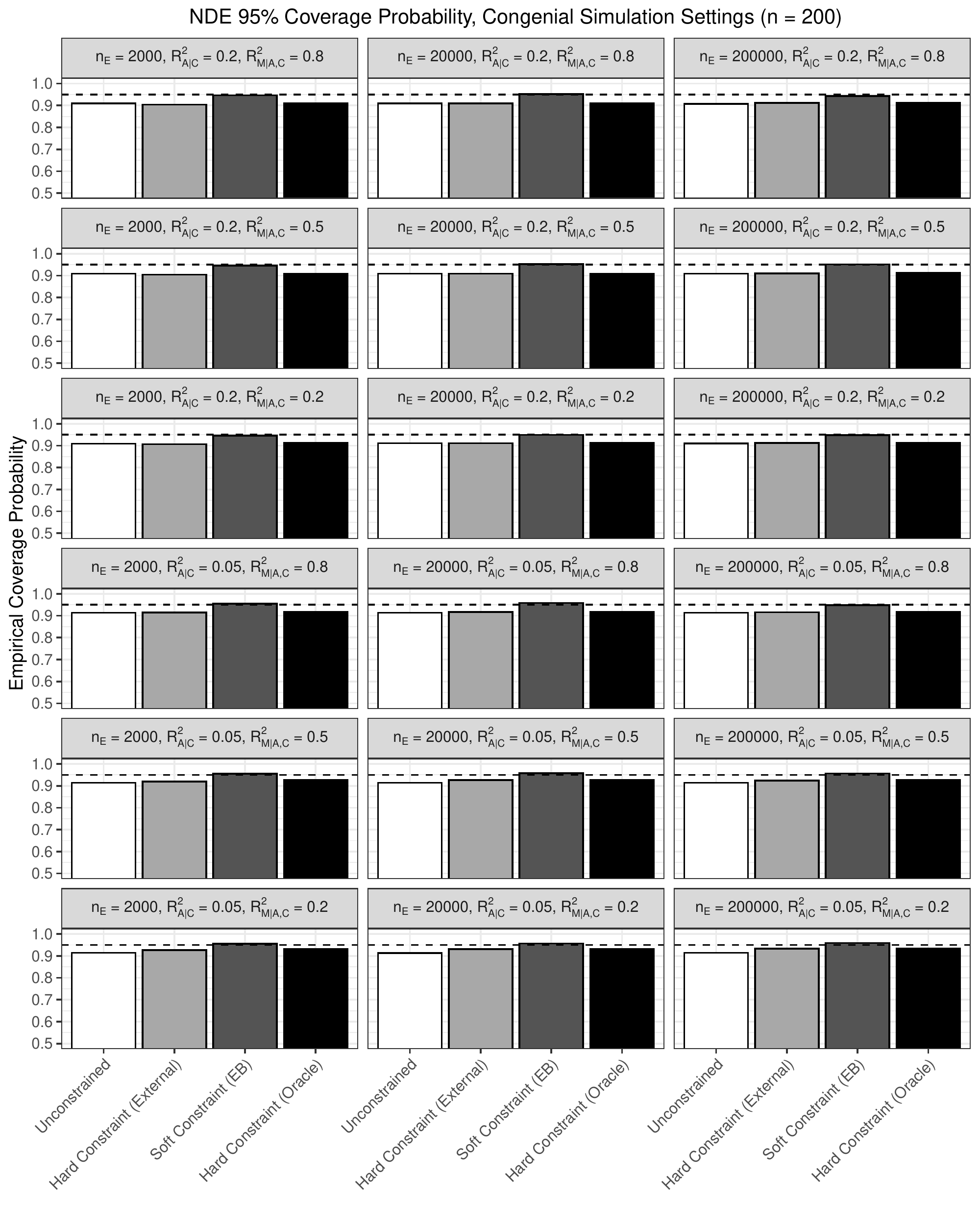}
    \caption{Empirical coverage probability corresponding to Natural Direct Effect (NDE) estimation for the congenial simulation scenarios ($n = 200$). The horizontal dashed line indicates the nominal coverage rate of 0.95.}
    \label{fig:cp95_n200_nde_unpenalized_correct}
\end{figure}

\newpage

\begin{figure}[!ht]
    \centering
    \includegraphics[scale=1.0, height = 0.9\textheight, width = 1.0\linewidth]{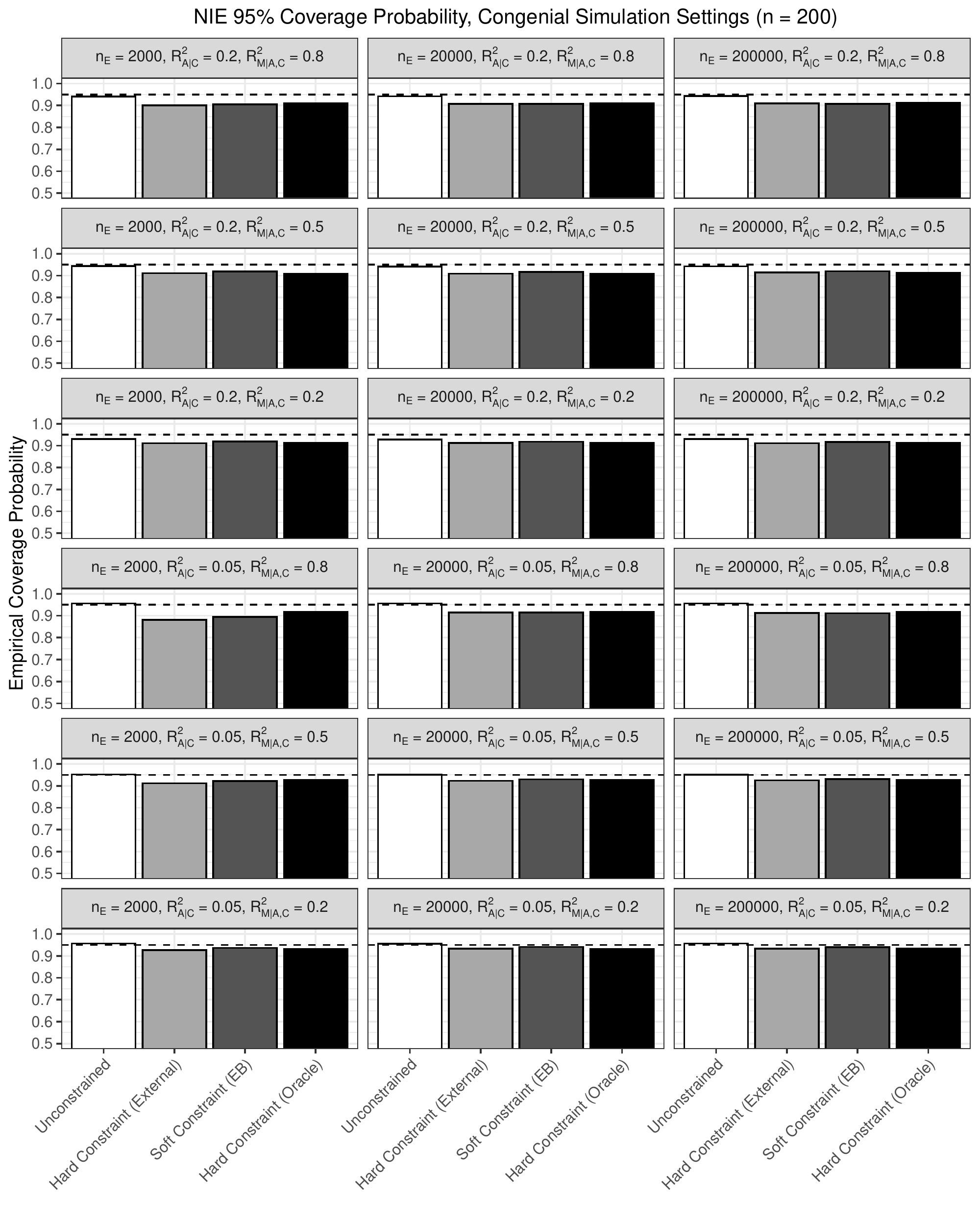}
    \caption{Empirical coverage probability corresponding to Natural Indirect Effect (NIE) estimation for the congenial simulation scenarios ($n = 200$). The horizontal dashed line indicates the nominal coverage rate of 0.95.}
    \label{fig:cp95_n200_nie_unpenalized_correct}
\end{figure}

\newpage

\begin{figure}[!ht]
    \centering
    \includegraphics[scale=1.0, height = 0.9\textheight, width = 1.0\linewidth]{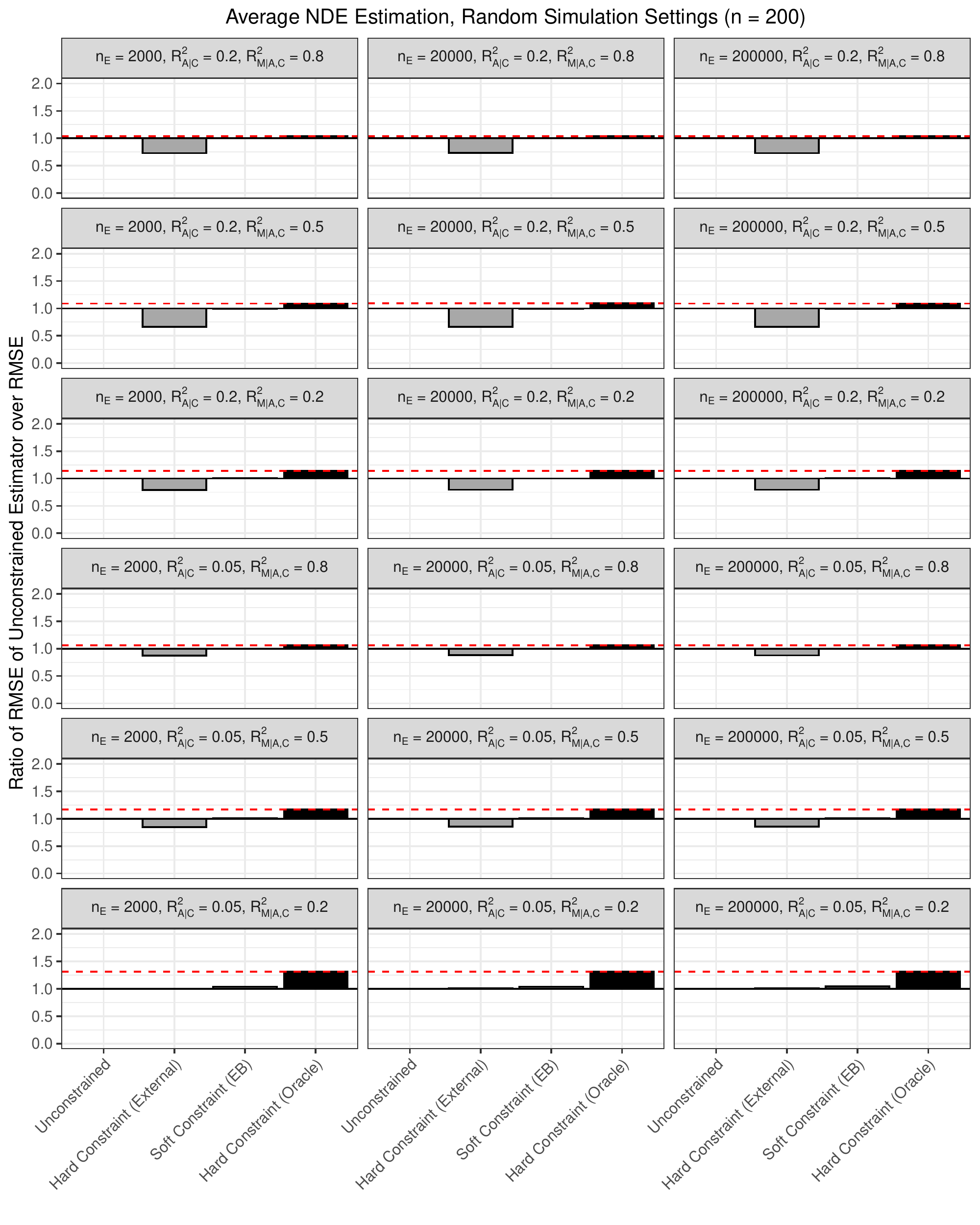}
    \caption{Relative root mean-squared error (RMSE) corresponding to Natural Direct Effect (NDE) estimation for the random simulation scenarios ($n = 200$). The red, horizontal dashed line indicates the upper bound on the possible gain in estimation efficiency, as determined by the hard constraint estimator with the oracle constraint.}
    \label{fig:rmse_n200_nde_unpenalized_incorrect_random}
\end{figure}

\begin{figure}[!ht]
    \centering
    \includegraphics[scale=1.0, height = 0.9\textheight, width = 1.0\linewidth]{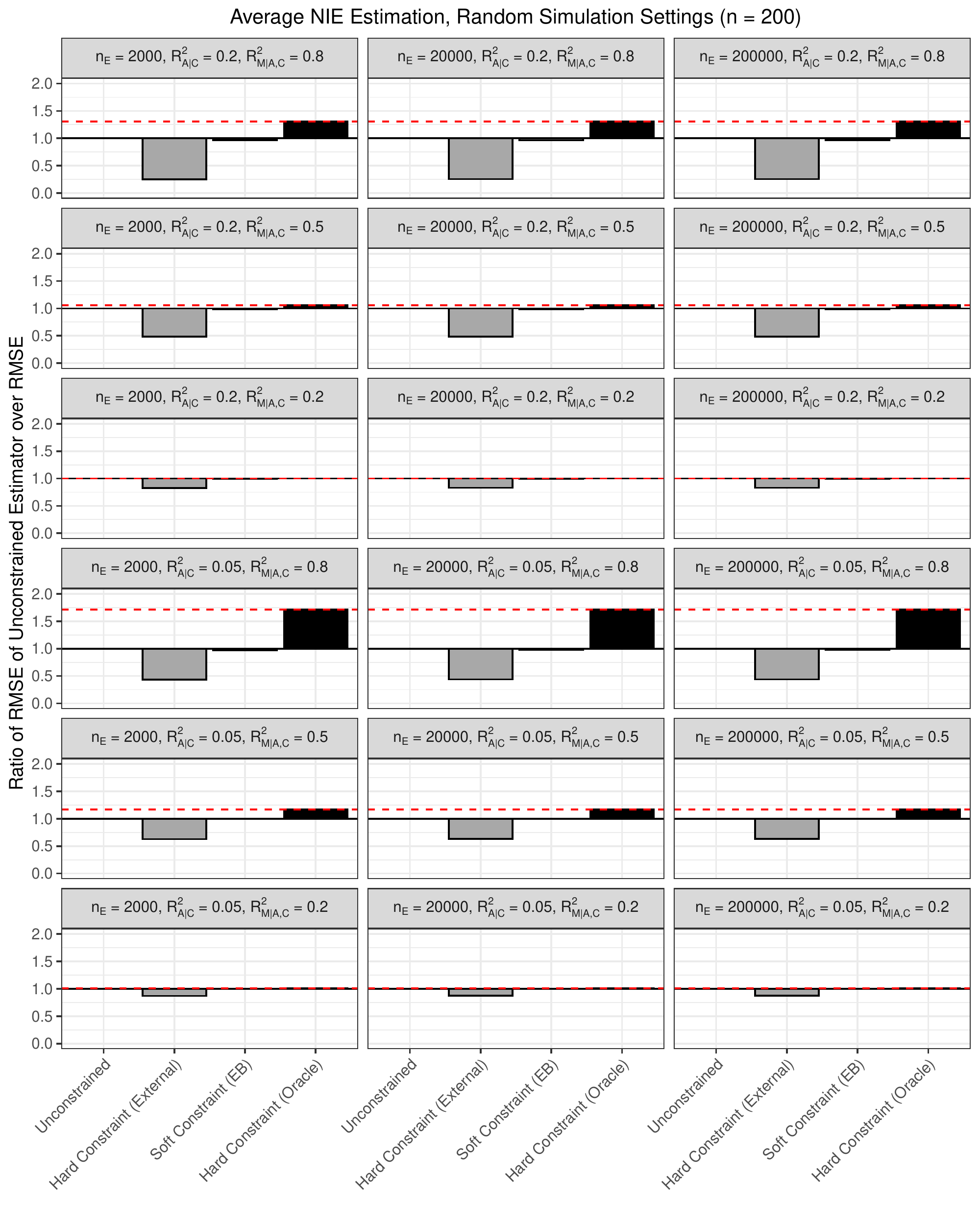}
    \caption{Relative root mean-squared error (RMSE) corresponding to Natural Indirect Effect (NIE) estimation for the random simulation scenarios ($n = 200$). The red, horizontal dashed line indicates the upper bound on the possible gain in estimation efficiency, as determined by the hard constraint estimator with the oracle constraint.}
    \label{fig:rmse_n200_nie_unpenalized_incorrect_random}
\end{figure}

\newpage

\begin{figure}[!ht]
    \centering
    \includegraphics[scale=1.0, height = 0.9\textheight, width = 1.0\linewidth]{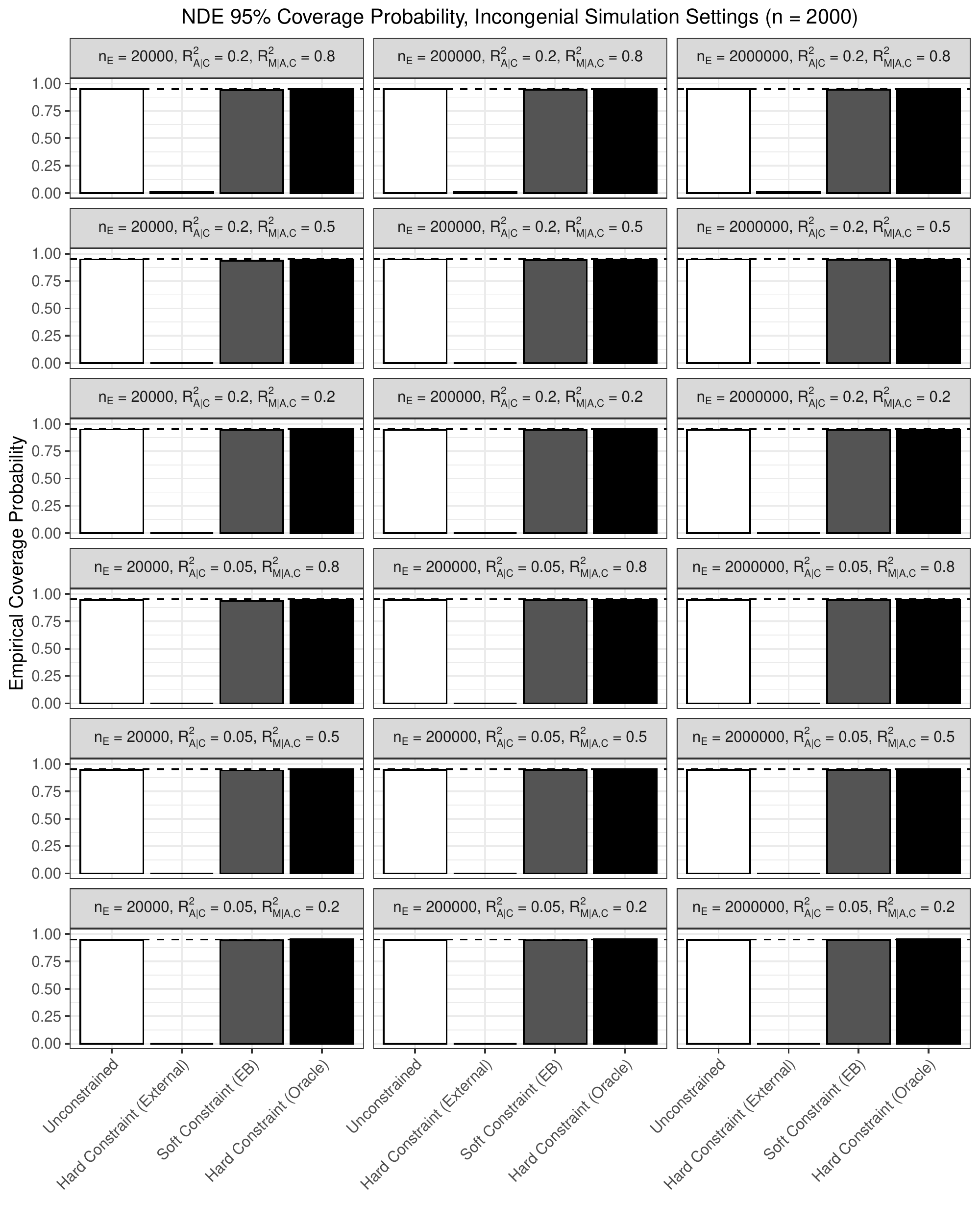}
    \caption{Empirical coverage probability corresponding to Natural Direct Effect (NDE) estimation for the incongenial simulation scenarios ($n = 2000$). The horizontal dashed line indicates the nominal coverage rate of 0.95.}
    \label{fig:cp95_n2000_nde_unpenalized_incorrect_fixed}
\end{figure}

\newpage

\begin{figure}[!ht]
    \centering
    \includegraphics[scale=1.0, height = 0.9\textheight, width = 1.0\linewidth]{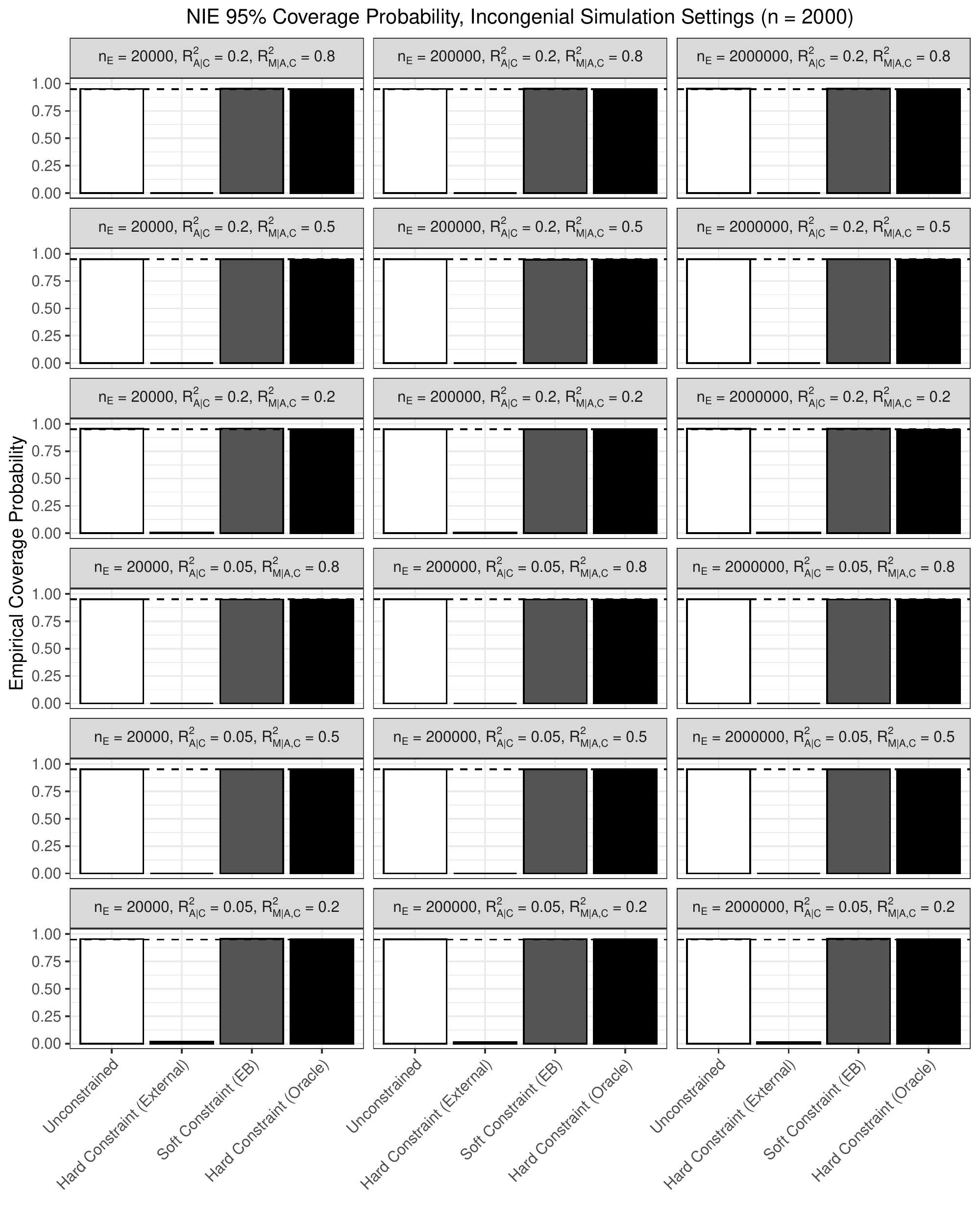}
    \caption{Empirical coverage probability corresponding to Natural Indirect Effect (NIE) estimation for the incongenial simulation scenarios ($n = 2000$). The horizontal dashed line indicates the nominal coverage rate of 0.95.}
    \label{fig:cp95_n2000_nie_unpenalized_incorrect_fixed}
\end{figure}

\newpage

\begin{figure}[!ht]
    \centering
    \includegraphics[scale=1.0, height = 0.9\textheight, width = 1.0\linewidth]{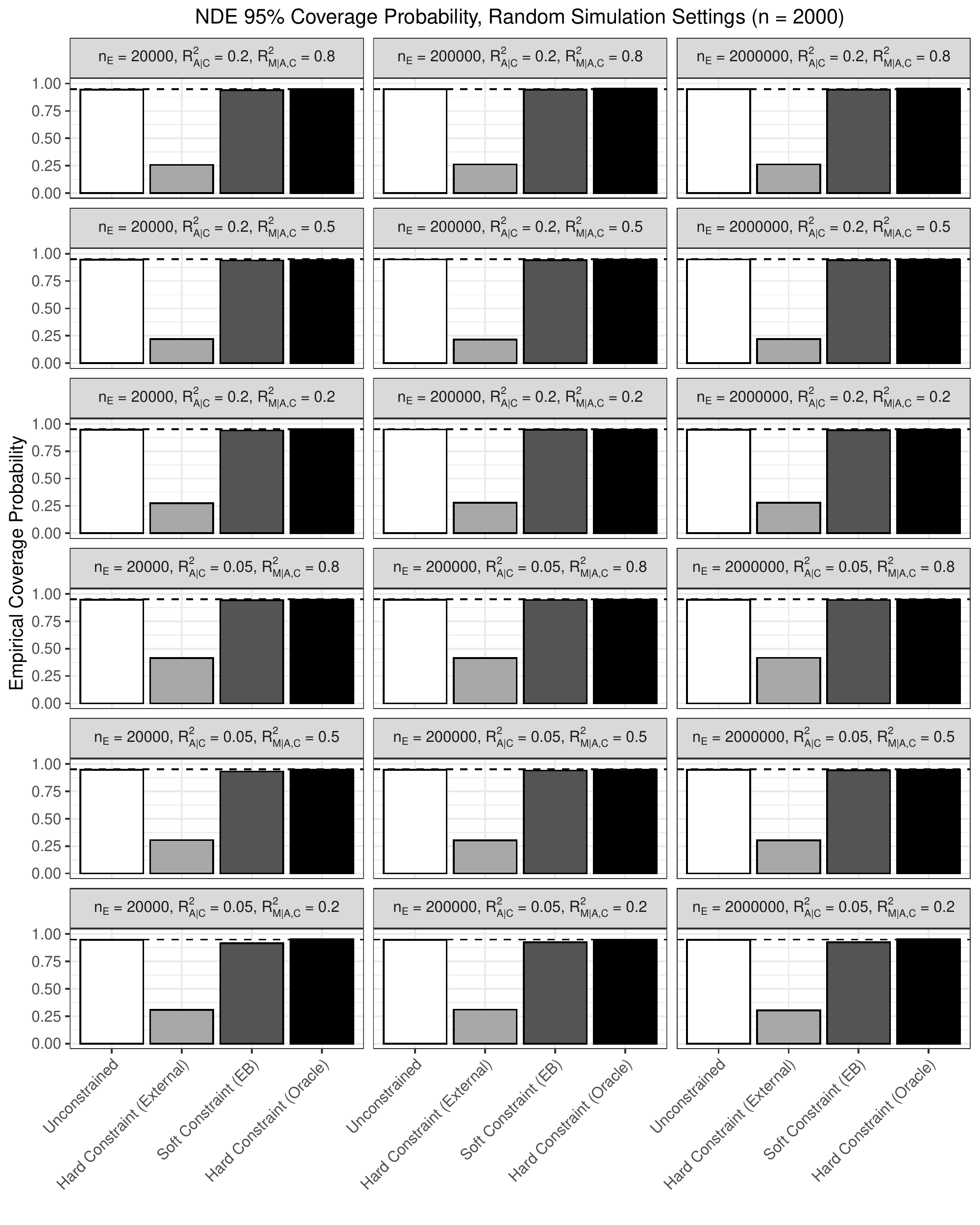}
    \caption{Empirical coverage probability corresponding to Natural Direct Effect (NDE) estimation for the random simulation scenarios ($n = 2000$). The horizontal dashed line indicates the nominal coverage rate of 0.95.}
    \label{fig:cp95_n2000_nde_unpenalized_incorrect_random}
\end{figure}

\newpage

\begin{figure}[!ht]
    \centering
    \includegraphics[scale=1.0, height = 0.9\textheight, width = 1.0\linewidth]{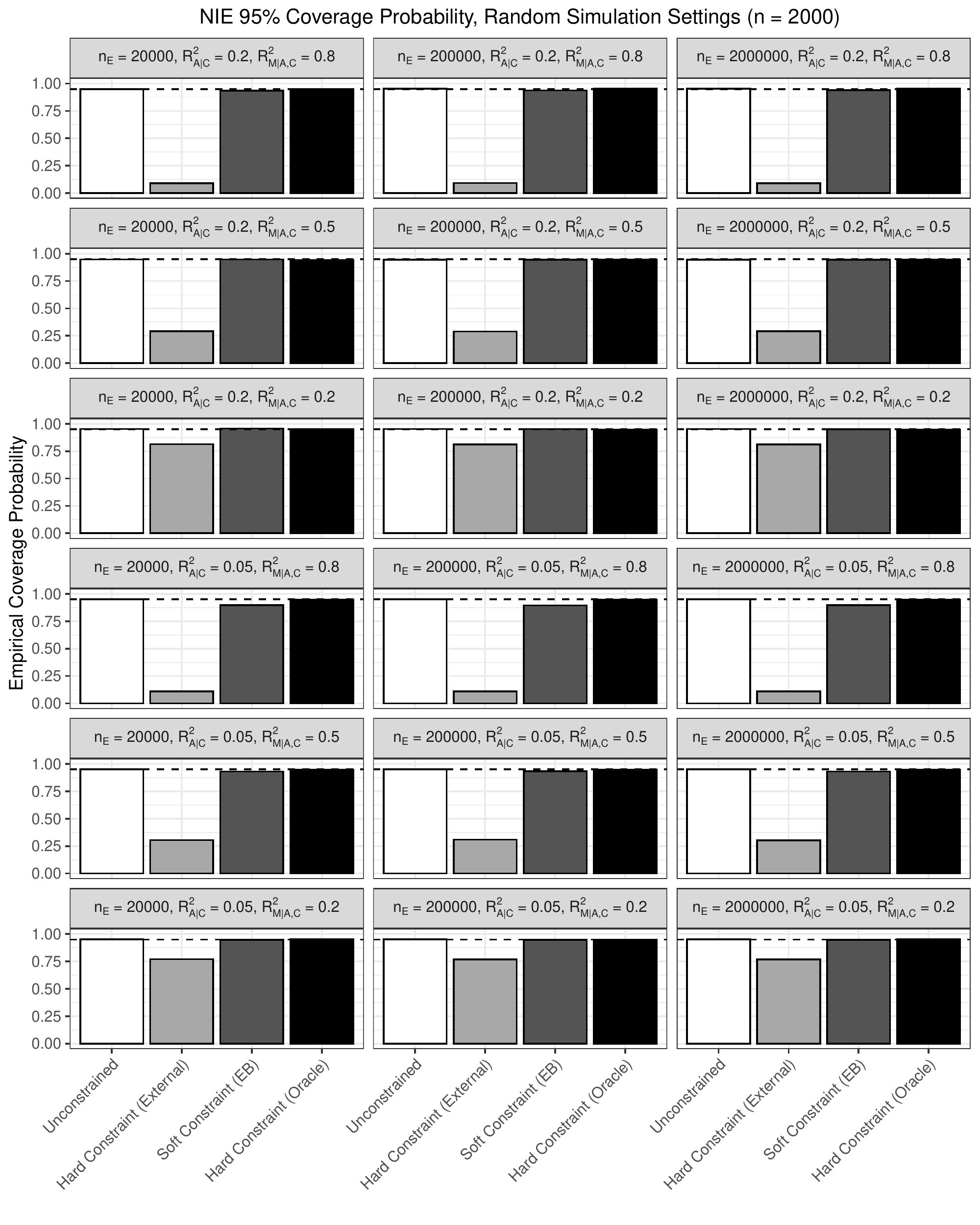}
    \caption{Empirical coverage probability corresponding to Natural Indirect Effect (NIE) estimation for the random simulation scenarios ($n = 2000$). The horizontal dashed line indicates the nominal coverage rate of 0.95.}
    \label{fig:cp95_n2000_nie_unpenalized_incorrect_random}
\end{figure}